\documentstyle[eqsecnum,psfig,aps,12pt]{revtex}

\def\beq{\begin{equation}}
\def\eeq{\end{equation}}

\def\bea{\arraycolsep .1em \begin{eqnarray}}
\def\eea{\end{eqnarray}}

\def\vp{{\bf p}}
\def\vx{{\bf x}}
\def\vX{{\bf X}}
\def\vy{{\bf y}}
\def\vv{{\bf v}}
\def\vk{{\bf k}}

\def\Tr{{\rm Tr}}

\def\grgl{\:\hbox to -0.2pt{\lower2.5pt\hbox{$\sim$}\hss}{\raise3pt\hbox{$>$}}\:}
\def\klgl{\:\hbox to -0.2pt{\lower2.5pt\hbox{$\sim$}\hss}{\raise3pt\hbox{$<$}}\:}
\def \lta {\mathrel{\vcenter
     {\hbox{$<$}\nointerlineskip\hbox{$\sim$}}}}

\let\de=\delta
\let\eps=\epsilon

\def\zeile{\\[.8ex]}
\def\step{}

\def\bpi{\mbox{\boldmath$\pi$}}
\def\bphi{\mbox{\boldmath$\phi$}}

\def\eq#1{(\ref{#1})}
\def\Eq#1{Eq.~(\ref{#1})}
\def\Eqs#1{Eqs.~(\ref{#1})}

\def\s0#1#2{\mbox{\small{$ \frac{#1}{#2} $}}}
\def\0#1#2{\frac{#1}{#2}}

\def\llangle{\left\langle}
\def\rrangle{\right\rangle}

\makeatletter
\renewcommand{\theequation}{\arabic{section}.\arabic{equation}}
\makeatother

\begin{document}

\thispagestyle{empty}

\begin{flushright}
{CERN-TH--2001--252}${}\,$ \\[10ex] 
\end{flushright}

\begin{center}

\mbox{\large \bf Semi-classical transport theory for non-Abelian plasmas} 

${\ }$\\[4ex]

{Daniel F. Litim and Cristina Manuel} \\
{\it Theory Division, CERN, CH-1211 Geneva 23.}

${\ }$\\[4ex]
 
{\small \bf Abstract}\\[2ex]
\begin{minipage}{14cm}{\small    
We review a semi-classical transport theory for non-Abelian
plasmas based on a classical picture of coloured point particles.
Within this formalism, kinetic equations for the mean particle
distribution, the mean fields and their fluctuations are obtained
using an ensemble-average in phase space. The framework permits
the integrating-out of fluctuations in a systematic manner. This
leads to the derivation of collision integrals, noise sources and
fluctuation-induced currents for the effective transport equations
of QCD. Consistency with the non-Abelian gauge symmetry is
established, and systematic approximation schemes are worked out.
In particular, the approach is applicable to both in- and
out-of-equilibrium plasmas.  The formalism is applied explicitly
to a hot and weakly coupled QCD plasma slightly out of
equilibrium. The physics related to Debye screening, Landau
damping or colour conductivity is deduced in a very simple manner.
Effective transport equations are computed to first and second
order in moments of the fluctuations. To first order, they
reproduce the seminal hard-thermal-loop effective theory. To
second order, the fluctuations induce collisions amongst the
quasi-particles, leading to a Langevin-type transport equation. A
complementary Langevin approach is discussed as well. Finally, we
show how the approach can be applied to dense quark matter
systems. In the normal phase, the corresponding kinetic equations
lead to the hard-dense-loop effective theory. At high density and
low temperature diquark condensates are formed, changing the
ground state of QCD. In the superconducting phase with two
massless quark flavours, a transport equation for coloured
excitations is given as well. Possible future applications are
outlined.\\[1ex] {\footnotesize PACS: 12.38.Mh, 11.10.Wx} \\[4ex]}
\end{minipage}

(submitted to Physics Reports)

\end{center}

\newpage

\pagestyle{plain}
\setcounter{page}{1}
\renewcommand{\thepage}{\roman{page}}

\tableofcontents

\newpage

\pagestyle{plain}
\setcounter{page}{1}
\renewcommand{\thepage}{\arabic{page}}

\section{Introduction}

\subsection{Motivation}

In recent years, there has been an increasing interest in the dynamics
of non-Abelian plasmas at both very high temperature and density.
One of the most spectacular predictions of quantum chromodynamics is
asymptotic freedom, which implies that quarks and gluons behave as
free particles in such extreme conditions because their coupling
becomes very weak at short distances. It is expected that a specific
state of matter with quarks and gluons unconfined -- the so-called
quark-gluon plasma -- can exist.  Many efforts for its experimental
detection in the core region of heavy-ion collisions will be made at
RHIC and LHC within the next years.

Other possible applications concern relativistic non-Abelian plasmas
in extreme cosmological and astrophysical conditions, like the
electro-weak plasma in the early universe, the physics of dense
neutron stars or the physics of supernovae explosions.  If
baryogenesis can be understood within an electroweak scenario, an
understanding of the physics of the electroweak plasma in the unbroken
phase is essential for a computation of the rate of baryon number
violation.  In different astrophysical settings, the density reached
can be such that the hadrons melt into their fundamental constituents,
which gives rise to a very rich phenomenology.

It is therefore mandatory to devise reliable and maniable theoretical
tools for a quantitative description of non-Abelian plasmas both in
and out-of equilibrium. While some progress has been achieved in the
recent years, we are still far away from having a satisfactory
understanding of the dynamics of non-Abelian plasmas.  There are
different approaches in the literature in studying non-Abelian
plasmas, ranging from thermal quantum field theory to transport
equations or lattice studies. Every approach has its advantages and
drawbacks, and each one appears to be suited to answer a specific
subset of questions. A quantum field theoretical description should be
able to describe all possible aspects of the plasma.  Most
applications have concerned to the weakly coupled plasma close to
equilibrium. But even there, the situation is complicated due to the
non-perturbative character of long-wavelength excitations. Lattice
simulations, which in principle can handle large gauge couplings, have
proven particularly successful for non-perturbative studies of static
quantities like equal-time correlation functions. However, it seems
very difficult to employ them for the dynamical case. Furthermore,
the standard Monte Carlo simulations of QCD with a
finite quark chemical potential fail.  A kinetic or transport theory
approach has proven most efficient for the computation of macroscopic
properties of the plasma, like transport coefficients such as
viscosities or conductivities. In turn, a direct evaluation of
transport coefficients in the framework of quantum field theory is
quite involved. Already the leading order result at weak coupling
requires the resummation of infinitely many perturbative loop diagrams,
an analysis which at present has only been performed for a scalar
theory.

A new semi-classical transport theory, based on a classical point
particle picture, has been introduced recently \cite{LM,LM2,LM3,LM4}.
There are several motivations for such a formalism. First of all, it
is expected that the main characteristics of non-Abelian plasmas can
be understood within simple semi-classical terms. Secondly, this
approach allows for the study of transport phenomena without the
difficulties inherent to a full quantum field theoretical analysis.
Finally, the formalism is even applicable to hot out-of-equilibrium
plasmas, which are of particular relevance for the
case of heavy ion collisions, or to dense quark matter in the
superconducting phase.

This article reviews in detail the conceptual framework for the
semi-classical approach. In addition, we discuss applications to
the case of a hot and weakly coupled non-Abelian plasma, and to dense
quark matter systems. In the remainder of the introduction, we review
the ideas behind the semi-classical approximation which are at the
basis of the present formalism. We also review the present
understanding for constructing effective theories of hot and weakly
coupled non-Abelian plasmas, where the formalism is finally put to
work. For an application to the physics of dense quark matter, we
briefly summarise the present understanding about superconducting
phases of matter at large baryonic density. A detailed outline of the
review is given at the end of the introduction.

\subsection{Semi-classical approach for hot plasmas}

There are several reasons for considering a semi-classical
approximation. A heuristic argument is that the occupation number per
mode vector $|\vp|$ of `soft' $(\hbar |\vp|\ll T)$ gauge fields in a
hot plasma at temperature $T$ is very high, due to the Bose-Einstein
enhancement. This suggests that the long wave-length limit corresponds
to the classical limit where the fundamental constant $\hbar$
vanishes. Therefore, one has reasons to believe that the soft {\it
  quantum} fields are, to leading order, well approximated by soft
{\it classical} ones. Such a reasoning has been substantiated by
various workers in the field (see \cite{Bodeker:2001pa} for a recent
review). On the other hand, the `hard' modes of the plasma cannot be
approximated this way as their occupation number is of order unity.
However, it has been established that weakly coupled hard modes behave
like quasi-particles \cite{Blaizot:2001nr}. They can be described, to
leading order in a gradient expansion, by an ensemble of coloured {\it
  classical point particles} moving on world lines. Therefore it is
conceivable that the main characteristics of such plasmas can be
understood within a purely semi-classical language.

For QED plasmas, semi-classical methods have been known and applied
for a long while \cite{Landau5,K}. They consist in describing the
charged constituents of the plasma as classical point particles moving
on world lines. Their interactions are determined self-consistently
through the Maxwell equations induced by the current of the particles.
On the mean field level, the resulting Boltzmann equation for the
one-particle distribution function is known as the Vlasov equation.
Beyond the mean field approximation, several approaches for the
construction of a full kinetic theory for such plasmas are known in
the literature, the most famous one being the BBGKY hierarchy for
correlator functions within non-relativistic statistical mechanics
\cite{Balescu1975}.  Alternative approaches have been put forward as
well. Of particular interest is the approach by Klimontovitch, who
constructed a kinetic theory on the basis of the one-particle
distribution function and the correlators of fluctuations about them
\cite{K}. This leads naturally to a description in terms of mean
fields and fluctuations. Conceptually, the new ingredient in his
approach is that the plasma is {\it not} considered as a continuous
medium. Instead, the stochastic fluctuations of the particles are
taken into account, and the dissipative character of effective
long-range interactions enters naturally in this framework. The
procedure leads to a Boltzmann-Langevin type of effective transport
equations, which, on phenomenological grounds, have been already
proposed for non-charged particles by Bixon and Zwanzig
\cite{Bixon1969}. This approach has been extended for the study of
nuclear collisions \cite{Randrup:1990bb,Ayik:1990bb,Chomaz:1991yn}
(see \cite{Abe:1996yw} for a recent review of applications to nuclear
dynamics). It also appeared that systematic approximations within the
Klimontovitch approach are better behaved than those based on the
BBGKY hierarchy \cite{K}.  Finally, this approach permits the
derivation of collision integrals, like the Balescu-Lenard one for
Abelian plasmas \cite{Landau10,K}.

For non-Abelian plasmas, semi-classical transport equations can be
obtained in essentially two distinct manners. The first one starts
from a quantum field theoretical framework, which is used to construct
a (quantum) transport theory, for example in terms of Wigner
functions, or for hierarchies of Schwinger-Dyson equations
\cite{Kadanoff1962,Balescu1975,Elze:1989un}. The semi-classical
approximation is performed in a second step on the level of the
transport theory, that is, on the `macroscopic' level (see, for
example \cite{Elze:1989un,Mrowczynski:1989np,Blaizot:1993zk}).
Alternatively, one may perform the semi-classical approximation
already on the `microscopic' level, in analogy to the Abelian case
outlined above. In this case the concept of an ensemble of classical
{\it coloured} point particles moving on world lines has to be
invoked. The new ingredients are the $SU(N)$ colour charges of the
particles. Their classical equations of motion for high dimensional
representations were first given by Wong \cite{Wong} and can be
understood as equations of motion for expectation values of quantum
wave packets, as shown by Brown and Weisberger \cite{Brown:1979bv}.

A transport equation based on the classical point particle
approximation has been given by Heinz
\cite{Heinz:1983nx,Heinz:1984my,Heinz:1985yq,Heinz:1986qe}. It
consists in a Boltzmann equation for a one-particle distribution
function for gluons, quarks or anti-quarks with a -- yet unspecified
-- collision term. On the mean field level, neglecting collision
terms, these equations are known as the non-Abelian Vlasov equations.
These are intimately linked to a gradient expansion of the Wigner
transform, as pointed out by Winter \cite{Winter}, and to the quantum
Boltzmann equation, as discussed by Elze and Heinz \cite{Elze:1989un}.
An important step forward in the understanding of the semi-classical
transport theory has been achieved by Kelly et al.~\cite{KLLM,KLLM2}.
They noticed that a gauge-consistent solution of non-Abelian Vlasov
equations to leading order in the gauge coupling reproduces precisely
the hard thermal loop effective kinetic theory. In addition, the
authors clarified the role of the non-Abelian colour charges as
non-canonical phase space variables and their explicit link to
canonical Darboux variables. This formalism has also been applied to
magnetic screening \cite{Manuel:1998is}, and to cold dense plasmas
\cite{Manuel:1996td} which are characterised by a large chemical
potential.

The effects of non-Abelian fluctuations have to be considered to go
beyond the Vlasov approximation. In the context of QCD transport
theory this was pointed out by Selikhov, who, motivated by the earlier
work of Klimontovitch, derived a collision term of the Balescu-Lenard
type for non-Abelian Boltzmann equations \cite{Selikhov}. This method,
applied by Selikhov and Gyulassy to the problem of colour
conductivity, uncovered a logarithmic sensitivity of the colour
relaxation time scale \cite{Selikhov2,SG}. However, in their
considerations only the local part of the corresponding collision term
has been identified, which implies that the corresponding colour
current is not covariantly conserved. Along similar lines, Markov and
Markova applied the procedure of Klimontovitch to a classical
non-Abelian plasma and formally derived a Balescu-Lenard collision
integral \cite{Markov}. The strategy is similar to Selikhov's
approach, except that it embarks from a purely classical starting
point. This approach overlooked the important point that the colour
charges are non-standard phase space variables, which is crucial for a
definition of an ensemble average. Also, neither the non-linear
higher-order effects due to the non-Abelian interactions have been
considered, nor the requirements implied due to gauge symmetry.

A fully self-contained approach, aimed at filling this gap in the
literature of classical non-Abelian plasmas was presented recently and
is the subject of the review
\cite{LM,LM2,LM3,LM4,Litim:Habil,Litim:2000uj,Litim:2001je}. It is
based on a classical point particle picture and uses the Klimontovitch
procedure, extended to the non-Abelian case, to describe non-Abelian
fluctuations. The essential contribution is considering the
non-Abelian colour charges as dynamical variables and introducing the
concept of ensemble average to the non-Abelian kinetic equations.
Equally important is the consistent treatment of the intrinsic
non-linearities of non-Abelian gauge interactions. The fundamental
role of fluctuations in the quasi-particle distribution function has
been worked out, and results in a recipe as to how effective
semi-classical transport equations can be {\it derived} in a
systematic manner. This set of coupled dynamical equations for mean
fields and correlators of fluctuations should be enough to consider
all transport phenomena in the plasma, at least in the domain where
the underlying point particle picture is applicable. This procedure
could even be applicable for out-of-equilibrium situations, since the
derivation of the transport equation does not depend on the system
being in equilibrium or not. Although we are not applying the
formalism to plasmas out-of-equilibrium, this observation could open a
door for interesting further applications in a domain relevant for
future experiments. It would be very interesting to investigate
out-of-equilibrium situations and plasma instabilities within the
present transport theory.

\subsection{Hot QCD plasmas}\label{IntroWeak}

The case of a hot and weakly coupled non-Abelian plasma close to
thermal equilibrium has already proven quite rich and involved in
structure due to the non-perturbative character of the long-wavelength
fluctuations \cite{Linde:1980tu,Gross}. We shall apply the
aforementioned formalism in detail to the weakly coupled plasma to
show how the physics related to Debye screening, Landau damping and
colour relaxation can be understood very efficiently within this
simple semi-classical framework.

Let us briefly summarise the present status for constructing effective
(transport) theories for hot and weakly coupled non-Abelian plasmas
close to thermal equilibrium at temperature $T$ (see Fig.~1). The
physics for the whole range of momentum modes of the non-Abelian
fields requires the framework of thermal QCD
\cite{Kapusta:1989tk,LeBellac:1996,Smilga:1997cm}. {\it Effective}
theories for the long-range modes are obtained when high-momentum
modes are integrated-out. There is one conceptual limitation for the
use of perturbative methods due to the non-perturbative magnetic
sector of QCD, which corresponds to momentum scales about $\sim g^2T$,
the magnetic mass scale. The interactions of modes with smaller
momenta are strictly non-perturbative \cite{Gross}, hence, any
perturbative scheme for integrating-out modes with momenta $\sim
|\vp|$ relies on $\sim g^2T/|\vp|$ as the effective expansion
parameter \cite{Bodeker:1998hm}.

The first step towards obtaining an effective theory for the long
wave-length excitations has been made by Pisarski
\cite{Pisarski:1989vd} and Braaten and Pisarski
\cite{Braaten:1990mz,Braaten:1990az}. Standard thermal perturbation
theory is plagued by severe infrared divergences due to massless
modes. Braaten and Pisarski proposed the resummation of all 1-loop
diagrams with hard internal momenta and soft external ones, the
seminal Hard Thermal Loops (HTL). Here, `hard' refers to momenta of
the order of the temperature $|\vp|\sim T$. We denote momenta with
$|\vp|\sim gT$ as `soft' (sometimes also referred to as `semi-hard' in
the literature). The HTL-resummed gluon propagator has its poles not
on the light cone and the dispersion relation yields, apart from a
complicated momentum dependence, the Debye (screening) mass $\sim gT$
for the chromo-electric fields. The HTL polarisation tensor also has
an imaginary part, which describes the emission and absorption of soft
gluons by the hard modes, known as Landau damping.  The resulting
effective theory for the soft modes contains highly non-local
interactions in space and time, induced as HTL corrections to the
propagator and to the vertices. It leads to gauge-invariant results
for physical observables like the soft gluon damping rate
\cite{Braaten:1990kk,Braaten:1990it}.  An effective action for the
HTLs was given by Taylor and Wong \cite{Taylor:1990ia}. Further
aspects of the HTL effective theory, for example their link to
Chern-Simons theory \cite{Efraty:1992gk,Efraty:1993pd,Jackiw:1993zr},
to Wess-Zumino-Novikov-Witten actions \cite{Nair:1993rx} and their
Hamiltonian structure \cite{Nair:1994xs} have been considered
subsequently.

A local formulation of the HTL effective theory was given by Blaizot
and Iancu \cite{Blaizot:1994be,Blaizot:1994da,Blaizot:1994am} and by
Nair \cite{Nair:1993rx,Nair:1994xs}. Blaizot and Iancu managed to
reformulate the HTL effective theory within the language of kinetic
theory. Such equations are similar to those for the HTL effective
theory of QED plasmas as considered by Silin \cite{Silin:1960}. The
derivation has been achieved invoking a truncation to a
Schwinger-Dyson hierarchy. This has lead to a transport equation for
the distribution function describing the hard or particle-like degrees
of freedom. The advantage of a kinetic description is that the {\it
  non-local} interactions in the HTL effective theory are replaced by
a {\it local} transport theory.  The crucial step is to consider the
quasi-particle distribution function as independent degrees of
freedom, describing the hard excitations of the plasma.  This also has
lead to a local expression for the HTL energy in terms of the soft
gauge fields and the colour current density
\cite{Nair:1993rx,Blaizot:1994am}. It is worthwhile pointing out that
the HTL effective theory can be derived within a semi-classical
transport theory based on a point particle picture \cite{KLLM,KLLM2}.

\begin{figure}[t]
\begin{center}
\unitlength0.001\hsize
\begin{picture}(800,550)
\put(0,520){\large $|{\bf p}|$}
\psfig{file=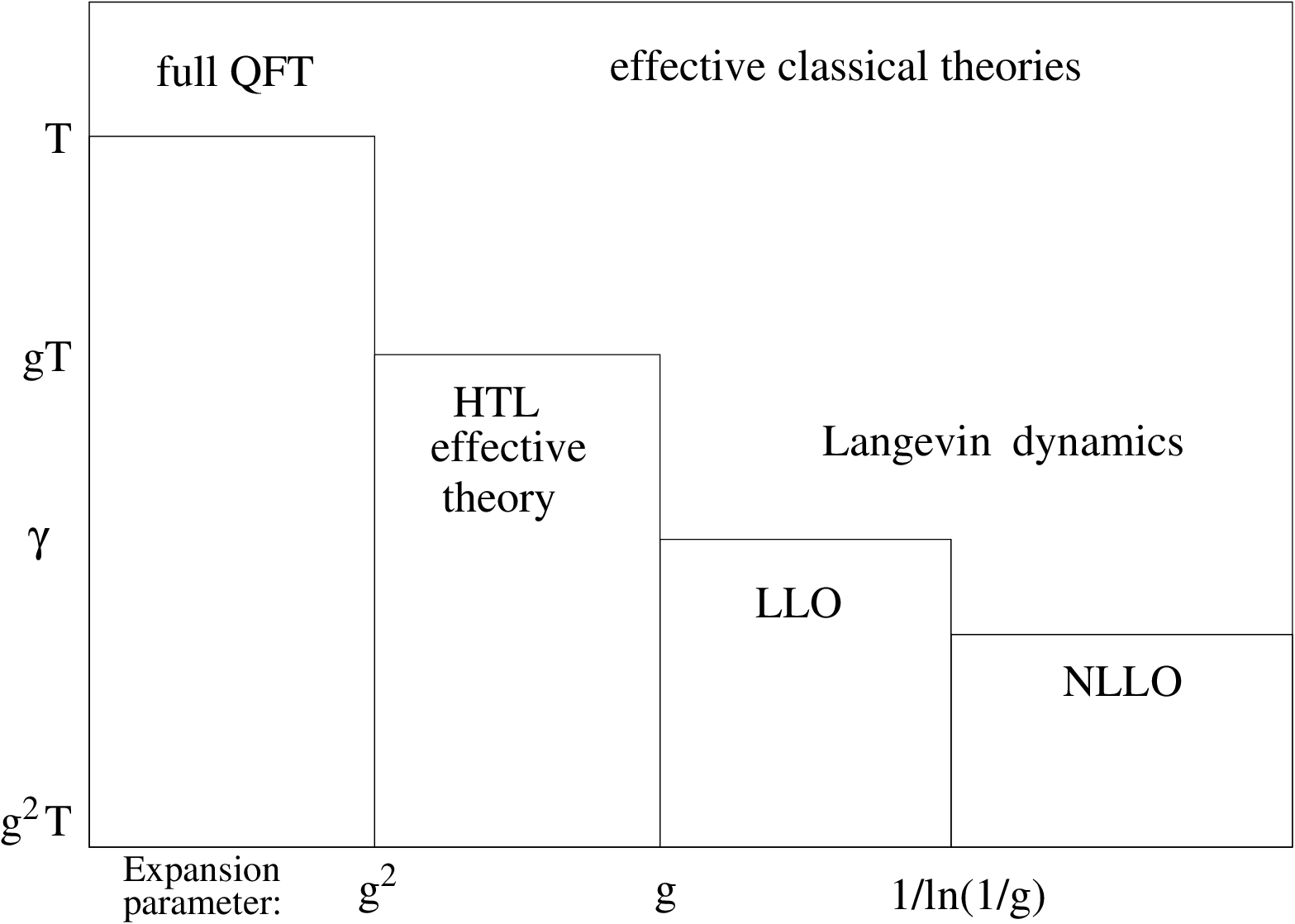,width=.75\hsize}
\end{picture}
\vskip.8cm
\begin{minipage}{\hsize}
  {\label{Fig1}\small {\bf Figure 1:} Schematic diagram for the series
    of effective theories for hot and weakly coupled non-Abelian
    plasmas close to thermal equilibrium. The physics of the hard
    modes with momenta $|\vp|\sim T$ or larger needs the full thermal
    QCD.  Effective {\it classical} theories are found for modes
    $|\vp|\ll T$.  The hard-thermal-loop (HTL) effective theory
    integrates-out the hard modes, and is effective for modes at about
    the Debye mass, $|\vp|\sim gT$. It can be written as a
    collisionless Boltzmann equation. The effective expansion
    parameter is $\sim g^2$. A collisional Boltzmann equation is found
    after integrating-out the modes $|\vp|\sim gT$ to leading
    logarithmic order (LLO), which is an expansion in $g$. The
    effective theory for the ultra-soft gauge fields with spatial
    momenta $|\vp|\ll gT$ is a Langevin-type dynamical equation. The
    next step integrates-out the modes with $|\vp|\sim \gamma\sim
    g^2T\ln(1/g)$, which is an expansion in $1/\ln(1/g)$ and yields
    next-to-leading-logarithmic order (NLLO) corrections without
    changing the qualitative form of the effective theory
    \cite{Bodeker:1999}.}
\end{minipage}
\end{center}
\end{figure}

Some attempts have been made for obtaining effective theories beyond
the HTL approximation \cite{Huet:1997sh,Arnold:1998gh}. Eventually,
B\"odeker showed the way how the Debye scale can be integrated-out
within a quantum field-theoretical framework
\cite{Bodeker:1998hm,Bodeker:2000ud,Bodeker:1999ey}.
His derivation relied on a semi-classical approximation which
treats the soft modes as classical fields, and made use of the local
expression for the HTL effective energy which allowed him to define a
weight function and to perform classical thermal averages over initial
conditions. This procedure has also been understood as an appropriate
resummation of certain classes of Feynman diagrams
\cite{Bodeker:2000ud,Blaizot:2000fq}. To leading logarithmic order
(LLO), the resulting effective theory corresponds to a Langevin-type
Boltzmann equation, including a collision term and a source for
stochastic noise. The physics behind it describes the damping of
colour excitations due to their scattering with the hard particles in
the plasma. A dynamical scale $\gamma$ appeared, which corresponds to
the damping rate of hard gluons in the plasma. It is of the order of
$\sim g^2T\ln(1/g)$.

These findings initiated further developments in the field. Arnold,
Son and Yaffe \cite{Arnold:1999cy,Arnold} interpreted the kinetic
equation in terms of Lenz Law and gave an alternative derivation of
B\"odeker's collision term and the related noise source. The very same
effective kinetic theory has also been obtained within the
semi-classical approach which will be discussed in the present article
\cite{LM,LM2}. Subsequently, and making use of an additional
fluctuation-dissipation relation, Valle-Basagoiti presented an
equivalent set of transport equations \cite{Valle}. Finally, Blaizot
and Iancu extended their earlier work to higher order and derived the
collision term from a truncated Schwinger-Dyson hierarchy
\cite{Blaizot:1999xk}.

The Boltzmann-Langevin equation, when solved to leading order in the
overdamped limit $p_0\ll |\vp|\ll \gamma$ results in a very simple
Langevin equation for the ultra-soft gauge fields only
\cite{Bodeker:1999ey}. This effective theory is, quite remarkably,
ultra-violet finite \cite{Arnold:1999cy} and has been used for
numerical simulations to determine the hot sphaleron transition rate
\cite{Moore:2000zk}. Some consideration beyond LLO have been made in
\cite{Bodeker:2000ud,Blaizot:2000fq}. A non-local Langevin equation,
valid to leading order in $g$ and to all orders in $\ln(1/g)$ has been
given by Arnold \cite{Arnold:1999uz}. It is valid for frequencies
$p_0\sim g^4T$, and has been used by Arnold and Yaffe
\cite{Arnold:1999ux,Arnold:1999uy} to push the computation of the
colour conductivity to the next-to-leading logarithmic order (NLLO).
They made use of the stochastic quantisation method
\cite{Zinn-Justin:1989mi} to convert the stochastic dynamical
equations into path integrals \cite{Arnold:1999uz}, which can be
treated with standard techniques.  It is interesting that the
effective theory for the gauge fields remains of the Langevin-type
even at NLLO.  A local Langevin equation, valid for frequencies
$p_0\lta g^2T$, has been given by B\"odeker \cite{Bodeker:2000da}. It
is expected that the UV divergences are local as well, in contrast to
those associated to the original Boltzmann-Langevin equation
\cite{Bodeker:1998hm}.

Thus, it is fair to say that the physics related to Debye screening,
Landau damping, and colour relaxation in the close-to-equilibrium
plasma is by now well understood. A variety of different and
complementary approaches have lead to identical effective transport
equations. All approaches made use of some semi-classical
approximation in the course of their considerations.  Interestingly,
these characteristics of a hot plasma can be understood within a
simple semi-classical language straight away.

\subsection{Dense quark matter}\label{IntroWeakDense}

So far we have discussed effective theories for the long distance
physics of QCD plasmas at very high temperature. It is also
interesting to consider situations when the baryonic density is very
high, while the temperature is low. Usually such a QCD plasma is
called dense quark matter, or simply quark matter. This state of
matter could be realized in various astrophysical settings, such as in
the core of neutron stars, collapsing stars and supernova explosions.
In the presence of strange quarks, and at zero pressure, quark matter
might even be stable, in which case quark stars could exist in nature.

It has been known for a long time that cold dense quark matter should
exhibit the phenomenon of colour superconductivity
\cite{Barrois:1977xd,Barrois:1979pv,Bailin:1984bm}.  The present
microscopic understanding of colour superconductivity relies on
techniques of BCS theory \cite{Schrieffer} adapted to dense quark
matter (see Refs.~\cite{Rajagopal:2000wf,Alford:2001dt} for recent
reviews and related literature).  At asymptotically large baryonic
densities, and because of asymptotic freedom, the strong gauge
coupling constant becomes small. Then, the relevant degrees of freedom
of the system are those of quarks, filling up their corresponding
Fermi seas up to the value of the Fermi energy $E_F = \mu$, where
$\mu$ is the quark chemical potential. These highly degenerate
fermionic systems are very unstable to attractive interactions.  In
dense quark matter the attractive interaction among quarks is provided
by one-gluon exchange in an antitriplet colour channel. This leads to
the formation of diquark condensates, in analogy to the Cooper pairs
of electromagnetic superconductors.

In QCD a diquark condensate cannot be colour neutral, and thus the
colour symmetry is spontaneously broken and gluons acquire a mass
through the Anderson-Higgs mechanism. Because the gauge symmetry is
$SU_c(3)$ and there are some flavour symmetries for massless or light
quarks, the possible patterns of symmetry breaking are richer than in
the Abelian case. One finds different phases of quark matter,
according to the number of quark flavours that participate in the
condensation. The form of the diquark condensate is dictated by
Pauli's principle and by the fact that it should minimise the free
energy of the system.  The different colour superconducting phases of
quark matter are characterised by the total or partial breaking of the
gauge group, and by the possible existence of Nambu-Goldstone modes
associated to the breaking of the global symmetries. Using standard
techniques of BCS theory \cite{Schrieffer} it is possible to study the
microscopic properties of the colour superconductor in the weak
coupling regime.

While our knowledge of the microscopic behaviour of quark matter is
increasing, little is known about its macroscopic behaviour. Is quark
matter a dissipative or dissipativeless system? Is it such a good heat
and electricity conductor as an electromagnetic superconductor? To
answer those questions, it is mandatory to compute the transport
coefficients of quark matter in their different possible phases.
Kinetic theory provides the perfect framework for such a computation.

A classical transport equation for the gapped quasiparticles of a
two-flavour colour superconductor has been proposed recently
\cite{Litim:2001je}. When the temperature is increased,
thereby melting the diquark condensates, the transport equation
reduces to the classical transport equation valid for a non-Abelian
plasma in the unbroken phase. In the close-to-equilibrium case and in
the Vlasov approximation, the leading-order solution to the transport
equation reproduces the one-loop gluon polarisation tensor for small
gluon energy and momenta, as found within a quantum field theoretical
computation. It is also worth mentioning that the microscopic dynamics
of the gapped quasiparticles is not governed by the Wong equations, as
in the normal phase.

\subsection{Further applications}
\label{Applications}

The starting point of this approach are the classical equations of
motion obeyed by coloured point particles, which, in the unbroken
phase of a non-Abelian gauge theory, are the Wong equations
\cite{Wong}.  In the past, these equations have also been studied for
other purposes, and we briefly summarise the main applications here.
We also comment on other uses of classical methods for studying the
quark-gluon plasma.

After Wong proposed the equations of motion for classical Yang-Mills
particles, a number of publications have been concerned with a more
fundamental understanding of them, either by providing corresponding
point particle Lagrangians or by establishing a link between point
particle Lagrangians and one-loop effective actions in quantum field
theories. Balachandran et
al.~\cite{Balachandran:1977ya,Balachandran:1978ub} and Barducci et
al.~\cite{Barducci:1977xq} proposed different Lagrangians which lead
to the Wong equations.  Upon quantisation, the several different
choices describe particles which belong to reducible or irreducible
representations of the Lie group. A unified description of the
different choices was given in \cite{Balachandran:1978ub}.
Balachandran et al.~showed that upon quantisation, some of the
parameters which appear in the Lagrangian are restricted to a certain
set of values. This reflects the fact that the spectrum of the Casimir
invariants of the Lie group is discrete.

A different line of research concerned the link between the point
particle Lagrangian, on one side, and quantum field theory on the
other. Brown and Weisberger \cite{Brown:1979bv} argued that the Wong
equations can be interpreted as classical equations of motion for
expectation values of quantum fields. Strassler
\cite{Strassler:1992zr} showed that the one-loop effective action in
quantum field theory can be expressed in terms of a quantum mechanical
path integral over a point particle Lagrangian.  In the case of
non-Abelian gauge theories, D'Hoker and Gagne
\cite{D'Hoker:1996ax,D'Hoker:1996bj} gave the world-line Lagrangian
for a non-Abelian gauge field theory.  Pisarski \cite{Pisarski:1997cp}
noted that this Lagrangian is identical to the one for Wong particles.
It has been suggested that this intimate link may provide a deeper
explanation for the applicability of the semi-classical approximation.
Jalilian-Marian et al.~\cite{Jalilian-Marian:1999xt} followed this
line of research. They derived, within a real time many-body formalism
for the world line action, a set of effective transport equations
which closely resemble those studied here.

Gibbons et al.~\cite{Gibbons:1982}, and Holm and Kupershmidt
\cite{Holm:1984hg} employed the Wong equations to derive a set of
chromohydrodynamic transport equations. These equations are the
non-Abelian analogues of the magnetohydrodynamic equations for charged
fluids.  These authors did not attempt, however, to link their
approach with a quantum field theoretical analysis.  At present, it is
not clear whether consistent chromohydrodynamic equations can be
derived from QCD in the first place, due to the (yet) unknown dynamics
of the chromomagnetic fields in the plasma. In this respect, the
Abelian and non-Abelian cases are qualitatively different, because
magnetic fields are never screened in a Coulomb plasma, while
chromomagnetic fields are supposed to be screened by a
non-perturbative magnetic mass.

The transport equations associated to the Wong particles have
been used to study non-Abelian dynamics, both analytically and
numerically, in combination with lattice simulations. Based on
assumptions linked to the colour flux model, some aspects of the
quark-gluon plasma during the very early stages of an
ultrarelativistic heavy-ion collision have been studied in
\cite{Nayak:1997ex,Nayak:1998kp,Dietrich:2000ex}.  This concerns the
production and evolution of a quark~\cite{Nayak:1997ex} or a
quark-gluon plasma \cite{Nayak:1998kp}, created in a constant
colour-electric field. The production of gluons from a space-time
dependent chromofield has been discussed in \cite{Dietrich:2000ex}.
The use of lattice simulations in combination with the Wong point
particle degrees of freedom have first been pointed out by Hu and
M\"uller \cite{Hu:1997sf}, and Hu, M\"uller and Moore
\cite{Moore:1998sn}.  Here, the classical Yang-Mills equation is
formulated on a spatial lattice, following the standard Kogut-Susskind
implementation. Then one adds the classical point particle degrees of
freedom. This technique has been used to construct a lattice
implementation of the HTL effective theory \cite{Moore:1998sn}, which
allowed for the computation of the Chern-Simons diffusion rate, a
quantity which is essential for the evaluation of baryogenesis in an
electroweak scenario.

A few further applications and extensions have been considered in the
literature. In a quantum mechanical framework, the non-Abelian charges
have been used to describe the non-Abelian analogue of the
Aharanov-Bohm effect \cite{Bak:1994dj}. The dual of the Wong equations
have been studied in \cite{Chan:1995zw}. These are the equations
obeyed by particles which are the non-Abelian analogues of the Dirac
magnetic monopole of electromagnetism.  Explicit analytical solutions
to the Wong equations for several coloured point particles, have been
found in \cite{Kosyakov:1998qi}.  The extension of Wong's equations
for QCD to curved space time has been studied by Brandt, Frenkel and
Taylor \cite{Brandt:1995mv}. They constructed the corresponding
effective action and obtained an exact, but implicit, solution of the
classical Boltzmann equation. Semi-classical methods have been
applied in the context of small $x$ physics. Here, the Wong equations
have been used to construct a small $x$ effective action
\cite{Jalilian-Marian:2001ad}, which opens an interesting door to
future applications.

An important domain of research concerns the computation of transport
coefficients of hot and dense non-Abelian plasmas, such as the shear
and bulk viscosities, heat and electrical conductivities, and baryon,
lepton and flavour diffusion. In the past, a derivation of shear and
bulk viscosity from quantum field theory has only been performed for a
scalar theory \cite{Jeon:1995if,Jeon:1996zm}. Based on classical
transport theory, a number of leading order computations have been
done, though not necessarily within the Wong particle picture. The
first computations made use of a relaxation time approximation, which
allowed a correct determination of the functional dependence of
transport coefficients on the gauge coupling \cite{Hosoya:1985xm}. The
considerations in
\cite{Baym:1990uj,Heiselberg:1994vy,Heiselberg:1994px,Joyce:1996zt,Moore:1995si,Baym:1997gq}
improved on the relaxation time approximation in that the relevant
collision terms were deduced from the scattering amplitudes resulting
from the particle interactions. All earlier computations have been
recently reviewed in \cite{Arnold:2000dr}, where some numerical errors
in the existing literature were detected and corrected. At present,
the results of these transport coefficients are only known to leading
logarithmic order in the non-Abelian gauge coupling.  While
computations of transport coefficients are typically based on linear
response, the case of non-linear response has recently been emphasised
in \cite{Carrington:2001ms}, for the example of the quadratic shear
viscosity of a weakly coupled scalar field. It has been argued that an
intimate link between classical transport theory and response theory
ensures that the non-linear response is correctly described by
classical transport theory.

\subsection{Outline}

This review presents an approach to semi-classical transport theory
for non-Abelian plasmas based on a classical point particle picture.
Both conceptual and computational issues are considered.  The first
part, Sections \ref{Micro} -- \ref{Consistent}, addresses the various
conceptual aspects of the approach, while the second part, Sections
\ref{Scales} -- \ref{DenseMatter}, presents an application to hot
plasmas close to thermal equilibrium, and to dense quark matter.  We
summarise, whenever appropriate, the main results at the end of the
sections.

In Section \ref{Micro}, the microscopic formalism is reviewed. This
starts with the classical equations of motions for coloured point
particles carrying a non-Abelian colour charge
(Section~\ref{MicroWong}) and the derivation as equations of motions
of expectation values for quantum wave packets (Section
\ref{WongDerivation}). The basic definitions of microscopic kinetic
functions are given (Section~\ref{Microscopic}) and the phase space
variables associated to the colour charges introduced
(Section~\ref{Measure}). The dynamical equations for the distribution
function (Section~\ref{MicroConservation}) and the microscopic gauge
symmetry (Section~\ref{MicroGauge}) are discussed.

In Section~\ref{Ensemble}, the step from a microscopic to a
macroscopic formulation is performed. The general assumptions made
when switching to an effective description are explained
(Section~\ref{EnsembleGeneral}), followed by a brief description of
the Gibbs ensemble average which is the starting point for the
subsequent applications (Section~\ref{EnsembleGibbs}). Finally, the
basic equal-time correlation functions for the quasi-particle
distribution function are given explicitly
(Section~\ref{EnsembleBasic})

In Section~\ref{Effective}, the ensemble average is performed for the
transport equations itself. The split of the distribution function and
the gauge fields into mean and fluctuating parts is considered next
(Section~\ref{EffectiveMean}), followed by a derivation of the
effective transport equations in their most general form for mean
fields and correlators (Section~\ref{EffectiveDynamical}). The new
terms in the effective kinetic equations are interpreted as collision
integrals, sources for stochastic noise, and fluctuation-induced
currents (Section~\ref{EffectiveCollisions}). Systematic approximation
schemes, able to truncate the infinite hierarchy of coupled
differential equations, are detailed
(Section~\ref{EffectiveSystematic}). This is followed by a discussion
of the basic macroscopic conservation laws
(Section~\ref{EffectiveConservation}) and the kinetic entropy
(Section~\ref{EffectiveEntropy}). The section is finished by a brief
discussion (Section~\ref{EffectiveDiscussion}).

All aspects connected to the requirements of gauge symmetry in the
effective transport theory are discussed in Section~\ref{Consistent}.
The intimate relationship to the background field method and the
invariance under both the background and the fluctuation field gauge
transformations are discussed (Section~\ref{ConsistentGauge}). Current
conservation for the mean and the fluctuation field implies
non-trivial cross-dependences amongst different correlation functions.
Their consistency is shown for the general case
(Section~\ref{ConsistentCurrent}), and for approximations to it
(Section~\ref{ConsistentApprox}).

The remaining part is dedicated to applications of the method to hot
non-Abelian plasmas close to thermal equilibrium. In
Section~\ref{Scales}, the relevant physical scales and parameters for
classical (Section~\ref{ScalesClassical}) and quantum plasmas
(Section~\ref{ScalesQuantum}) are discussed.

In Section~\ref{HTL} we discuss how the HTL effective theory is
recovered within the present formalism. To leading order in the gauge
coupling one obtains the non-Abelian Vlasov equation
(Section~\ref{HTL-Vlasov}). Their solution
(Section~\ref{HTL-Amplitudes0}) allows us to identify all HTL
amplitudes, including the HTL polarisation tensor
(Section~\ref{HTL-Amplitudes-1}). As an application, it is shown how a
local expression for the Hamiltonian and the Poynting vector is
obtained (Section~\ref{HTL-Energy}).

Non-Abelian fluctuations have to be taken into account beyond the HTL
approximation. This is done in Section~\ref{Beyond}. All
approximations are controlled by a small gauge coupling, and the
leading order dynamical equations are given
(Section~\ref{BeyondLeading}). The dynamics of fluctuations is solved
explicitly in terms of initial fluctuations of the quasi-particle
distribution function (Section~\ref{BeyondIntegrating}). The basic
equal-time correlators are obtained, and the example of Landau damping
is discussed (Section~\ref{BeyondCorrelators}).  The domain of
validity is derived from the two-particle correlators and the
associated correlation length (Section~\ref{Validity2on}).  The
correlators in the effective transport equation are evaluated, and the
relevant collision integral (Section~\ref{BeyondCollision}) and the
corresponding noise source (Section~\ref{BeyondStochastic}) are
identified to leading logarithmic accuracy.  The resulting transport
equation is discussed.  Iterative solutions allow the computation of
the ultra-soft amplitudes (Section~\ref{BeyondUltra}).  In the
over-damped limit, B\"odeker's Langevin-type dynamical equation for
the ultra-soft gauge fields is recovered
(Section~\ref{BeyondLangevin}).

In Section~\ref{Langevin}, a phenomenological approach to non-Abelian
fluctuations is discussed. It is based on the idea of coarse-graining
the microscopic transport equations (Section~\ref{LangevinCoarse}). We
describe the line of reasoning for the example of classical
dissipative systems (Section~\ref{LangevinClassical}). This is
extended to the case of non-Abelian plasmas, where the basic spectral
functions and equal-time correlators for stochastic fluctuations are
derived from the kinetic entropy (Section~\ref{LangevinFluctuations}).
As an application, B\"odeker's effective kinetic theory is
re-considered and shown to be compatible with the
fluctuation-dissipation theorem (Section~\ref{LangevinApplication}).
We close with a discussion of the results and further applications
(Section~\ref{LangevinDiscussion}).

In Section~\ref{DenseMatter} we consider dense quark matter. When the
effects of quark pairing can be neglected, the transport equations are
the same as for the hot non-Abelian plasma
(Section~\ref{HDL-subsection}). In the superconducting phase, the
ground state is given by a diquark condensate
(Section~\ref{Superconducting}). For two massless quark flavours, the
thermal colour excitations of the condensate are described by
quasiparticles (Section~\ref{Quasiparticles}). The corresponding
transport equation is given and solved to leading order
(Section~\ref{HSL}). We close with a brief discussion of the results
(Section~\ref{Super-Discussion}).

Two Appendices contain technical details.

\newpage

\section{Microscopic approach} 
\label{Micro}

The microscopic approach to semi-classical transport theory considers
an ensemble of classical point particles. In the Abelian case, these
are simply electrons or ions, interacting self-consistently through
the Maxwell equations. Based on this picture, a complete effective
theory for classical Coulomb or Abelian plasmas has been worked out in
the literature (see for example \cite{K0,K,K2,Landau10,Sitenko1982}).

For the non-Abelian case, the concept of an electro-magnetically
charged classical point particle is replaced by a {\it coloured} point
particle, where `colour' stands for a non-Abelian colour charge. The
classical equations of motion for such particles have been given by
Wong \cite{Wong}, and a transport theory based on it has been
discussed in
\cite{Heinz:1983nx,Heinz:1984my,Heinz:1985yq,Heinz:1986qe,Heinz:1988fg,Heinz:1989cq,Elze:1989un}.
In this section, we review the microscopic approach to non-Abelian
plasmas based on classical equations of motions for such `particles'.
We also introduce the basic notation to be used in the following
sections for the construction of a kinetic theory \cite{GLW:1980}.

\subsection{Wong equations}
\label{MicroWong}

Let us consider a system of particles carrying a non-Abelian colour
charge $Q^a$, where the colour index runs from $a=1$ to $N^2-1$ for a
$SU(N)$ gauge group.  Within a microscopic description, the
trajectories in phase space are known exactly.  The trajectories
${\hat x}(\tau), {\hat p}(\tau)$ and ${\hat Q}(\tau)$ for every
particle are solutions of their classical equations of motions, known
as the Wong equations \cite{Wong}
\begin{mathletters}\label{Wong}
\step
\bea
m\0{d{\hat x}^\mu}{d\tau} &=& {\hat p}^\mu \ , 
\zeile
\step
m\0{d{\hat p}^\mu}{d\tau} &=& g {\hat Q}^a \, F_a^{\mu\nu}({\hat x}) 
                                \, {\hat p}_\nu \ ,
\zeile
\step
m\0{d{\hat Q}^a}{d\tau} &=& -g f^{abc} {\hat p}^\mu\, 
                             A^{b}_{\mu}({\hat x})\,{\hat Q}^c\ .
\label{dQ}
\eea
\end{mathletters}%
Here, $A_\mu$ denotes the microscopic gauge field. A dependence on
spin degrees of freedom, which can be incorporated as well
\cite{Heinz:1985yq}, is not considered in the present case. The
microscopic field strength $F^{a}_{\mu\nu}$ and the energy momentum
tensor of the gauge fields $\Theta^{\mu \nu}$ are given by
\bea
F^{a}_{\mu\nu}[A]&=&
\partial_\mu A^{a}_\mu -\partial_\nu A^{a}_\mu
+ gf^{abc}A^{b}_\mu A^{c}_\nu\ ,\label{NA}
\zeile 
\label{NA-EM}
\Theta^{\mu \nu}[A] &=&
\s014 g^{\mu \nu}F_{\rho \sigma}^a F^{\rho \sigma}_a 
+F^{\mu \rho}_{a} F^{a\, \,\nu}_{\rho}
\eea
and $f^{abc}$ are the structure constants of $SU(N)$. 
We set $c=k_B=\hbar =1$ 
and work in natural units, unless otherwise indicated.
Note that the non-Abelian charges are also subject to  
dynamical evolution. 
\Eq{dQ} can
be rewritten as $D_\tau Q = 0$, where
$D_\tau = \frac{ d \hat x^\mu}{d \tau} D_\mu$ is the covariant 
derivative along the world line, and 
$D_\mu^{ac}[A]=\partial_\mu \de^{ac} + g f^{abc} A^b_\mu$ the
covariant derivative in the adjoint representation. 

The colour current can be constructed once the
solutions of the Wong equations are known. For every particle it
reads
\beq\label{j-particle}
j^\mu_a(x) = g \int {d \tau} \frac{d {\hat x}^\mu}{d \tau} \ {\hat Q}_a
(\tau)\ \delta^{(4)}[x -{\hat x}(\tau)]  \ .
\eeq
Employing the Wong equations \eq{Wong} we find that the current is
covariantly conserved, $D_\mu j^\mu = 0$ \cite{Wong}.
Similarly, the energy momentum tensor of the particles 
is given by \cite{Wong}
\beq
\label{em-particle}
t^{\mu \nu}(x) = \int d \tau \,
\frac{ d{\hat x}^\mu}{d \tau}  \, {\hat p}^\nu(\tau)  
\, \delta^{(4)}[x -{\hat x}(\tau)] \ .
\eeq
The Wong equations couple to classical non-Abelian gauge fields. 
The Yang-Mills equations are
\beq\label{YM}
D_\mu F^{\mu\nu}(x)=J^\nu(x)\ .
\eeq
The source for the Yang-Mills fields 
\beq\label{ji}
J^\nu(x)=\sum_{\rm particles} j^\nu(x)
\eeq
is given by the sum of the currents of all particles.

\subsection{Classical limit in a non-Abelian quantum field theory}
\label{WongDerivation}

One may wonder under which conditions the point particle picture and
Wong equations are a good approximation to the full quantum field
theory. Originally, the Wong equations have been derived as the
non-Abelian generalisation of equations of motion for electrically
charged point particles. Here, we shall outline how the Wong equations
are obtained as equations of motions for wave packets, in a gauge
field theory coupled to matter.  The derivation is valid for a system
with matter interacting with non-Abelian gauge fields, and it does not
apply for the gluons themselves.  This line of reasoning is due to
Brown and Weisberger \cite{Brown:1979bv}. It is argued that these
equations derive from the conservation laws
\begin{mathletters}
\label{ConLaw}
\step
\bea
\partial_\nu  t^{\mu\nu} &=&  F_a^{\mu\nu}J_\nu^a \ ,
\zeile
\step
D_\mu\,J^\mu            &=& 0\ ,
\eea
\end{mathletters}%
as classical equations of motion for sufficiently localised quantum
states (or `wave packets'), provided that the gauge fields are `soft',
{\it i.e.}~vary only slowly over typical scales associated to the
particles.  We begin with the following definitions,
\begin{mathletters}
\label{WongDef}
\step
\bea
X^\mu(t)&=&\0{\int d^3x \, x^\mu \,
\langle  t^{00}(\vx,t)\rangle}{\int d^3x\, \langle
 t^{00}(\vx,t)\rangle}
\equiv \0{1}{P^0}\int d^3x\, x^\mu \langle  t^{00}(\vx,t)\rangle \ ,
\zeile
\step
P^\mu(t) &=& \int d^3x\, \langle  t^{0\mu}(\vx,t)\rangle \ ,
\zeile
\step
g\, Q_a(t)&=&  \int d^3x\, \langle J^0_a(\vx,t)\rangle\ .
\label{Qdef}
\eea
\end{mathletters}%
The variable $P^\mu(t)$ describes the particle's mechanical
four-momentum at the time $t\equiv x_0\equiv X_0$ as an expectation
value of the quantum fields, $X^i(t)$ the center-of-energy expectation
value of the quantum fields, and $Q_a(t)$ the expectation value for
the associated colour charge.  Making use only of the conservation
laws \Eqs{ConLaw}, and partial integrations, one derives the following
equation of motion for the particles,
\begin{mathletters}
\label{eom1}
\step
\bea
\0{dX^i}{dt}&=& 
\0{{P^i}}{P^0}
+\0{1}{(P^0)^2}
\int d^3x\, d^3y\, \langle F^{0k}(\vx,t)\, J^k(\vx,t)\rangle
                 \langle  t^{00}(\vy,t)\rangle (x^i-y^i)\ ,
\zeile
\step
\0{dP^\mu}{dt}&=& 
\int d^3x\, \langle F^{\mu\nu}_a(\vx,t)J^a_\nu(\vx,t)\rangle \ ,
\zeile
\step
g\,\0{dQ_a}{dt}&=& -
g\,f_{abc}\int d^3x\, \langle \,A_\mu^b(\vx,t)\,J^{\mu,c}(\vx,t)\rangle\ .
\eea
\end{mathletters}%
These equations of motion still involve integrals over all space on
the right-hand side. To obtain a classical limit, these equations can
be simplified in the case where the characteristic length-scales of
the particles are much smaller than those associated to the gauge
fields.  In this case, the particle current $J(\vx,t)$ is localised
close to the location of the particle, and equally $ t^{00}(\vx,t)$.
If the gauge fields do vary very slowly over these short distance
scales, we can perform the following approximation,
\begin{mathletters}
\label{eomApprox}
\step
\beq
\int d^3x\, \langle F^{\mu\nu}(\vx,t)J_\rho(\vx,t)\rangle
\approx F^{\mu\nu}(\vX,t)\int d^3x\, \langle J_\rho(\vx,t)\rangle\ ,
\eeq
by replacing the gauge fields in the integrand through their values at
the location of the particle. Equally, we approximate
\beq
\step
\int d^3x\, \langle A^{\mu}(\vx,t)J_\rho(\vx,t)\rangle
\approx A^{\mu}(\vX,t)\int d^3x\, \langle J_\rho(\vx,t)\rangle\ .
\eeq
This implies in addition that
\step
\beq
\int d^3x\,  d^3y\, \langle F^{0k}(\vx,t)\, J^k(\vx,t)\rangle
                    \langle  t^{00}(\vy,t)\rangle (x^i-y^i )\approx 0 \\
\eeq
\end{mathletters}%
to leading order. The approximation \Eq{eomApprox} corresponds to the
leading order in a gradient expansion. Employing \Eq{eomApprox}, the
equations of motion become
\begin{mathletters}
\label{eom2}
\step
\bea
\0{dX^i}{dt}&=& 
\0{P^i}{P^0}\ , 
\zeile
\step
\0{dP^\mu}{dt}&=& 
 F^{\mu\nu}_a(\vX,t)\, \int d^3x\, \langle J^a_\nu(\vx,t)\rangle \ ,
\label{dPdt}
\zeile
\step
g\,\0{dQ_a}{dt}&=& -
g\,f_{abc}\,A_\mu^b(\vX,t)\,\int d^3x\, \langle J^{\mu,c}(\vx,t)\rangle\ .
\eea
\end{mathletters}%
We shall now exploit the fact that the particle's mass $m$ with
\beq
m^2=P^\mu\,P_\mu 
\eeq
is a constant of motion, $dm^2/dt=0$. With \Eq{dPdt}, this yields the
constraint
\beq
0= F^{\mu\nu}_a(\vX,t)\, P_\mu\, P_0\, \int d^3x\, \,
\langle J^a_\nu(\vx,t)\rangle 
\eeq
which has to hold for any field configuration. The field strength is
an arbitrary antisymmetric tensor, therefore, the constraint implies
\beq\label{sym}
P_\mu\, \int d^3x\, \langle J^a_\nu(\vx,t)\rangle \,
=
P_\nu\, \int d^3x\, \langle J^a_\mu(\vx,t)\rangle \ .
\eeq
This is automatically the case if the current density is proportional
to the momentum density. Evaluating \eq{sym} for $\nu =0$ and using
the definition \eq{Qdef} we indeed find
\beq
P_0\, \int d^3x\, \langle J^a_\mu(\vx,t)\rangle 
=  g \, P_\mu\, Q^a\ ,
\eeq
and the approximate equations of motion read
\begin{mathletters}
\label{eom3}
\step
\bea
\0{dX^i}{dt}&=& 
\0{P^i}{P^0}\label{dXidt3}\ ,
\zeile
\step
\0{dP^\mu}{dt}&=& 
g\,F^{\mu\nu}_a\,\0{P^\nu}{P^0} \,Q^a\ ,
\zeile
\step
\0{dQ_a}{dt}&=& 
-g\,f_{abc}\,A_\mu^b\,\0{P^\mu}{P^0}\,Q^c\ .
\eea
\end{mathletters}%
Let us finally introduce the proper time $\tau$ for the particles,
which serves as a normalisation condition for \Eq{dXidt3}. The proper
time relates the mass of the particles to the 00-component of the
energy-momentum tensor through the requirement $m\,dX/d\tau=P$, hence
\beq\label{tau}
m\0{d}{d\tau}
=P^0\, \0d{dt}\ .
\eeq
Using \Eqs{eom3} and \eq{tau}, the equations of motion become
\begin{mathletters}
\label{eom4}
\step
\bea
m\0{d{X}^\mu}{d\tau}&=&{P}^\mu \ ,
\zeile
\step
m\0{d{P}^\mu}{d\tau}&=&g {Q}^a \, F_a^{\mu\nu}\,  {P}_\nu \ ,
\zeile
\step
m\0{d{Q}^a}{d\tau}&=&-g f^{abc}\, {P}^\mu A^{b}_{\mu}\,{Q}^c \ ,
\eea
\end{mathletters}%
and agree with the equations found by Wong, \Eqs{Wong}, if the
replacements $X\to \hat x$, $P\to \hat p$ and $Q\to \hat Q$ are made.
We conclude that the Wong equations are the leading order approximate
equations of motions for point particles, if the induced gauge fields
are soft. This scale separation between hard particles and soft fields
is at the root of the present approach.

\subsection{Microscopic distribution functions}
\label{Microscopic}

Instead of describing every particle individually, it is convenient to 
introduce a phase space density for the ensemble of particles, that is 
a distribution function which depends on the whole set of coordinates 
$x^\mu, p^\mu$
and $Q_a$.  To that end, we introduce two functions $n(x, p, Q)$ and 
$f(x, p, Q)$, which only differ by appropriately chosen normalisation 
factors. We begin with  the function 
\beq
\label{n-def}
n(x, p, Q) = \sum_{i} \int d \tau \,
\de^{(4)}[x-{\hat x}_i(\tau)]\
\de^{(4)}[p-{\hat p}_i(\tau)]\
\de^{(N^2-1)}[Q-{\hat Q}_i(\tau)]\ ,
\eeq
where 
the index $i$ labels the particles. This
distribution function is constructed in such a way that the colour 
current
\beq\label{current-n}
J^\mu _a (x) = g \int d^4p \, d^{(N^2-1)} Q \, 
                 \frac{p^\mu}{m}\, Q_a \, n(x,p,Q) 
\eeq
coincides with the sum over all currents associated to the individual
particles $J^\mu_a=\sum_ij^\mu_a$, \Eq{ji}. The current is covariantly
conserved, $D_\mu J^\mu =0$.  It is convenient to make the following
changes in the choice of the distribution function.  For convenience,
we introduce a momentum and a group measure such that the physical
constraints like the on-mass shell condition, positive energy and
conservation of the group Casimirs are factored out into the phase
space measure. Consider the momentum measure
\beq
\label{Pmeasure}
dP = d^4 p \, 2 \theta(p_0)\, \de(p^2 -m^2) \ ,
\eeq
which accounts for the on-mass-shell constraint.
The measure for the colour charges is
\beq
dQ = d^3 Q \, c_R \,\de (Q_a Q_a -q_2) \ ,
\label{col-mes}
\eeq
in the case of $SU(2)$. For $SU(3)$ the measure is
\beq
dQ = d^8 Q \, c_R \,\de (Q_a Q_a -q_2)\, \de(d_{abc} Q^a Q^b Q^c - q_3) \ .
\label{col-mes-3}
\eeq
For $SU(N)$, $N-1$ $\delta$-functions ensuring the conservation of the
set of $N-1$ Casimirs have to be introduced into the measure for the
colour charges.  We have also introduced the representation-dependent
normalisation constant $c_R$ into the measure, which is fixed by the
normalisation condition $\int dQ =1$.  Furthermore, we have $\int
dQQ_a=0$. The quadratic Casimir $C_2$ is defined as
\beq
\label{quadraticQ}
 \int dQ Q_a Q_b = C_2 \delta_{a b} \ ,
\eeq
and depends on the group representation of the particles. For
particles in the adjoint representation of $SU(N)$ we have $C_2=N$
(gluons). For particles in the fundamental representation, $C_2=\012$
(quarks). Notice that the colour charges have to be quantised within a
quantum field theoretical approach.

We will define a second distribution function $f(x,p,Q)$ such that the
physical constraints within $n(x,p,Q)$ have been factored out,
\beq
dP\,dQ\,f(x,p,Q)=d^3p\,\0{dp_0}m\,\,d^{(N^2-1)}Q\,n(x,p,Q)\ .
\eeq
With this convention the colour current of the particles \Eq{current-n} 
now reads
\beq
\label{j-f}
{J}^\mu_a (x) = g \int dP dQ\, p^\mu \,Q_a \, f(x,p,Q) \ . 
\eeq
The energy-momentum tensor associated to the particles is
\beq\label{em-particles}
t^{\mu \nu}(x) = \int dP dQ\ p^\mu p^\nu\ f(x,p,Q) \ 
\eeq
when expressed in terms of the distribution function $f$.

\subsection{Phase space}\label{Measure}

We have introduced the distribution function $f(x,p,Q)$ to further a
description in phase space. While $\vx$ and $\vp$ are standard phase
space variables with a canonical Poisson bracket, the colour charges
$Q_a$ are {\it not}.  However, it is always possible to define the set
of canonical (Darboux) variables associated to the $Q_a$ charges. For
$SU(N)$, there are $N(N-1)/2$ pairs of canonical variables which we
denote as ${\bphi} = ( \phi_1, \ldots, \phi_{N(N-1)/2})$ and ${\bpi} =
(\pi_1, \ldots, \pi_{N(N-1)/2})$. The canonical variables define the
canonical Poisson bracket
\beq\label{canonical}
\{A,B\}_{PB} = \0{\partial A}{\partial x_i}\0{\partial B}{\partial p_i}
- \0{\partial A}{\partial p_i}\0{\partial B}{\partial x_i} + 
\0{\partial A}{\partial \phi_a}\0{\partial B}{\partial \pi_a}
-\0{\partial B}{\partial \phi_a}\0{\partial A}{\partial \pi_a}\ ,
\eeq
and obey trivially
\beq
\label{canonical-2}
\{x_i,p_k\}_{PB} = \delta_{ik} \ , \qquad
\{\phi_a,\pi_b\}_{PB} = \delta_{ab}\ .
\eeq
The colour charges $Q_a$ are a representation of $SU(N)$. When
expressed as functions of the canonical variables, their Poisson
bracket reads
\beq\label{darstellung}
\{Q_a,Q_b\}_{PB} = f_{abc} Q_c\ ,
\eeq
where $f_{abc}$ are the structure constants of $SU(N)$. 

The explicit construction of Darboux variables for $SU(2)$ and $SU(3)$
has been performed in \cite{KLLM}. Let us first consider the
$SU(2)$-case. We define the set of variables $\phi_1, \pi_1$ and $J$
by the implicit transformation
\beq
\label{Qsu2}
Q_1  =   \cos{\phi_1} \sqrt{J^2 - \pi_1^2} \ , \qquad
Q_2   =  \sin{\phi_1} \sqrt{J^2 - \pi_1^2} \ , \qquad   
 Q_3  =  \pi_1   \ ,
\eeq
where the variable $\pi_1$ is bounded by $-J \leq \pi_1 \leq J$.  The
variables $\phi_1, \pi_1$ form a canonically conjugate pair and obey
\Eq{canonical-2}, while $J$ is fixed by the value of the quadratic
Casimir, which is constant under the dynamical evolution. One confirms
that \Eq{Qsu2} obey \Eq{darstellung} with $f_{abc}=\epsilon_{abc}$.
The phase space volume element \Eq{col-mes} becomes
\beq
\step
dQ=d\pi_1\,d\phi_1\,dJ\,J\,c_R\,\delta(J^2-q_2)
\eeq
in terms of the Darboux variables.  With the above change of
variables, one can fix the value of the representation-dependent
normalisation constant $c_R$ introduced in (\ref{col-mes}).  From the
condition $\int dQ =1$ one finds $c_R = 1/{2 \pi \sqrt{q_2}}$.  From
the condition $\int dQ Q_a Q_b = C_2 \delta_{ab}$ one gets $q_2 = 3
C_2$. This entirely fixes the value of $c_R$ as a function of $C_2$.

The group $SU(3)$ has eight charges, $(Q_{1}, \ldots , Q_{8})$ and two
conserved quantities, the quadratic and the cubic Casimirs,
$Q^{a}Q_{a}$ and $d_{abc}Q^{a}Q^{b}Q^{c}$, respectively. The
phase-space colour measure is quoted above in (\ref{col-mes-3}).  As in
the $SU(2)$ case, new coordinates
$(\phi_{1},\phi_{2},\phi_{3},\pi_{1},\pi_{2},\pi_{3},J_1,J_2)$ may be
introduced by means of the following
transformations~\cite{Johnson:1989qm}:
\begin{equation}
\label{Qsu3}
\begin{array}{rclcrcl}
Q_{1} &=& \cos\phi_{1}\, \pi_{+}\,\pi_{-} \ , &&
Q_{2} &=& \sin\phi_{1}\, \pi_{+}\,\pi_{-} \ , \nonumber\\
Q_{3} &=& \pi_{1} \ , &&&&\nonumber\\
Q_{4} &=& C_{++}\,\pi_{+}\,A + C_{+-}\,\pi_{-}\,B \ , &\qquad&
Q_{5} &=& S_{++}\,\pi_{+}\,A + S_{+-}\,\pi_{-}\,B \ , \label{gens-
su3}\\
Q_{6} &=& C_{-+}\,\pi_{-}\,A - C_{--}\,\pi_{+}\,B \ , &\qquad&
Q_{7} &=& S_{-+}\,\pi_{-}\,A - S_{--}\,\pi_{+}\,B \ , \nonumber\\
Q_{8} &=& \pi_{2} \ , &&&&\nonumber
\end{array}
\end{equation}
in which we have used the definitions:
\begin{equation}
\begin{array}{rclcrcl}
\pi_+ &=& \sqrt{\pi_3 +\pi_1}\ ,&\qquad&
\pi_- &=& \sqrt{\pi_3 -\pi_1}\ , \nonumber\\[2mm]
C_{\pm\pm} &=& \cos\left[\frac{1}{2}(\pm\phi_1
+\sqrt{3}\phi_{2}\pm\phi_{3})\right]\ , &\qquad&
S_{\pm\pm} &=& \sin\left[\frac{1}{2}(\pm\phi_1
+\sqrt{3}\phi_2\pm\phi_3)\right] \ , \nonumber
\end{array}
\end{equation}
and $A,\ B$ are given by
\begin{eqnarray}
A &=& \frac{1}{2\pi_{3}} \sqrt{
\left(\frac{J_1 - J_2}{3} + \pi_{3} + \frac{\pi_{2}}{\sqrt{3}}\right)
\left(\frac{J_1 + 2J_2}{3} + \pi_{3} + \frac{\pi_{2}}{\sqrt{3}}\right)
\left(\frac{2J_1 + J_2}{3} - \pi_{3} - \frac{\pi_{2}}{\sqrt{3}}\right) }
\quad , \nonumber\\[3mm]
B &=& \frac{1}{2\pi_{3}} \sqrt{
\left(\frac{J_2 - J_1}{3} + \pi_{3} - \frac{\pi_{2}}{\sqrt{3}}\right)
\left(\frac{J_1 + 2J_2}{3} - \pi_{3} + \frac{\pi_{2}}{\sqrt{3}}\right)
\left(\frac{2J_1 + J_2}{3} + \pi_{3} - \frac{\pi_{2}}{\sqrt{3}}\right) }
\quad .\nonumber
\end{eqnarray}
Note that in this representation, the set $(Q_{1},Q_{2},Q_{3})$ forms
an $SU(2)$ subgroup with quadratic Casimir $Q_{1}^{2} + Q_{2}^{2} +
Q_{3}^{2} = \pi_{3}^{2}$.  It can be verified, using the values of the
structure constants given in Tab.~1, that the expressions above for
$Q_{1}, \ldots, Q_{8}$ form a representation of the group $SU(3)$.
\begin{center}
\vspace{5mm}
\begin{tabular}{c||ccccccccc}
$f_{abc}$
&$f_{123}$
&$f_{147}$
&$f_{156}$
&$f_{246}$
&$f_{257}$
&$f_{345}$
&$f_{367}$
&$f_{458}$
&$f_{678}$
\\[.5ex] \hline\\[-2.5ex]
&$1$
&${1\over 2}$
&$-{1\over 2}$
&${1\over 2}$
&${1\over 2}$
&${1\over 2}$
&$- {1\over 2}$
&${\sqrt{3}\over 2}$
&${\sqrt{3}\over 2}$
\\[1ex] 
\end{tabular}
\end{center}
\begin{center}
\begin{minipage}{.5\hsize}
  {\label{Tab1}\small {\bf Table 1:} The non-zero constants $f_{abc}$ for
    $SU(3)$.}
\end{minipage}
\end{center}

\begin{center}
\begin{tabular}{c||cccccccccccccccc}
$d_{abc}$
&$d_{118}$
&$d_{146}$
&$d_{157}$
&$d_{228}$
&$d_{247}$
&$d_{256}$
&$d_{338}$
&$d_{344}$
&$d_{355}$
&$d_{366}$
&$d_{377}$
&$d_{448}$
&$d_{558}$
&$d_{668}$
&$d_{778}$
&$d_{888}$
\\[.5ex] \hline\\[-2.5ex]
&$1\over\sqrt{3}$
&$1\over 2$
&$1\over 2$
&$1\over\sqrt{3}$
&$-{1\over 2}$
&$1\over 2$
&$1\over\sqrt{3}$
&$1\over 2$
&$1\over 2$
&$-{1\over 2}$
&$-{1\over 2}$
&$-{1\over 2\sqrt{3}}$
&$-{1\over 2\sqrt{3}}$
&$-{1\over 2\sqrt{3}}$
&$-{1\over 2\sqrt{3}}$
&$-{1\over\sqrt{3}}$
\\ 
\end{tabular}
\end{center}
\begin{center}
\begin{minipage}{.5\hsize}
  {\label{Tab2}\small {\bf Table 2:} The non-zero constants $d_{abc}$ for
    $SU(3)$.}
\end{minipage}
\end{center}
As is implicit in the above, the two Casimirs depend only on $J_1$ and
$J_2$. They can be computed, using the values given in the Tab.~2,
as:
\begin{mathletters}
\label{cas-su3}
\begin{eqnarray}
Q^{a}Q_{a} &=& \frac{1}{3} (J_1^2+J_1J_2+J_2^2) \ ,
\label{c2-su3} \\[3mm]
d_{abc}Q^{a}Q^{b}Q^{c} &=& \frac{1}{18}(J_1-J_2)
(J_1+2J_2)(2J_1+J_2) \ .
\label{c3-su3}
\end{eqnarray}
\end{mathletters}
The phase-space colour measure for $SU(3)$, given in
(\ref{col-mes-3}), may be transformed to the new coordinates
through use of (\ref{Qsu3}) and evaluation of the Jacobian
\begin{equation}
\left|\frac{\partial (Q_{1},Q_{2},\ldots,Q_{8})}
{\partial
(\phi_{1},\phi_{2},\phi_{3},\pi_{1},\pi_{2},\pi_{3},J_1,J_2)}\right|
= \frac{\sqrt{3}}{48}\ J_1\,J_2\,(J_1+J_2) \ .\label{jac-su3}
\end{equation}
The measure reads:
\begin{eqnarray}
dQ = c_R\,d\phi_{1}\,d\phi_{2}\,d\phi_{3}\,d\pi_{1}\,d\pi_{2}
\,d\pi_{3}\,dJ_1\,dJ_2\ &&\frac{\sqrt{3}}{48}\ J_1\,J_2\,(J_1+J_2) \
\delta\Bigl(\frac{1}{3} (J_1^2+J_1J_2+J_2^2)- q_{2}\Bigr)\
\times\nonumber\\
&&\delta\Bigl(\frac{1}{18}(J_1-J_2)(J_1+2J_2)(2J_1+J_2) - q_{3}
\Bigr)\ . \label{foo2}
\end{eqnarray}
Since the two Casimirs are linearly independent, the delta-functions
uniquely fix both $J_1$ and $J_2$ to be representation-dependent
constants. Upon integrating over $J_1$ and $J_2$, (\ref{foo2}) reduces
to a constant times the proper canonical volume element
$\prod_{i=1}^3\,d\phi_i\,d\pi_i$. The value of the normalisation
constant $c_R$ will now depend both on $q_2$ and $q_3$.

For $SU(N)$, the canonical variables can be constructed along similar
lines \cite{AFS}. For the quadratic and cubic Casimir, one finds
$q_2=(N^2-1)C_2$, and $C_2=\012$ for particles in the fundamental
(quarks), and $C_2=N$ for particles in the adjoint (gluons). The
constant $q_3$ reads $q_3=(N^2-4)(N^2-1)/4N$ for particles in the
fundamental, and $q_3=0$ for particles in the adjoint. We also comment
in passing that in the pure classical framework, the quadratic Casimir
$C_2$ carries the dimensions of $\hbar c$.  After quantisation, the
quadratic Casimirs should take quantised values proportional to
$\hbar$. The Poisson brackets then have to be replaced by commutators.

The microscopic phase space density, expressed in terms of the real
phase space variables, is given by
\beq
\label{NA-PD}
 \hat n (\vx, \vp, {\bphi}, {\bpi} ) = 
\sum_{i} \de^{(3)}[\vx-{\hat \vx}_i(t)]\
\de^{(3)}[\vp-{\hat \vp}_i(t)]\ 
\de[{\bphi} -{\hat {\bphi}}_i(t)] \
\de [{\bpi} - {\hat {\bpi}}_i (t)] \ ,
\eeq
where the sum runs over all particles of the system, and
$(\hat\vx_i,\hat\vp_i, \hat{\bphi}_i ,\hat{\bpi}_i)$ refers to the
trajectory of the $i$-th particle in phase space. Then $ \hat n\, d
\vx\, d \vp\, d {\bphi} d{\bpi}$ gives the number of particles at time
$t$ in an infinitesimal volume element of phase space around the point
$z=(\vx,\vp, {\bphi} ,{\bpi})$.  The function $\hat n(\vx, \vp,
{\bphi}, {\bpi})$ agrees with the microscopic function $f(x,p,Q)$
introduced above, except for a representation-dependent normalisation
constant.

\subsection{Dynamical equations and conservation laws}
\label{MicroConservation}

Now we come to the dynamical equation of the microscopic distribution
functions $n(x,p,Q)$, $\hat n(\vx, \vp, {\bphi}, {\bpi})$ and
$f(x,p,Q)$, which will serve as the starting point for the subsequent
formalism. Although the independent degrees of freedom are given by
the phase space variables $(\vx, \vp, {\bphi}, {\bpi})$, it is more
convenient to derive the dynamical equations in terms of the variables
$(x,p,Q)$.  The Darboux variables will become important when an
ensemble average is defined in the following section. Secondly, we
note that the dynamical equation for $n(x,p,Q)$ is the same as for
$f(x,p,Q)$.  This is so because the physical constraints which we have
factored out to obtain $f(x,p,Q)$ are not affected by the Wong
equations.  Employing \Eqs{Wong}, we find
\begin{mathletters}
\label{NA-Micro}
\step
\beq
\label{NA-f}
p^\mu\left(\0{\partial}{\partial x^\mu}
- g f^{abc}A^{b}_\mu Q^c\0{\partial}{\partial Q^a}
-gQ_aF^{a}_{\mu\nu}\0{\partial}{\partial p_\nu}\right) f(x,p,Q)=0 \ , 
\eeq
which can be checked explicitly by direct inspection of \Eq{n-def}
into \Eq{NA-f} \cite{KLLM2}. Equivalently, one could have made use of
Liouville's theorem $df/d\tau=0$, which states that the phase space
volume is conserved. In combination with \Eqs{Wong}, one obtains
\Eq{NA-f}. In a self-consistent picture this equation is completed
with the Yang-Mills equation,
\beq
\label{NA-J}
 (D_\mu F^{\mu\nu})_a(x) =J_a^{\nu}(x) \ ,
\eeq
\end{mathletters}%
and the current being given by \Eq{j-f}. It is worth noticing that
\Eqs{NA-Micro} are exact in the sense that no further approximations
apart from the quasiparticle picture have been made. This Boltzmann
equation looks formally as {\it collisionless}. However, it
effectively contains collisions inasmuch as the Wong equations account
for them, that is, due to the long range interactions between the
particles.

For the microscopic energy-momentum tensor of the gauge fields
\Eq{NA-EM} we find
\beq
\partial_\mu \Theta^{\mu \nu}(x) = - F^{\nu\mu}_a(x) J_\mu^a(x) \ .
\label{cons-en-mo}
\eeq
On the other hand, using \Eq{NA-f} and the definition \Eq{em-particles}
we find 
\beq
\partial_\mu t^{\mu \nu}(x) =\  F^{\nu\mu }_a(x) J_\mu^a(x)  
\eeq
for the energy-momentum tensor of the particles, hence 
\beq 
\partial_\mu \,T^{\mu \nu}(x) = 0,\,\quad
T^{\mu \nu}(x) =\Theta^{\mu \nu}(x) +t^{\mu \nu}(x)
\eeq
which establishes that the combined energy-momentum tensor of the
particles and the fields is conserved.

\subsection{Gauge symmetry}
\label{MicroGauge}

To finish the discussion of the microscopic description of the system,
let us recall the gauge symmetry properties of the Wong \Eqs{Wong} and
the set of microscopic dynamical equations \eq{NA-Micro} (a detailed
discussion is given in Section~\ref{Consistent}). With $Q_a$ and
$F^a_{\mu\nu}$ transforming in the adjoint representation, the Wong
equations are invariant under gauge transformations.  The equation
(\ref{dQ}) ensures that the set of $N-1$ Casimir of the $SU(N)$ group
is conserved under the dynamical evolution. For $SU(2)$, it is easy to
verify explicitly the conservation of the quadratic Casimir $Q_a Q_a$.
For $SU(3)$, both the quadratic and cubic Casimir $d_{abc} Q_a Q_b
Q_c$, where $d_{abc}$ are the symmetric structure constants of the
group, are conserved under the dynamical evolution. The last
conservation can be checked using (\ref{dQ}) and a Jacobi-like
identity which involves the symmetric $d_{abc}$ and antisymmetric
$f_{abc}$ constants \cite{KLLM2}.

From the definition of the distribution function $f(x,p,Q)$ we
conclude that it transforms as a scalar under (finite) gauge
transformations, $f'(x,p,Q')=f(x,p,Q)$.  This implies the gauge
covariance of \Eq{NA-J} because the current \Eq{j-f} transforms like
the vector $Q_a$ in the adjoint representation.  The non-trivial
dependence of $f(x,p,Q)$ on the non-Abelian colour charges implies
that the partial derivative $\partial_\mu f(x,p,Q)$ does not transform
as a scalar. Instead, its covariant derivative $D_\mu f(x,p,Q)$, which
is given by
\beq
\label{Df}
D_\mu[A] f(x,p,Q)\equiv 
[\partial_\mu-g f^{abc}Q_c A_{\mu,b}{\partial ^Q_a}]f(x,p,Q)\ ,
\eeq
does. Notice that \Eq{Df} combines the first two terms of \Eq{NA-f}.
Here and in the sequel we use the shorthand notation
$\partial_\mu\equiv\partial/\partial x^\mu$,
$\partial_\mu^p\equiv\partial/\partial p^\mu$ and
$\partial^Q_a\equiv\partial/\partial Q^a$.  The invariance of the
third term in \Eq{NA-f} follows from the trivial observation that
$Q_aF^a_{\mu\nu}$ is invariant under gauge transformations, which
establishes the gauge invariance of \Eq{NA-f}.  This terminates the
review of the basic microscopic quantities.

\newpage

\section{Macroscopic approach}
\label{Ensemble}

\subsection{General considerations}
\label{EnsembleGeneral}

Within the semi-classical approach introduced in Section~\ref{Micro},
all information about properties of the non-Abelian plasma is given by
the microscopic dynamical equations as written down in the previous
section. However, for most situations not all the microscopic
information is of relevance. Of main physical interest are the
characteristics of the system at large length scales. This includes
quantities like damping rates, colour conductivities or screening
lengths within the kinetic regime, or transport coefficients like
shear or bulk viscosities within the hydrodynamic regime. The
microscopic length scales, like typical inter-particle distances, are
much smaller than such macroscopic scales.

There are two closely related aspects worth noticing when performing
the transition from a microscopic to an effective, or macroscopic,
description.  We first observe that the classical problem as described
in the previous section is well-posed only if all initial conditions
for the particles are given. If the system under study contains a
large number of particles it is impossible to follow their individual
trajectories. A natural step to perform is to switch to a {\it
  statistical} description of the system. In this way, the stochastic
character of the initial conditions are taken into account. It follows
that the microscopic distribution function can no longer be considered
a deterministic, but rather a {\it stochastic} quantity. This program
is worked out in detail in the following two sections.

Given the statistical ensemble which represents the state of the
system, the macroscopic properties should be given as functions of the
fundamental parameters and the interactions between the particles.
This requires an appropriate definition of macroscopic quantities as
ensemble averages. Within kinetic theory, the basic `macroscopic'
quantity is the one-particle distribution function, from which all
further macroscopic observables can be derived. The aim of a kinetic
theory is to construct, with as little restrictions or assumptions as
possible, a closed set of transport equations for this distribution
function \cite{GLW:1980}. Such an approach assumes implicitly that the
`medium', described by the distribution function, is continuous. If
the medium is {\it not} continuous, stochastic fluctuations due to the
particles can be taken into account as well, and their consistent
inclusion leads to {\it effective} transport equations for correlators
of fluctuations and the one-particle distribution function
\cite{K0,K,K2}. The random fluctuations of the distribution function
are at the root of the dissipative character of the effective
transport theory.

An alternative reading of the above invokes the notion of {\it
  coarse-graining}. This amounts to an averaging of both the
microscopic distribution function and of the non-Abelian fields over
characteristic physical volumes. The resulting effective kinetic
equations dissipative due to the coarse-graining over microscopic
quantities, and require the consistent inclusion of a corresponding
noise term. This is very similar to the phenomenological Langevin
approach to dissipative systems \cite{Landau5}. In the regime where
fluctuations can be taken as linear these two approaches are
equivalent \cite{K}. We come back to this point of view in
Section~\ref{Langevin}, where its application to the theory of
non-Abelian fluctuations in plasmas is discussed \cite{LM4}.

In this section, we work out the first line of thought. The basics
related to the Gibbs ensemble average in phase space are discussed,
and the basic correlators reflecting the stochastic fluctuations are
derived. In the following section, this procedure is applied to the
microscopic transport equations, ultimately resulting in a closed set
of macroscopic transport equations.

\subsection{Ensemble average}
\label{EnsembleGibbs}

As we are studying classical point particles in phase space, the
appropriate statistical average corresponds to the Gibbs ensemble
average for classical systems \cite{K,Landau5}.  We will review the
main features of this procedure as defined in phase space.  Let us
remark that this derivation is completely general, valid for any
classical system, and does not require equilibrium situations.

We introduce two basic functions. The first one is the phase space
density function $n(z)$ which gives, after integration over a phase
space volume element, the number of particles contained in that
volume.  Microscopically the phase space density function reads
\beq
n(z) = \sum_{i=1}^L \delta [z - z_i(\tau)]\ ,
\eeq 
where $z$ are the phase space coordinates, and $z_i$ the trajectory of
the particle $i$ in phase space. Let us also define the distribution
function ${\bf \rho}$ of the microstates of a system of $L$ identical
classical particles. Due to Liouville's theorem, $d{\bf \rho}/dt=0$.
Thus, it can be normalised as
\beq
\int dz_1 dz_2 \ldots d z_L\ {\bf \rho}(z_1,z_2, \ldots , z_L, t) = 1 \ .
\eeq 
For simplicity we have considered only one species of particles.  The
generalisation to several species of particles is straightforward.

The statistical average of any function $G$ defined in phase space is
given by
\beq
\langle G \rangle = 
\int dz_1 dz_2 \ldots d z_L \, G(z_1,z_2,\ldots , z_L) \ 
{\bf \rho}(z_1,z_2, \ldots , z_L, t) \ .
\eeq
A particularly important example is the one-particle distribution
function, which is obtained from ${\bf \rho}$ as
\beq
 f_1 (z_1,t) = 
V \int dz_2 \ldots d z_L\ 
{\bf \rho}(z_1,z_2, \ldots , z_L, t) \ .
\eeq
Here $V$ denotes the phase space volume.  Correspondingly, the
two-particle distribution function is
\beq
 f_2 (z_1,z_2,t) = 
V^2 \int dz_{3} \ldots d z_L\ 
\, {\bf \rho}(z_1,z_2, \ldots, z_L, t) \ ,
\eeq
and similarly for the $k$-particle distribution functions.  A complete
knowledge of ${\bf \rho}$ would allow us to obtain all the set of $(
f_1, f_2, \ldots , f_L)$ functions; this is, however, not necessary
for our present purposes.

Notice that we have allowed for an explicit dependence on the time $t$
of the function ${\bf \rho}$, as this would typically be the case in
out-of-equilibrium situations. We will drop this $t$ dependence from
now on to simplify the formulas.

Using the above definition one can obtain the first moment (mean
value) of the microscopic phase space density.  The statistical
average of this function is
\beq
\langle n(z) \rangle = 
\int dz_1 dz_2 \ldots d z_L \
{\bf \rho}(z_1,z_2,\ldots, z_L)
\sum_{i=1}^L \delta (z - z_i) = \s0{L}{V}\, f_1 (z) \ .
\eeq 
The second moment $\langle n (z)\, n (z') \rangle$ can similarly
be computed, and it is not difficult to see that it gives
\beq
\langle n (z)\, n (z') \rangle = 
\s0{L}{V} \delta (z-z')  f_1(z) +\s0{L (L-1)}{V^2} f_2(z,z')\ .
\eeq
Let us now define a deviation of the phase space density from its mean
value
\beq
\de n (z) \equiv n (z) - \langle n(z) \rangle \ .
\eeq
By definition $\langle \de n(z) \rangle = 0$, although the second
moment of this statistical fluctuation does not vanish in general,
since
\beq
\langle \de n (z)\, \de n (z') \rangle = \langle  n (z)
n (z') \rangle
-  \langle n(z) \rangle \langle n(z') \rangle \ .
\eeq
If the number of particles is large, $L \gg 1$, we have
\beq\label{BasicGibbs}
\langle \de n (z)\, \de n (z') \rangle =
\left(\s0{L}{V}\right)\delta (z-z') f_1(z) +
\left(\s0{L}{V}\right)^2 g_2(z,z') \ ,
\eeq
where the function 
\beq
\label{2-point-cor}
g_2(z,z') = f_2(z,z') - f_1(z) f_1(z')
\eeq
measures the two-particle correlations in the system.  Notice that the
above statistical averages are well defined in the thermodynamic
limit, $L, V \rightarrow \infty$ but $L/V$ remaining constant.

Similarly, one can define the $k$-point correlator of fluctuations,
and the $k$-point correlation function $g_k(z,z', \ldots)$.  In an
ideal (non-interacting) system, the $k$-particle distribution function
factorizes $f_k = \prod_{i=1}^k f_1$, and hence $g_k \equiv 0$, simply
because all particles in the system are statistically independent from
each other.  Interactions induce correlations among particles.
Typically, higher order correlations depend on the distance between
particles and have a characteristic (finite) range or correlation
length. Exceptions are met close to critical points of phase
transitions, where correlation lengths tend to diverge.

Starting from the Liouville equation, obeyed by the distribution
function $\rho$, it is possible to deduce a set of chained equations
for the $k$-point distribution functions $f_k$. For non-relativistic
systems this is the BBGKY hierarchy. These equations exhibit a
hierarchical structure: the determination of the $k$-particle
distribution function requires the knowledge of the $(k+1)$ particle
function. Alternatively, one can describe the set of equations obeyed
by the correlation functions $g_k$. These equations are non-linear due
to the non-linear relationship between $f_k$ and $g_k$. In the
following section we will describe a different approach.  It is based
on deducing the equations for the statistical fluctuations, and their
correlators. When using some approximate methods, this approach is
more effective in a number of cases, as we will explicitly illustrate
in the following sections.

\subsection{Basic equal-time correlators}
\label{EnsembleBasic}

We now return to the case of our concern.  The statistical averages
have to be performed in phase space. The phase space density function
\Eq{NA-PD} is a function of the time $t$, the vectors $\vx $ and
$\vp$, and the set of canonical variables ${\bphi}$ and ${\bpi}$.  We
scale for later convenience the density factors $L/V$ into the mean
functions $\langle f\rangle $. Those small changes in the
normalisation simplify slightly the notations of the equations. Also,
to adopt a unified description of both the classical and quantum
plasmas, from now on we will use dimensionless distribution functions,
replacing the measure $d^3xd^3p$ by $d^3xd^3p/(2\pi\hbar)^3$ (although
working in natural units $\hbar =1$). This change in the measure also
affect the normalisation of the basic correlators, as we will show
below.

We now turn to the basic correlators which will be of relevance for
later applications. For a classical plasma, the basic equal-time
correlator obtained from averaging over initial conditions follows
from \Eq{BasicGibbs} after the redefinitions as indicated above as
\bea
\langle \delta f_{{\bf x},p,Q}   \, 
        \delta f_{{\bf x}',p',Q'} \rangle_{t=0} 
&=& (2 \pi)^3 \delta^{(3)}({\bf x}-{\bf x}')
    \delta^{(3)}({\bf p}-{\bf p}') 
    \delta (Q- Q')\, \bar f
\nonumber\\ &&
  + {\tilde g}_2({\bf x},p,Q;{\bf x}',p',Q')\      \label{average}
\eea
for each species of particles and each internal degree of freedom.
The function ${\tilde g}_2$ comes from the two-particle correlator,
and
\beq\label{deltaQQ}
\delta (Q -Q') = \frac{1}{c_R} \de({\bphi} - {\bphi}') \
                               \de ({\bpi} -  {\bpi}) \ ,
\eeq
and $\bphi$, ${\bpi}$ are the Darboux variables associated to the
colour charges $Q_a$.  The appearance of the factor $1/c_R$ in the
above expression is due to the change of normalisation factors
associated to the functions $n$ and $f$.

Within the semi-classical approach, the quantum statistical properties
of the particles are taken into account as well. For bosons, and for
every internal degree of freedom, this amounts to replacing
\Eq{average} by
\bea
\langle\delta f_{{\bf x},p,Q} \ \delta f_{{\bf x}',p',Q'}\rangle_{t=0} 
&=& (2\pi)^3 \delta^{(3)}({\bf x}-{\bf x}')
             \delta^{(3)}({\bf p}-{\bf p}') 
             \delta (Q- Q') \bar f_{\rm B} (1 + \bar f_{\rm B})
\nonumber\\ &&
             + {\tilde g}^{\rm B}_2({\bf x},p,Q;{\bf x}',p',Q') \ ,
\label{averageB}
\eea
for the quadratic correlator of fluctuations. For fermions, the
corresponding equal-time correlator is
\bea
\langle\delta f_{{\bf x},p,Q} \ \delta f_{{\bf x}',p',Q'}\rangle_{t=0} 
&=& (2\pi)^3\delta^{(3)}({\bf x}-{\bf x}')
            \delta^{(3)}({\bf p}-{\bf p}') 
            \delta (Q- Q') \bar f_{\rm F} (1 - \bar f_{\rm F})
\nonumber\\ &&
             + {\tilde g}^{\rm F}_2({\bf x},p,Q;{\bf x}',p',Q') \ . 
\label{averageF}
\eea
The functions ${\tilde g}^{\rm B}_2$ or ${\tilde g}^{\rm F}_2$ are the
bosonic or fermionic two-particle correlation function, up to a
normalisation factor. The above relations could be derived from first
principles in a similar way as \Eq{average}.  In the limit $\bar
f_{\rm B/F}\ll 1$ they reduce to the correct classical value.  We
present a justification of the use of the above correlators in
Appendix B.  It also has to be pointed out that the correlators
\eq{averageB} and \eq{averageF} have been derived for the cases of
both an ideal gas of bosons and ideal gas of fermions close to
equilibrium. Hence, the above correlators can be taken as the correct
answer in the case where the non-Abelian interactions are
perturbative.

\newpage

\section{Effective transport theory}
\label{Effective}

The goal of transport theory is to derive a closed system of dynamical
equations based on the one-particle distribution function from which
all macroscopic characteristics can be derived \cite{GLW:1980}.  Given
the prescription as to how statistical averages over the particles in
phase space have to be performed, we apply this formalism to the
quasi-particle distribution function, the non-Abelian gauge fields and
the dynamical equations themselves. This first step results in a set
of transport equations for the distribution function coupled to
correlators of statistical fluctuations \cite{LM2}.  The dynamical
equations are worked out in their most general form.  Integrating-out
the fluctuations, in a second step, yields the seeked-for effective
kinetic theory for the mean fields only.  Such a procedure amounts to
a derivation of collision terms, noise sources and fluctuation-induced
currents. As usual, applications are tied to certain systematic
approximations, which are discussed as well. We mainly follow the
lines of reasoning as first outlined in \cite{LM,LM2}. For a brief
summary, see \cite{LM4}.

\subsection{Mean fields vs.~fluctuations}
\label{EffectiveMean}

To perform the step from the microscopic to the macroscopic
formulation of the problem, we take the ensemble average of the
microscopic equations \eq{NA-Micro}. As argued above, this implies
that the distribution function $f(x,p,Q)$, which in the microscopic
picture is a deterministic quantity, now has a probabilistic nature
and can  be considered as a random function, given by its mean value
and statistical (random) fluctuation about it. Let us define the
quantities
\begin{mathletters}
\label{NA-delta}
\bea
f(x,p,Q)&=& {\bar f}(x,p,Q) + \de f(x,p,Q) \ ,
\zeile
J^\mu_a(x)&=&\bar J^\mu _a (x)+\de J^\mu _a(x) \ ,
\eea
\end{mathletters}%
where the quantities carrying a bar denote the mean values, e.g.
${\bar f} = \langle f \rangle$, ${\bar J} = \langle J \rangle$,  while
the mean value of the statistical fluctuations vanish  by definition,
$\langle \de f \rangle =0$ and   $\langle \de J \rangle =0$.  This
separation into the mean distribution function and the mean current on
the one side, and their fluctuations on the other, takes into account in
particular the stochastic (or source) fluctuations of the one-particle
distribution function. These fluctuations in the quasi-particle
distribution function and in the induced current 
are responsible for fluctuations in the gauge fields as well, and we
therefore split the gauge fields accordingly as
\begin{mathletters}
\bea
A^{a}_\mu(x)&=& {\bar A}^a_\mu(x) + a^a_\mu(x) \ ,
\label{NA-delta-a}
\zeile
\langle A \rangle &=& {\bar A}\ , \quad
\langle a \rangle  = 0 \ .
\label{constraint}
\eea
\end{mathletters}%
Notice that the split of the gauge fields \Eq{NA-delta-a} has to be
seen on a different footing as the split for the one-particle
distribution function.  These gauge field degrees of freedom are not
defined in phase space. Their fluctuations are induced by those of the
particles.  We postpone a detailed discussion on further implications
due to gauge symmetry until Section~\ref{Consistent}.

Effectively, such a split corresponds to a separation of the low
frequency or long wavelength modes associated to the mean quantities
from the high frequency or short wavelength modes associated to the
fluctuations.  As we shall see below, the relevant momentum scales
depend on the approximations employed. They are identified explicitly
for a plasma close to thermal equilibrium (see
Sections~\ref{Scales}-\ref{Beyond}).

The induced fluctuations in the gauge fields \Eq{NA-delta-a} require
additionally the split of the field strength tensor as
\begin{mathletters}
\label{NA-deltaF}
\bea
F^{a}_{\mu\nu}     &=&   {\bar F}^a_{\mu\nu} + f^a_{\mu\nu}\ ,
\zeile
{\bar F}^a_{\mu\nu}&=&   F^{a}_{\mu\nu}[\bar A]\ ,
\zeile
f^a_{\mu\nu}       &=&   (\bar D_\mu a_\nu-\bar D_\nu a_\mu)^a
                         + g f^{abc}a^b_\mu a^c_\nu\ .
\label{NA-deltaF-c}
\eea
\end{mathletters}%
We used $\bar D_\mu\equiv D_\mu[\bar A]$. The term $f^a_{\mu\nu}$
contains terms linear and quadratic in the fluctuations. Note that the
statistical average of the field strength $\langle
F^a_{\mu\nu}\rangle$ is not only given by $\bar F^a_{\mu\nu}$, but
rather by
\beq 
\langle F^a_{\mu\nu}\rangle
= \bar F^a_{\mu\nu}+ g f^{abc}\langle a^b_\mu a^c_\nu\rangle\ , 
\eeq 
due to quadratic terms contained in $f^a_{\mu\nu}$.

\subsection{Effective transport equations}
\label{EffectiveDynamical}

We now perform the step from the microscopic to the macroscopic
Boltzmann equation by taking the statistical average of
\Eqs{NA-Micro}. This yields the dynamical equation for the mean
values,
\begin{mathletters}
\label{NA-Macro}
\begin{equation}
\label{NA-1}
p^\mu\left(\bar D_\mu- gQ_a\bar F^a_{\mu\nu} 
 \partial _p^\nu\right)\bar f=\left\langle\eta\right\rangle
+ \left\langle\xi\right\rangle \ . 
\end{equation}
We have made use of the covariant derivative of $f$ as introduced in
\Eq{Df}. The macroscopic Yang-Mills equations are 
\begin{equation}
\label{NAJ-1} 
\bar D_\mu \bar F^{\mu\nu}  + \left\langle J_{\mbox{\tiny fluc}}^{\nu}
\right\rangle =\bar J^\nu \ .
\end{equation}
\end{mathletters}%
In \Eqs{NA-Macro}, we collected all terms quadratic or cubic in the
fluctuations into the functions $\eta(x,p,Q), \xi(x,p,Q)$ and
$J_{{\mbox{\tiny fluc}}}(x)$. These terms are qualitatively new as
they are not present in the original set of microscopic transport
equations. Their physical relevance is discussed in
Section~\ref{EffectiveCollisions} below.  Written out explicitly, they
read
\begin{mathletters}
\label{NA-func}
\begin{eqnarray}
\eta(x,p,Q) 
&\equiv & 
\ \ gQ_a\,p^\mu
\left(\bar D_\mu a_\nu-\bar D_\nu a_\mu\right)^a\,
\partial_p^\nu\,\delta f(x,p,Q)
\nonumber \zeile 
&&
+ g^2\,Q_a\,p^\mu
\,f^{abc}\, a^b_\mu a^c_\nu\,\partial_p^\nu \,\delta f(x,p,Q)\ ,
\label{NA-eta}
\\[1ex]
\xi(x,p,Q) 
&\equiv & 
\ \ g\,p^\mu f^{abc}Q^c\, a_\mu^b\,\partial_Q^a \delta f(x,p,Q)\ 
\nonumber \zeile 
&&
+g^2 p^\mu f^{abc}Q^c\,a_\mu^a\, a_\nu^b\,\partial^\nu_p\bar f(x,p,Q) ,    
\label{NA-xi}
\zeile
J_{\mbox{\tiny fluc}}^{a\nu}(x)
&\equiv & 
\ \ g \left[f^{dbc} \bar D^{\mu}_{ad}  a_{b\mu} a_c^\nu  
 + f^{abc} a_{b\mu}\, (\bar D^\mu a^\nu-\bar D^\nu a^\mu)_c\right]
\nonumber \zeile 
&&
    + g^2 f^{abc} f^{cde}\, a_{b\mu}\, a^\mu_d \, a^\nu_e \, .   
\label{NA-Jfluc}
\end{eqnarray}
\end{mathletters}%
We remark in passing that the split of the fluctuation-induced terms
into $\eta$ and $\xi$ is, to some extend, arbitrary. The term $\eta$
stems entirely from the fluctuations of the field strength tensor
\Eq{NA-deltaF-c} and the fluctuation of the distribution function.  In
turn, $\xi$ contains two contributions of different origin: the
fluctuation fields from the covariant derivative term, and the
fluctuation gauge fields to quadratic order of the field strength
tensor multiplied with the mean value of the distribution function.
The first term is due to fluctuations of the `drift term' in the
Boltzmann equation. The second term can be seen as a
fluctuation-induced force. Both vanish identically in the Abelian case
(see Section~\ref{EffectiveCollisions} below).

The effective transport equations (\ref{NA-Macro}) are not yet a
closed system of differential equations involving only the mean
fields. They still contain correlators of fluctuations, for which the
appropriate transport equations have to be studied separately. They
are obtained by subtracting \Eqs{NA-Macro} from \Eqs{NA-Micro}. The
result is
\begin{mathletters}
\label{NA-fluc}
\begin{eqnarray}
p^\mu\left(\bar D_\mu-gQ_a\bar F^a_{\mu\nu}\partial _p^\nu\right)\delta f
&=&
\ \ g Q_a(\bar D_\mu a_\nu-\bar D_\nu a_\mu)^a p^\mu \partial^p_\nu\bar f
\nonumber \zeile 
&& 
+ gp^\mu a_{b\mu} f^{abc} Q_c\partial ^Q_a \bar f 
\nonumber \zeile 
&& 
+ \eta 
+ \xi 
- \left\langle \eta + \xi \right\rangle
\label{NA-2}
\zeile
\left[\bar D^2 a^\mu-\bar D^\mu(\bar D_\nu a^\nu)\right]^a
+2 gf^{abc} \bar F_b^{\mu\nu}a_{c\nu}
&=& 
\delta J^{a\mu} -J_{{\mbox{\tiny fluc}}}^{a\mu}
+\left\langle J_{{\mbox{\tiny fluc}}}^{a\mu}\right\rangle
\ .
\label{NAJ-2}
\end{eqnarray}
\end{mathletters}%
The above set of dynamical equations -- in addition to the initial
conditions as derived in the previous section from the Gibbs ensemble
average -- is at the basis for a description of all transport
phenomena in the plasma.

While the dynamics of the mean fields \Eqs{NA-Macro} depends on
correlators quadratic and cubic in the fluctuations, the dynamical
equations for the fluctuations \Eqs{NA-fluc} also depend on higher
order terms (up to cubic order) in the fluctuations themselves.  The
dynamical equations for the higher order correlation functions are
contained in \Eqs{NA-fluc}.  To see this, consider for example the
dynamical equation for the correlators $\langle\delta f\, \delta
f\rangle$.  After multiplying \Eq{NA-2} with $\delta f'$ and taking
the statistical average, we obtain
\begin{eqnarray}
p^\mu\left(\bar D_\mu - gQ_a\bar F^a_{\mu\nu}\partial _p^\nu\right)
\langle\delta f\, \delta f' \rangle
&=&g Q_a p^\mu \partial ^p_\nu\bar f\ 
\llangle (\bar D_\mu a_\nu-\bar D_\nu a_\mu)^a \delta f' \rrangle
\nonumber 
\zeile && 
+gp^\mu  f^{abc} Q_c\partial ^Q_a \bar f \ \llangle a_{b,\mu}\,
\delta f'\rrangle
\nonumber 
\zeile && 
\label{quad-corr}
+\llangle (\eta  + \xi-\langle \eta  + \xi\rangle )\, \delta f'\rrangle \ .
\end{eqnarray}
(To simplify the notation, we have not given the arguments of all
fields explicitly. In particular, $\langle\delta f\, \delta f'
\rangle$ means $\langle\delta f(x,p,Q)\, \delta f(x',p',Q') \rangle$,
and the derivatives act only on the $(x,p,Q)$ dependences.) In the
same way, we find for $\langle\delta f\, \delta f'\, \delta
f''\rangle$ the dynamical equation
\begin{eqnarray}
p^\mu\left(\bar D_\mu - gQ_a\bar F^a_{\mu\nu}\partial _p^\nu\right)
\langle\delta f\, \delta f'\, \delta f''\rangle
&=&g Q_a p^\mu \partial ^p_\nu\bar f\ 
\llangle 
(\bar D_\mu a_\nu-\bar D_\nu a_\mu)^a \, \delta f'\, \delta f''
\rrangle
\nonumber 
\zeile &&
+gp^\mu  f^{abc} Q_c\partial ^Q_a \bar f \ 
\llangle a_{b,\mu}\, \delta f'\, \delta f''\rrangle
\nonumber 
\zeile && 
+\llangle (\eta  + \xi- \langle \eta  + \xi\rangle)\,
\delta f'\, \delta f''\rrangle  
\ ,
\label{cubic-corr}
\end{eqnarray}
and similarly for higher order correlators.  Typically, the dynamical
equations for correlators of $n$ fluctuations will couple to
correlators ranging from the order $(n-1)$ up to order $(n+2)$ in the
fluctuations.  From cubic order onwards, the back-coupling contains
terms non-linear in the correlation functions.

The correlators of gauge field fluctuations, like $\langle a\, \de
f'\rangle$ and $\langle a\, a'\rangle$ or higher order ones, are
related to those of the one-particle distribution function through the
Yang-Mills equations. For example, from \Eq{NAJ-2}, we deduce for the
quadratic correlators
\bea
\left[\left(\bar D^2\de^{\mu\nu}-\bar D^\mu\bar D^\nu\right)^{ab} 
+ 2g f^{acb}\bar F^{\mu\nu}_c\right]
\llangle a_{\nu b} \, \de f'\rrangle
&=&
\int dPdQ\,p^\mu\,Q^a\,\llangle \de f\, \de f' \rrangle
\nonumber
\zeile &&
-\llangle (J^{\mu a}_{{\mbox{\tiny fluc}}}
  -\langle J^{\mu a}_{{\mbox{\tiny fluc}}}\rangle)\, \de f' \rrangle
\eea
and 
\bea
\left[\left(\bar D^2\de^{\mu\nu}-\bar D^\mu\bar D^\nu\right)^{ab} 
+ 2g f^{acb}\bar F^{\mu\nu}_c\right]
\llangle a_{\nu b} \, a'_{\rho d}\rrangle
&=&
\int dPdQ\,p^\mu\,Q^a\,\llangle \de f\, a'_{\rho d} \rrangle
\nonumber
\zeile &&
-\llangle (J^{\mu a}_{{\mbox{\tiny fluc}}}
  -\langle J^{\mu a}_{{\mbox{\tiny fluc}}}\rangle)\, a'_{\rho d}\rrangle\ .
\eea
In this manner, the full hierarchy of coupled dynamical equations for
all $n$-point correlation functions are obtained. The initial
conditions are the equal-time correlation functions as derived from
the ensemble average.

The resulting hierarchy of dynamical equations for the correlators is
very similar to the BBGKY hierarchy within non-relativistic
statistical mechanics \cite{Balescu1975}.  A decisive difference stems
from the fact that the present set of dynamical equations is
dissipative even in their un-approximated form, while the complete
BBGKY hierarchy remains time-inversion invariant
\cite{Balescu1975,K2}.

\subsection{Collision integrals, noise and induced currents}
\label{EffectiveCollisions}

Let us comment on the qualitatively new terms $\eta, \xi$ and
$J_{\mbox{\tiny fluc}}$ as defined in \Eqs{NA-func}.  In the effective
Boltzmann equation, the functions $\langle\eta\rangle$ and
$\langle\xi\rangle$ appear only after the splitting \Eqs{NA-delta} has
been performed. These terms are qualitatively different from those
already present in the initial transport equation.  The correlators
$\langle\eta\rangle$ and $\langle\xi\rangle$ are interpreted as
effective collision integrals of the macroscopic Boltzmann equation.
The fluctuations in the distribution function of the particles induce
fluctuations in the gauge fields, while the gauge field fluctuations,
in turn, induce fluctuations in the motion of the quasi-particles. In
the present formalism, the correlators of statistical fluctuations
have the same effect as collisions. This yields a precise recipe for
obtaining collision integrals within semi-classical transport theory.

The term $\langle\eta\rangle$ contains the correlator $\langle
f^a_{\mu\nu}\delta f\rangle$ between the fluctuations of the field
strength and the fluctuations of the distribution function.  In the
Abelian limit, only the collision integral $\langle\eta\rangle$
survives and \Eqs{NA-Macro} and \eq{NA-fluc} reduce to the known set
of kinetic equations for Abelian plasmas \cite{K}. Then,
$\langle\eta\rangle$ can be explicitly expressed as the Balescu-Lenard
collision integral \cite{Balescu1960,Lenard1960} after solving the
dynamical equations for the fluctuations and computing the correlators
involved \cite{K,Landau10}. This proves in a rigorous way the
correspondence between fluctuations and collisions in an Abelian
plasma.

The term $\langle\xi\rangle$ contains two contributions. The term
proportional to $\langle\delta D \delta f\rangle$ leads to a collision
integral due to the fluctuations of the `drift covariant derivative'.
The term proportional to $\langle\de f^a_{\mu\nu}\rangle\bar f$ is
interpreted as a fluctuation-induced force term, because $\langle\de
f^a_{\mu\nu}\rangle$ is to be seen as a fluctuation-induced field
strength in the effective transport equation. Both terms describe a
purely non-Abelian effect, they vanish identically in the Abelian
limit.

At the same time we observe the presence of stochastic noise in the
effective equations. The noise originates in the source fluctuations
of the particle distributions and induces {\it field-independent}
fluctuations to the gauge fields. The corresponding terms in the
effective Boltzmann equations are therefore $\eta$, $\xi$ and
$J_{\mbox{\tiny fluc}}$ at vanishing mean field or mean current. An
explicit example is given in Section~\ref{BeyondStochastic}.

Finally, we observe the presence of a fluctuation-induced current
$\langle J_{\mbox{\tiny fluc}} \rangle$ in the effective Yang-Mills
equation for the mean fields. This current, due to its very nature,
stems from the induced correlations of gauge field fluctuations. It
vanishes identically in the Abelian case. While the collision
integrals are linear in the quasi-particle fluctuations, the induced
current only contains the gauge field fluctuations. As the
fluctuations of the one-particle distribution function are the basic
source for fluctuations, we expect that a non-vanishing induced
current will appear as a subleading effect.

In order to find explicitly the collision integrals, noise sources or
the fluctuation-induced currents for non-Abelian plasmas, one has to
solve first the dynamical equations for the fluctuations in the
background of the mean fields.  This step amounts to incorporating the
fluctuations within the mean particle distribution function
(`integrating-out' the fluctuations). In general, this is a difficult
task, in particular due to the non-linear terms present in
\Eqs{NA-fluc}. As argued above, this will only be possible when some
approximations have been performed.

\subsection{Systematic approximations}
\label{EffectiveSystematic}

The coupled set of dynamical equations, as derived and presented here
within a semi-classical point particle picture, are exact.  No further
approximations apart from the original assumption have been made. In
order to solve the fluctuation dynamics, it is necessary to apply some
systematic approximations, or to find a reasonable truncation for the
hierarchy of dynamical equations for correlator functions. With
`solving' we have in mind finding explicit solutions to the dynamics
of fluctuations. When reinserted into the mean field equations it
should be possible to obtain explicit expressions for them. Such a
procedure amounts to incorporating the physics at larger scales, as
described by the fluctuations, into the mean quasi-particle
distribution function.

Here, two systematic approximation schemes are outlined: an expansion
in moments of the fluctuations and an expansion in a small gauge
coupling.  Although they have distinct origins in the first place, we
will see below (Section~\ref{Consistent}) that they are intimately
linked due to the requirements of gauge invariance.

\subsubsection*{1. Expansion in moments of the distribution function}

An expansion in moments of the fluctuations has its origin in the
framework of kinetic equations. Effectively, the kinetic equations
describe the coherent behaviour of the particles within some
physically relevant volume. This coherent behaviour is described by
the {\it plasma parameter} $\epsilon$, the inverse of which measures
the number of particles within a physically relevant volume element as
described by the one-particle distribution function.  In a
close-to-equilibrium plasma, the plasma parameter is given by the
ratio between the cube of the mean particle distance and the Debye
radius (see Section~\ref{Scales}). The fluctuations in the number of
particles become arbitrarily small if the physical volume -- or the
number of particles contained in it -- can be made arbitrarily large.
For realistic situations, both the physical volume and the particle
number are finite. Still, the fluctuations remain at least
parametrically small and suppressed by the plasma parameter
\cite{K,Landau5}.  Hence, the underlying expansion parameter for an
expansion in moments of fluctuations is a small {plasma parameter}
\beq \epsilon \ll 1\ .  
\eeq 
The leading order approximation in an expansion in moments is the {\it
  first moment approximation}. It consists in imposing \beq\label{1st}
f=\bar f\ ,\quad {\rm or}\quad \de f\equiv 0\ , \eeq and corresponds
to neglecting all fluctuations throughout. Sometimes it is referred to
as the mean field or {\it Vlasov approximation}. It leads to a closed
system of equations for the mean one-particle distribution function
and the gauge fields. In particular, the corresponding Boltzmann
equation is dissipationless.  It remains time-inversion
invariant\footnote{This holds as long as the initial conditions do not
  violate explicitly time-inversion invariance (cf.~Landau damping)
  \cite{K}.} as does the original microscopic transport equation.

Beyond leading order, the {\it second moment approximation} takes into
account the corrections due to correlators up to quadratic order in
the fluctuations $\langle \de f\de f\rangle$. All higher order
correlators like
\beq\label{2nd}
\langle \de f_1\de f_2\ldots\de f_n\rangle=0
\eeq 
for $n>2$ are neglected within the dynamical equations for the mean
fields and the quadratic correlators. This approximation is viable if
the fluctuations remain sufficiently small (see also
Section~\ref{Scales}). We remark that the second moment approximation,
the way it is introduced here, and unlike the first moment
approximation, no longer yields a closed system of dynamical equations
for the one-particle distribution function and quadratic correlators.
The reason for this is that the initial conditions for the evolution
of correlators, which are given by the equal-time correlation
functions as derived from the Gibbs ensemble average, still do involve
the two-particle correlation functions. Hence, in addition to
\Eq{2nd}, we have to require that two-particle correlators remain
small as compared to products of one-particle distribution functions,
\beq\label{two-particle} g_2\ll f_1\,f_1\ , \eeq so that $f_2 \approx
f_1 f_1$ (see \Eq{2-point-cor}).  This is the case if the effective
dynamical equations are valid at scales larger than typical scales of
two-particle correlations. Once the method is put to work, it is
possible to check explicitly whether \Eq{two-particle} holds true or
not.  The combined approximations \Eqs{2nd} and \eq{two-particle} are
known as the {\it approximation of second correlation functions},
sometimes also referred to as the {\it polarisation approximation}
\cite{K}.

For the dynamical equations of the fluctuations \Eqs{NA-fluc} these
approximations imply that the terms non-linear in the fluctuations
should be neglected to leading order. This corresponds to setting
\begin{mathletters}\label{polarisation}
\bea
\eta-\langle\eta \rangle&=&0\ ,\zeile
\xi -\langle \xi \rangle&=&0\ ,\zeile
J_{{\mbox{\tiny fluc}}}-\langle J_{{\mbox{\tiny fluc}}}\rangle&=&0\ .
\eea
\end{mathletters}%
The essence of this step is that the dynamical equations for the
correlators become homogeneous. It is easy to see that \Eq{quad-corr}
or \Eq{cubic-corr} depend only on quadratic or cubic correlators,
respectively, once \Eqs{polarisation} are imposed. This approximation
permits truncating the infinite hierarchy of equations for the mean
fields and the correlators of fluctuations down to a closed system of
differential equations for both mean quantities and quadratic
correlators.  The mean fields then couple only to quadratic
correlators, and all higher order correlators couple amongst
themselves. This turns the dynamical equation for the fluctuations
\Eqs{NA-fluc} into a differential equation linear in the fluctuations.
Notice, however, that the approximation \Eq{polarisation} can be
improved even within the second moment approximation, if these
differences are found to be again linear in the fluctuations.

The polarisation approximation is the minimal choice necessary to
genuinely describe dissipative processes, because it takes into
account the feed-back of stochastic fluctuations within the particle
distribution function.
 
In the light of the discussion in Section~\ref{EffectiveCollisions},
the second moment approximation \Eq{2nd} can be interpreted as
neglecting three-particle collisions in favour of two-particle
collisions within the collision integrals for the mean fields. Notice
that no explicit correlators higher than cubic order appear in the
collision integral \Eq{NA-1}. Effective four- or more-particle
interactions will only appear due to the back-coupling of the
quadratic and cubic correlators to higher order ones.  The
approximation \Eq{polarisation} for the fluctuations can be
interpreted as neglecting the back-coupling of collisions to the
dynamics of the fluctuations. Beyond leading order, this approximation
is modified and the right-hand side of \Eqs{polarisation} will be
replaced by the effective collision terms as obtained to leading
order. If such terms turn out to be linear in the fluctuations we can
perform a `resummed' polarisation approximation, taking the higher
order effects iteratively into account.

\subsubsection*{2. Expansion in the gauge coupling}

A qualitatively different approximation scheme concerns the
non-Abelian sector of the theory, characterised by a small gauge
coupling $g$. It is possible to perform a systematic perturbative
expansion in powers of the gauge coupling $g$, keeping only the
leading order terms.  This can be done because the differential
operator appearing in the effective Boltzmann equation \Eq{NA-1}
admits such an expansion.  In a small coupling expansion, the force
term $g\,p^\mu Q_a\bar F^a_{\mu\nu}\partial _p^\nu$ is suppressed by a
power of $g$ as compared to the leading order term $p^\mu\bar D_\mu$.
Notice that expanding the covariant derivative term $p^\mu\bar D_\mu$
of \Eq{Df} into powers of $g$ is not allowed as it will break gauge
invariance. In this spirit, we expand
\bea
\bar f&=&      \bar f^{(0)}
         +g\,  \bar f^{(1)}
         +g^2\,\bar f^{(2)}
         +\ldots           \label{f-g}
\eea
and similarly for $\de f$. This is at the basis for a systematic
organisation of the dynamical equations in powers of $g$.

To leading order, this concerns in particular the cubic correlators
appearing in $\langle\eta\rangle$ and $\langle J_{\mbox{\tiny
    fluc}}\rangle$. They are suppressed by a power of $g$ as compared
to the quadratic ones. Hence, the second moment approximation and an
expansion in a small gauge coupling are mutually compatible. At the
same time, the quadratic correlator $\sim f_{abc}\langle
a^b\,a^c\,\rangle$ within $\langle\xi\rangle$ is also suppressed by an
additional power of $g$ and should be suppressed to leading order. We
shall show in the following section that such approximations are
consistent with the mean field gauge symmetry.

A word of caution is due at this point. While a small gauge coupling
appears to be at the basis for perturbative expansions, it cannot be
excluded that {\it another} dimensionless expansion parameter becomes
relevant due to particular dynamical properties of the system. Indeed,
as we shall see below in the close-to-equilibrium plasma, at higher
order the natural expansion parameter happens to be $[\ln (1/g)]^{-1}$
instead of $g$. This implies that expansions like \Eq{f-g} might be
feasible only for the first few terms.

In principle, after these approximations are done, it should be
possible to express the correlators of fluctuations  appearing in
\Eqs{NA-Macro} through known functions. This requires finding a
solution of the fluctuation dynamics first.

\subsection{Conservation laws}
\label{EffectiveConservation}

After this discussion of the basic set of dynamical equations we
return to the conservation laws for the energy-momentum tensor and
current conservation. 
In the same spirit as the splitting of the fundamental variables into
fluctuations and mean values we split the energy-momentum tensor of
the gauge fields into  the part from the mean fields and the
fluctuations,  according to
\begin{mathletters}

\bea
\Theta^{\mu \nu} &=&\bar \Theta^{\mu \nu} + \theta^{\mu \nu}\ ,
\zeile
\bar \Theta^{\mu \nu} & = & 
\s014 g^{\mu \nu} \bar F^a_{\rho \sigma}  \bar F_a^{\rho \sigma} +
 \bar F^{\mu \rho}_{a}  \bar F^{a\,\,\nu}_{\rho} \ , 
\zeile
\theta^{\mu \nu} & = & 
\s012 g^{\mu\nu}\bar F^a_{\rho\sigma}f^{\rho\sigma}_a
+\bar F^{\mu\rho}_af^a_{\rho\nu}
+\bar F^{\nu\rho}_a f^a_{\rho\mu}
+ \s014 g^{\mu \nu}  f_{\rho \sigma, a} f^{\rho \sigma, a} 
+  f^{\mu \rho}_{a}   f^{a\,\,\nu}_{\rho} \ .
\eea
\end{mathletters}%
The term $\theta^{\mu \nu}$ contains the fluctuations up to quartic order.
Due to the non-linear character of the theory, we find that the ensemble
average of the energy momentum tensor is not only given by 
$\bar \Theta^{\mu \nu} $, but
\beq
\langle\Theta^{\mu \nu}\rangle=\bar \Theta^{\mu \nu} 
+\langle\theta^{\mu \nu}\rangle\ .
\eeq
The dynamical equation for the energy momentum tensor of the 
gauge fields comes from the average of \Eq{cons-en-mo}. The corresponding
one for the particles is found after integrating \Eq{NA-f} over 
$dPdQ\, p^\mu$. The two of them read
\begin{mathletters}
\label{em-conservation}

\begin{eqnarray}
\partial_\nu \bar \Theta^{\mu \nu} 
+ \partial_\nu \llangle \theta^{\mu \nu}\rrangle
&=& - \bar F^{\mu \nu}_a \bar J_{\nu a} 
- \langle f^{\mu \nu}_a \delta J_{\nu a}\rangle 
-  \llangle f_a^{\mu\nu} \rrangle  \bar J_\nu^a \ ,
\zeile
\partial_\nu \bar t^{\mu \nu} 
&=&\ \ \, \bar F^{\mu \nu}_a \bar J_{\nu a} 
+ \langle f^{\mu \nu}_a \delta J_{\nu a}\rangle 
+  \llangle f_a^{\mu\nu} \rrangle  \bar J_\nu^a \ .
\end{eqnarray}
\end{mathletters}%
Hence, we confirm that the total mean energy-momentum tensor 
$\langle T^{\mu \nu}\rangle$ is conserved,
\beq
\partial_\nu \langle T^{\mu \nu}\rangle = 0\ .
\eeq
The condition for the microscopic current conservation 
translates, after averaging, into two equations, one for the mean 
fields, and another one for the fluctuation fields. From 
$\langle D_\mu J^ \mu\rangle=0$ we obtain
\beq\label{DJ}
  (\bar D_\mu \bar J^\mu)_a
+ g f_{abc}\langle a^b_\mu \de J^{c \mu}\rangle
= 0 \ .
\eeq
For the fluctuation current, 
we learn from $D_\mu J^ \mu-\langle D_\mu J^ \mu\rangle=0$ that
\beq\label{DdJ}
(\bar D_\mu \de J^\mu)_a 
+ gf_{abc}\left( a^b_\mu \bar J_c^\mu + a^b_\mu \de J_c^\mu- 
\langle a^b_\mu \de J_c^\mu\rangle\right)=0\ .
\eeq
Similar equations are obtained from the Yang-Mills equations
themselves. Here, we only remark that these two set of equations are
mutually consistent, which is shown in
Section~\ref{ConsistentCurrent}.

\subsection{Entropy}
\label{EffectiveEntropy}

Finally, we shall also introduce the kinetic entropy $S$ associated 
to these particles. The entropy density $S^\mu(x)$, as a function
of the one-particle distribution function, is defined as
\beq\label{Smicro}
S^\mu(x)=\int dPdQ\, p^\mu\, {\Sigma}[f](x,p,Q)\ ,
\eeq
from which the entropy obtains as
\beq
S(t)=\int d^3x\, S_0(x)\ .
\eeq
The function ${\Sigma}[f](x,p,Q)$ depends on the statistics of the particles. 
For classical (Maxwell-Boltzmann) statistics, we have \cite{GLW:1980}
\beq\label{Sigma-cl}
{\Sigma}_{\rm cl}[f]=-f\left(\ln f -1\right)
\eeq
while for quantum (Bose-Einstein or Fermi-Dirac) statistics, we use
\beq\label{Sigma-qm}
{\Sigma}_{\rm qm}[f]=-f\ln f \pm (1\pm f)\ln \left(1\pm f\right)
\eeq
instead. The `$-$' sign stands for bosonic degrees of freedom, and the
`$+$' sign for fermionic ones.  Microscopically, the entropy
\Eq{Smicro} is conserved, $dS/dt=0$. This follows from the vanishing
of
\beq
\partial_\mu S^\mu(x)=0\ ,
\eeq
which, for both classical or quantum plasmas, can be deduced from
inserting the microscopic Boltzmann equation \Eq{NA-f} into
\Eq{Smicro}. Ultimately, this is linked to the fact that the Boltzmann
equation contains no explicit collision term.

On the macroscopic level, and after separation into
mean field contributions and fluctuations, the entropy four-flow
reads
\begin{mathletters}\label{Smacro}
\bea
S^\mu(x)&=&\bar S^\mu(x)+\Delta S^\mu(x)
\zeile
\bar S^\mu(x)&=&\int dPdQ\, p^\mu\,\Sigma[\bar f]
\zeile
\Delta S^\mu(x)&=&\int dPdQ\, p^\mu\,
\left[\Delta \Sigma^{(1)} + \Delta \Sigma^{(2)}\right]\ .
\eea
\end{mathletters}%
We have separated the terms of linear order in $\de f$ into
$\Delta \Sigma^{(1)}$, and all the higher order terms into 
$\Delta \Sigma^{(2)}$. They read explicitly
\begin{mathletters}
\bea
\Delta \Sigma^{(1)}_{\rm cl} 
&=&  
-\de f\, \ln \bar f
\zeile
\label{Sigma2-cl}
\Delta \Sigma^{(2)}_{\rm cl} 
&=& 
-(\bar f + \de f)\ln\left(1+{\de f}/{\bar f}\right) 
+\de f
\eea
\end{mathletters}%
for the classical plasma, and
\begin{mathletters}
\label{MeanSmacro}

\bea
\Delta \Sigma^{(1)}_{\rm qm}
&=&
-\de f\, \left[\ln \bar f - \ln (1\pm \bar f)\right] 
\zeile
\label{Sigma2-qm}
\Delta \Sigma^{(2)}_{\rm qm} 
&=&
-(\bar f + \de f)\ln\left(1+{\de f}/{\bar f}\right)
\pm (1\pm \bar f \pm \de f)\ln\left[1\pm{\de f}/({1\pm \bar f})\right]
\eea
\end{mathletters}%
for quantum plasmas.
The presence of fluctuations is closely linked to dissipative
processes. In particular, the mean entropy density is no longer given by 
the entropy density of the mean particle distribution, but rather by
\beq
\label{Mean-Smu}
\langle S^\mu(x)\rangle
=
 \bar S^\mu(x)
+\int dPdQ\, p^\mu\,\langle \Delta \Sigma^{(2)}\rangle \ ,
\eeq
which involves arbitrarily high order correlation functions of the
fluctuations.
 
\subsection{Discussion}
\label{EffectiveDiscussion}

A self-contained semi-classical transport theory has been derived from
a microscopic point particle picture. The ensemble average transformed
the microscopic kinetic equation into a coupled set of dynamical
equations for mean distribution functions, mean fields, and correlator
functions of fluctuations. Usually, a kinetic description considers
the plasma as a continuous medium. Here, the stochastic fluctuations
are taken into account as well. The source of stochastic noise is
given by the fluctuations of the one-particle distribution function.
These enter the initial conditions for the dynamics of the correlation
functions. Fluctuations in the gauge fields are induced by the latter.

The split into mean quantities and fluctuations is convenient for
several reasons. First of all, it separates the short-scale
characteristics of the plasma, associated to the fluctuations, from
the large-scale ones, associated to the mean quantities.  Second, the
set of coupled dynamical equations can be reduced to an effective
transport equation for the mean fields only, at least within some
approximations. This amounts to the `integrating-out' of fluctuations.
Only then are the modes associated to the fluctuations incorporated in
the quasi-particle distribution function.

Two systematic approximation schemes have been discussed, an expansion
in the plasma parameter and an expansion in a small gauge coupling.
These schemes are mutually compatible and linked further by the
requirement of gauge invariance (Section \ref{Consistent}).  On a
technical level, this procedure corresponds to a recipe for {\it
  deriving} collision integrals and the corresponding noise sources
for the effective transport equation, and a fluctuation-induced
current for the Yang-Mills equation.
 
We stress that the present formalism is applicable for both in- and
out-of-equilibrium situations. This is due to the fact that the
statistical properties of the system are all encoded in the Gibbs
ensemble average, which in turn does not rely on a
close-to-equilibrium situation.  A detailed discussion of the gauge
symmetry, and in particular the consistency of the split
\Eq{NA-delta-a}, is given in Section \ref{Consistent}.

\newpage

\section{Gauge symmetry}
\label{Consistent}

The formalism developed in the two preceding sections is based on a
split of non-Abelian gauge fields into a mean field and a fluctuation
field.  Accordingly, their original dynamical equation, the Yang-Mills
equation, splits into two separate ones. Ultimately, one aims at
integrating-out the fluctuation fields such that the remaining
effective theory only involves mean fields. It remains to be shown
that such a procedure is consistent with the requirements of gauge
symmetry.

The idea of splitting gauge fields into two parts in order to
integrate-out the fluctuation part is not new. The {\it background
  field method} is precisely one such formalism based on a path
integral approach.\footnote{For a discussion of the background field
  method applied to QCD, see Abbott \cite{Abbott}.} Within the
background field method, the gauge fields in the path integral are
formally separated into a mean field piece and a quantum piece. The
original gauge symmetry splits accordingly into a {\it background
  gauge symmetry} under which the quantum field transforms
homogeneously, and a {\it quantum gauge symmetry}, under which the
mean field transforms trivially. The background field formalism allows
the derivation of an effective theory for the mean fields only, which
corresponds to the integrating-out of the quantum fluctuations. It is
to be noticed that the background field is an auxiliary field, which
is identified with the mean field only after the quantum field has
been integrated out. The quantum gauge symmetry is the physical gauge
symmetry, which, after the quantum field is integrated out, is
inherited by the mean gauge field symmetry. The converse is not true
\cite{Abbott}.  An application of the background field method within
the QCD transport equation for Wigner functions has been considered by
Elze \cite{Elze:1990gm}.

The present formalism is very similar to such a procedure. Here, we
aim at integrating-out induced stochastic fluctuations as opposed to
quantum ones. Furthermore, after having integrated-out these
fluctuations within a given approximation, the resulting effective
Boltzmann equation can be seen as the generating functional for the
mean gauge field interactions entering the effective Yang-Mills
equations.

In this section the requirements of gauge symmetry are exploited. It
is shown that the present approach is consistent within the background
field approach. This discussion will concern the consistency of the
general set of equations. The question of consistent approximations
will be raised as well. In this section, we shall for convenience
switch to a matrix notation, using the conventions $A\equiv A^a t_a$,
$Q\equiv Q^a t_a$ etc., as well as $[t_a,t_b]=f_{abc}t^c$ and $\Tr\ 
t_at_b = -\s012 \de_{ab}$.

\subsection{Background  gauge symmetry vs.~fluctuation gauge symmetry}
\label{ConsistentGauge}

To begin with, let us consider finite gauge transformations,
parametrised as
\begin{mathletters}
\step
\label{finite}
\bea
gA'_\mu &= &U(x)(\partial_\mu + gA_\mu)U^{-1}(x)\ ,
\zeile\step 
U(x)    &=&\exp \left[-g\eps^a(x)t_a \right]\ ,
\eea
\end{mathletters}%
with parameter $\eps^a(x)$. Under these transformations, we have the
transformation laws
\begin{mathletters}
\step
\bea
         Q' &= &U(x)\,      Q         \, U^{-1}(x)\ ,\zeile\step
\partial_Q' &=& U^{-1}(x)\, \partial_Q\, U(x)\      ,\zeile\step
F_{\mu\nu}' &=& U(x)\,      F_{\mu\nu}\, U^{-1}(x)\ .
\eea
\end{mathletters}%
From the definition of the microscopic distribution function
$f(x,p,Q)$ we conclude that it transforms as a scalar under (finite)
gauge transformations,
\beq 
f'(x,p,Q')=f(x,p,Q)\ ,
\eeq 
which has been shown in \cite{KLLM2} and establishes that the
microscopic set of equations \eq{NA-Micro} transform covariantly under
the gauge transformations \Eq{finite}.

When switching to a macroscopic description a statistical average has
to be performed. The averaging procedure $\langle\ldots \rangle$ as
defined in Section~\ref{Ensemble} is naturally invariant under gauge
transformations. It remains to be shown that the subsequent split of
the gauge field into a mean (or background) field and a fluctuation
field respects the gauge symmetry. We split the gauge field as
\begin{mathletters}
\label{split-gauge}
\step
\bea
\label{split-A}
A_\mu                &=& \bar A_\mu+a_\mu
\zeile\step
\label{split-a}
\langle A\rangle &=& \bar A + \langle a \rangle\ .
\eea 
\end{mathletters}%
For the time being, $\bar A$ is an arbitrary constant, and in
particular, we shall {\it not} yet require $\langle a \rangle=0$.  The
field $\bar A$ is identified as the mean field only when the
additional constraint $\langle a \rangle=0$ is employed. Only then
does the dynamical equation reduce to those discussed in the preceding
section.

The separation \Eq{split-gauge} is very similar to what is done in the
background field method \cite{Abbott,Elze:1990gm}.  Two symmetries are
left after the splitting is performed, the {\it background gauge
  symmetry},
\begin{mathletters}\label{BGS}\step
\bea
g \bar A'_\mu&=&U(x)(\partial_\mu + g\bar A_\mu)U^{-1}(x)\ ,
\zeile\step
       a_\mu'&=&U(x)\, a_\mu\, U^{-1}(x)\ ,
\eea
\end{mathletters}%
and the {\it fluctuation gauge symmetry},
\begin{mathletters}\label{QGS}\step
\bea
g \bar A'_\mu &=&  0\ ,
\zeile\step
      ga_\mu' &=&  U(x)\left(\partial_\mu 
               + g(\bar A_\mu+a_\mu)\right)U^{-1}(x)\ .
\eea
\end{mathletters}%
Under the background gauge symmetry, the fluctuation field transforms
covariantly (as a vector in the adjoint). 

In the first step, we split the microscopic Boltzmann equation
\eq{NA-f} according to \Eq{split-A}. It follows trivially that the
resulting equation is invariant under both the mean field symmetry
\Eq{BGS} and the fluctuation field symmetry \Eq{QGS}, if both $\bar f$
and $\delta f$ transform as $f$, that is, as scalars.

Turning to the macroscopic equations, we perform the statistical
average and split the transport equation into those for mean fields
and fluctuations. We employ the fundamental requirement $\langle \de
f\rangle =0$, but leave $\langle a\rangle$ unrestricted. It is useful
to rewrite the effective transport equations in matrix convention. We
have
\begin{mathletters}\label{M-NA-Macro}\step
\begin{eqnarray}
p^\mu\left(\bar D_\mu
  +2 g\ \Tr\,\big( Q\bar F_{\mu\nu}\big)\,
             \partial _p^\nu\right)\bar f
&=& 
  \left\langle\eta\right\rangle
+ \left\langle\xi\right\rangle 
+ \left\langle\zeta\right\rangle \ ,
\label{M-NA-1}
\zeile\step
\left[ \bar D_\mu,  \bar F^{\mu\nu} \right]  
+ \left\langle J_{\mbox{\tiny fluc}}^{\nu}\right\rangle 
+ \left\langle J_{\mbox{\tiny lin}}^{\nu}\right\rangle 
&=& 
  \bar J^\nu \ . \label{M-NAJ-1} 
\end{eqnarray}
\end{mathletters}%
Notice that \Eqs{M-NA-Macro} appears to be of the same form as
\Eqs{NA-Macro}, except for the new terms $\langle\zeta\rangle$ and
$\langle J_{\mbox{\tiny lin}}^{\nu}\rangle$, which contain the pieces
linear in $\langle a\rangle$. The functions $\eta$, $\xi$ and
$J_{\mbox{\tiny fluc}}$ read
\begin{mathletters}\label{M-NA-func}\step
\begin{eqnarray}
\eta(x,p,Q) &=&
   -2 g \ \Tr\, \left(Q\,[\bar D_\mu,a_\nu]-Q\,[\bar D_\nu,a_\mu]\right)\, 
p^\mu \partial _p^\nu \delta f(x,p,Q)  
\nonumber
\zeile 
&&
-2 g^2 \ \Tr\, \left(Q\,[a_\mu,a_\nu]\right)
p^\mu \partial _p^\nu \delta f(x,p,Q) \ , 
\label{M-NA-eta}
\zeile\step
\xi(x,p,Q)
&=& -2 g\ p^\mu\ \Tr\,\big([Q,\partial ^Q]\, a_\mu\big)\, \delta f(x,p,Q)\ 
\nonumber
\zeile 
&& -2 g^2p^\mu\ \Tr\,\big([a_\mu, a_\nu]Q\big)\,\partial^\nu_p\bar f(x,p,Q)\ ,
\label{M-NA-xi}
\zeile\step
J_{{\mbox{\tiny fluc}}}^{\nu}(x)&=& 
       g\left[\bar D^{\mu},[a_{\mu},a^\nu]\right]  
   +   g\left[a_{\mu},[\bar D^\mu, a^\nu]\right]
   -   g\left[a_{\mu},[\bar D^\nu, a^\mu]\right]
\nonumber\zeile 
&&   + g^2\left[a_\mu,[a^\mu,a^\nu]\right] \ .
\label{M-NA-Jfluc}
\end{eqnarray}
\end{mathletters}%
while the linear terms  $\zeta$ and $J_{\mbox{\tiny lin}}$ are given by
\begin{mathletters}\label{M-NA-func-lin}\step
\begin{eqnarray}
\zeta(x,p,Q) &=&
-2 g\ \Tr\,\left( Q[\bar D_\mu, \, a_\nu  ]
                   -Q[\bar D_\nu,  a_\mu  ]\right)
\,p^\mu \partial_p^\nu\bar f(x,p,Q) 
\nonumber
\zeile 
&&
-2 g\ p^\mu\ \Tr\,\left([Q,\partial^Q] a_{\mu} \right)\,
\bar f(x,p,Q) \ ,
\label{M-NA-zeta}
\zeile\step
J_{{\mbox{\tiny lin}}}^{\nu}(x)&=& 
g \left[ a_\mu ,\bar F^{\mu\nu}\right]
  + \left[\bar D_\mu,[\bar D^\mu, a^\nu]\right]
  - \left[\bar D_\mu,[\bar D^\nu, a^\mu ]\right] \ .
\label{M-NA-Jlin}
\end{eqnarray}
\end{mathletters}%
For $\langle a \rangle=0$, the terms $\langle \zeta \rangle$ and
$\langle J_{{\mbox{\tiny lin}}} \rangle$ vanish, and \Eqs{M-NA-Macro}
reduce to \Eqs{NA-Macro}. Along the same lines, the fluctuation
dynamics becomes
\begin{mathletters}\label{M-NA-fluc}\step
\begin{eqnarray}
p^\mu\left(\bar D_\mu 
           - g\ \Tr\,\big(Q\bar F_{\mu\nu}\big)\,\partial_p^\nu\right)
\delta f&=&-2 g \Tr\,\left(Q[\bar D_\mu, a_\nu]-Q[\bar D_\nu, a_\mu]\right)
\, p^\mu\partial_p^\nu\bar f 
\nonumber
\zeile 
&&
-2 g\, p^\mu \ \Tr\,\big([Q,\partial^Q]a_{\mu}\big)\,\bar f
- \left\langle\zeta\right\rangle
\nonumber
\zeile 
&&
+\eta  + \xi - \left\langle\eta+\xi\right\rangle \ , 
\label{M-NA-2}
\zeile\step
\left[\bar D_\nu,[\bar D^\nu,a^\mu]\right]-\left[\bar D^\mu,[\bar
D_\nu,a^\nu]\right]&+&2 g[\bar F^{\mu\nu},a_{\nu}]=
\delta J^{\mu} 
 - J_{\mbox{\tiny fluc}}^{\mu}
 + \left\langle 
       J_{\mbox{\tiny fluc}}^{\mu}
      +J_{\mbox{\tiny lin}}^{\mu}
  \right\rangle .
\label{M-NAJ-2}
\end{eqnarray}
\end{mathletters}%
Again, the vanishing of $\langle a \rangle$, and hence of $\langle
\zeta \rangle$ and $\langle J_{{\mbox{\tiny lin}}} \rangle$, reduces
\Eqs{M-NA-fluc} to \Eqs{NA-fluc}.

It is straightforward, if tedious, to confirm that this coupled set of
differential equations \eq{M-NA-Macro} to \eq{M-NA-fluc} is invariant
under both the fluctuation gauge symmetry \Eqs{QGS} and under the
background field symmetry \Eqs{BGS}. It suffices to employ the
cyclicity of the trace, and to note that $a_\mu$ and background
covariant derivatives of it transform covariantly. This establishes
that the full gauge symmetry of the underlying microscopic set of
equations is respected at the effective mean field level.

The next step,  to finally obtain the set of equations given in the
preceding section, involves the requirement that the statistical
average of the  gauge field fluctuation vanishes, $\langle
a\rangle=0$.  This additional constraint  is fully compatible with the
background gauge symmetry, as $\langle a\rangle=0$ is invariant under
\Eqs{BGS}. Any inhomogeneous transformation law for $a$, and in
particular \Eq{QGS},  can no longer be a symmetry of the macroscopic
equations as the constraint $\langle a\rangle=0$ is not invariant
under the  fluctuation gauge symmetry. This is similar to what happens
in the background field method, where the  fluctuation gauge symmetry
can no longer be seen once the expectation value of the fluctuation
field is set to zero. As we have just verified,  the symmetry
\Eqs{QGS} is observed in both \Eqs{NA-1} and \eq{NA-2}, 
as long as the terms linear in $\langle a\rangle$ are
retained. 

The value for $a$ is obtained when the dynamical equations for the
fluctuations are solved explicitly. This requires that some gauge for
the fluctuation field has to be fixed. For any (approximate) explicit
solution which expresses the fluctuation field $a$ as a functional of
the source fluctuations of the particle distribution function one has
to check for consistency that the intial constraint $\langle
a\rangle=0$ is satisfied.  If the solution $a$ turns out to be a {\it
  linear} functional of $\de f|_{t=0}$, this is automatically
satisified. An example for this is encountered in Section
\ref{Beyond}. This justifies the dynamical equations as given in
Section \ref{Effective}.

\subsection{Current conservation}
\label{ConsistentCurrent}

In \Eqs{DJ} and \eq{DdJ}, we have given the equations which imply the
covariant current conservation of the mean and the fluctuation
current.  However, this information is contained both in the transport
and in the Yang-Mills equation. It remains to be shown that these
equations are self-consistent. For the reasons detailed above it is
sufficient to consider from now on only the case $\langle a\rangle=0$.

We start with the mean current $\bar J$. Performing $g\int dPdQ\, Q$ of  
the transport equation  \Eq{M-NA-1}, we find
\beq\label{DJ1}
0=  [\bar D_\mu, \bar J^\mu] 
  + g \langle[a_\mu,\de J^\mu] \rangle\ .
\eeq 
This is \Eq{DJ}. In order to obtain \Eq{DJ1}, we made use of the following 
moments of the effective transport equation,
\begin{mathletters}\label{someintegrals}\step
\bea
   \int dP\ \eta(x,p,Q)                         &=&0 \ ,
\zeile\step
\int dP\ p^\mu F_{\mu\nu}\partial_p^\nu f(x,p,Q)&=&0 \ ,
\zeile\step
   \int dPdQ\, Q\ \xi(x,p,Q)                    &=& - [a_\mu,\de J^\mu] \ ,
\zeile\step
  g\int dPdQ\, Q\ p^\mu \bar D_\mu \bar f(x,p,Q)&=&[\bar D_\mu,\bar J^\mu]\ .
\eea
\end{mathletters}%
On the other hand we can simply take the background-covariant derivative
of the mean field Yang-Mills equation, \Eq{M-NAJ-1}, to find
\beq
0 = [\bar D_\mu, \bar J^\mu] 
   -[\bar D_\mu,\langle J_{{\mbox{\tiny fluc}}}^\mu \rangle]\ .
\eeq
This equation has to be consistent with \Eq{DJ1}.
Combining them, we end up with the {\it consistency condition}
\beq\label{consistencyM}
0= [\bar D_\mu,\langle J_{{\mbox{\tiny fluc}}}^\mu \rangle]
  + g \langle[a_\mu,\de J^\mu] \rangle\ .
\eeq
This consistency condition links the background covariant derivative of 
some correlator of induced gauge field fluctuations with the correlator 
between the current fluctuations and the gauge field fluctuations. Such 
a condition can hold because the gauge field fluctuations are induced 
by those of the current.

In order to prove the consistent current 
conservation for the mean fields \Eq{consistencyM}, and the corresponding
equation for the fluctuation current, it is useful to establish explicitly 
the following identity
\beq   \label{general-cb}
 0=   [\bar D_\mu,J_{{\mbox{\tiny fluc}}}^\mu ]
   +g [a_\mu,\de J^\mu]
   +g [a_\mu,\langle J_{{\mbox{\tiny fluc}}}^\mu \rangle] \ .
\eeq
The check of \Eq{general-cb} is algebraic, and it 
will  make use of symmetry arguments like the antisymmetry of the 
commutator and the tensors $\bar F_{\mu\nu}, f_{\mu\nu}$, and of the 
Jacobi identity $[t_a,[t_b,t_c]]+[t_b,[t_c,t_a]]+[t_c,[t_a,t_b]]=0$. The 
identity $[\bar D_\mu,\bar D_\nu]=g\bar F_{\mu\nu}$ is employed as well. 
To simplify the computation, we will separate the fluctuation part of 
the field strength \Eq{NA-deltaF-c} into the terms linear and quadratic 
in $a$, according to
\begin{mathletters}\step
\bea
f_{\mu\nu}   &=& f_{1,\mu\nu}+f_{2,\mu\nu}               \ ,
\zeile\step
f_{1,\mu\nu} &=&   [\bar D_\mu,a_\nu]-[\bar D_\nu,a_\mu] \ ,
\zeile\step
f_{2,\mu\nu} &=& g [a_\mu,a_\nu]\ .
\eea
\end{mathletters}%
Recall furthermore, using \Eq{M-NA-Jfluc} and \Eq{M-NAJ-2}, that
\bea
J_{{\mbox{\tiny fluc}}}^\mu&=&
               [\bar D_\nu,f^{\nu\mu}_2] 
             +g[a_\nu,f^{\nu\mu}_1+f^{\nu\mu}_2]  \ ,        \label{Jfluc}
\zeile 
\de J^\mu &=& [\bar D_\nu,f^{\nu\mu}_1]
             + g[a_\nu,\bar F^{\nu\mu}]
             +J_{{\mbox{\tiny fluc}}}^\mu
             -\llangle J_{{\mbox{\tiny fluc}}}^\mu\rrangle \label{deJ}
\eea
are functions of the fluctuation field $a$. The first term of 
\Eq{general-cb} reads, after inserting  $J_{{\mbox{\tiny fluc}}}$ 
from \Eq{Jfluc}, 
\bea
\label{lhs}
[\bar D_\mu,J_{{\mbox{\tiny fluc}}}^\mu]&=& 
               [\bar D_\nu,[\bar D_\mu,f^{\mu\nu}_{2}]]
             +g[\bar D_\nu,[a_\mu,f^{\mu\nu}_{1}]]
             +g[\bar D_\nu,[a_\mu,f^{\mu\nu}_{2}]]\ .
\eea
Using $\de J$ from \Eq{deJ}, it follows for the second term of 
\Eq{general-cb}
\bea
\label{rhs}
g[a_\mu,\de J^\mu]&=& g^2[a_\nu,[a_\mu,\bar F^{\mu\nu}]]
        +g[a_\nu,[\bar D_\mu,f^{\mu\nu}_{1}]]
        +g\left[a_\nu,J_{\mbox{\tiny fluc}}^{\nu}\right]
        -g\left[a_\nu,\langle J_{\mbox{\tiny fluc}}^{\nu}\rangle\right]\ .
\quad
\eea
The last term of \Eq{rhs} will be cancelled by the last term in 
\Eq{general-cb}. We show now that the first three terms of \Eq{lhs} 
and \Eq{rhs} do cancel one by one. The first term in \Eq{lhs} can be 
rewritten as
\beq
[\bar D_\nu,[\bar D_\mu,f^{\mu\nu}_{2}]] 
              =  [[\bar D_\nu,\bar D_\mu],f^{\mu\nu}_2] 
                -[\bar D_\nu,[\bar D_\mu,f^{\mu\nu}_2]]  
              =  \s012 g [\bar F_{\nu\mu},f^{\mu\nu}_2]\ .
\eeq
Similarly, the first term of \Eq{rhs} yields
\beq
g^2[a_\nu,[a_\mu,\bar F^{\mu\nu}]]  
              = -g^2[\bar F^{\mu\nu},[a_\nu,a_\mu]]
                -g^2[a_\nu,[a_\mu,\bar F^{\mu\nu}]] 
              = -\s012 g [\bar F_{\mu\nu},f^{\nu\mu}_2]\ .
\eeq
For the second term in \Eq{lhs} we have
\beq
g[\bar D_\nu,[a_\mu,f^{\mu\nu}_{1}]]  
              =  g[a_\mu,[\bar D_\nu,f^{\mu\nu}_{1}]]
                +g[[\bar D_\nu,a_\mu],f^{\mu\nu}_{1}] 
              = -g[a_\mu,[\bar D_\nu,f^{\nu\mu}_{1}]] \ ,
\eeq
which equals (minus) the second term of \Eq{rhs}. Finally, consider the
third term of \Eq{rhs},
\bea
g\left[a_\nu,J_{\mbox{\tiny fluc}}^{\nu}\right]
&=&    g^2[a_\nu,[\bar D_\mu,[a^\mu,a^\nu]]] 
     + g^2[a_\nu,[a_\mu,f^{\mu\nu}]]           \nonumber\zeile
&=&    \s012 g[f^{\mu\nu}_2,f_{1,\mu\nu}]
     - g[\bar D_\mu,[f_2^{\mu\nu},a_\nu]]
     -\s012 g[f^{\mu\nu}_2,f_{1,\mu\nu}]       \nonumber\zeile
&=&  -g[\bar D_\mu,[a_\nu,f_2^{\nu\mu}]]\ ,
\eea
which equals (minus) the third term of \Eq{lhs}. This establishes
\Eq{general-cb}.

Returning to our main line of reasoning we take the average of 
\Eq{general-cb} which reduces it to
\Eq{consistencyM} and establishes the self-consistent conservation 
of the mean current.
The analogous consistency equation for the fluctuation current follows
from \Eq{M-NA-2} after performing $g\int dPdQ\, Q$, and reads
\beq\label{consistency2}
 0=  [\bar D_\mu,\de J^\mu] 
   + g[a_\mu,\de J^\mu]
   + g[a_\mu,\bar J^\mu]
   - g\langle[a_\mu,\de J^\mu]\rangle\ .
\eeq
This is \Eq{DdJ}. Here, in addition to \Eq{someintegrals}, we made use of
\bea
 2\, g\int dPdQ\, Q\ \Tr\, \big([Q,\partial ^Q]\, a_\mu\big)\,
   \bar f(x,p,Q) &=& g[a_\mu,\bar J^\mu] \ .
\eea
The background covariant derivative of \Eq{M-NAJ-2} is
given as
\beq \label{consistency2b}
0=  [\bar D_\mu,\de J^\mu]
  + g\left[a_\nu,[\bar D_\mu,\bar F^{\mu\nu}]\right]
  - [\bar D_\mu,J_{{\mbox{\tiny fluc}}}^\mu ]
  + [\bar D_\mu,\langle J_{{\mbox{\tiny fluc}}}^\mu\rangle ] \ .
\eeq 
Subtracting these equations yields the consistency condition
\bea
0 &=& [\bar D_\mu,J_{{\mbox{\tiny fluc}}}^\mu ]
   + g[a_\mu,\de J^\mu]
   - [\bar D_\mu,\langle J_{{\mbox{\tiny fluc}}}^\mu\rangle ]
\nonumber\zeile &&   \label{consistencyF}
   -g \langle[a_\mu,\de J^\mu]\rangle
   + g[a_\mu,\bar J^\mu]-g\left[a_\nu,[\bar D_\mu,\bar F^{\mu\nu}]\right] \ .
\eea
\Eq{consistencyF} is a consistency condition which links different
orders of the fluctuations of gauge fields with those of the current.
Using \Eqs{M-NAJ-1}, \eq{consistencyM} and \eq{general-cb} we confirm
\Eq{consistencyF} explicitly. This establishes the self-consistent
conservation of the fluctuation current.

\subsection{Approximations}
\label{ConsistentApprox}

We close this section with a comment on the consistency of {\it
  approximate} solutions. The consistent current conservation can no
longer be taken for granted when it comes to finding approximate
solutions of the equations. On the other hand, finding an explicit
solution will require some type of approximations to be performed. The
relevant question in this context is to know which approximations will
be consistent with gauge invariance.

Consistency with gauge invariance requires that approximations have to
be consistent with the background gauge symmetry. From the general
discussion above we can already conclude that dropping any of the
explicitly written terms in \Eqs{NA-Macro} to \eq{NA-fluc} is
consistent with the background gauge symmetry \Eq{BGS}. This holds in
particular for the first and second moment approximations \Eq{2nd} as
well as for the polarisation  approximation \Eq{polarisation}.

The first moment approximation \Eq{1st} is automatically consistent
with the covariant current conservation, simply because the
approximate equations are structurally the same as the microscopic
ones.

Consistency of the polarisation 
approximation \Eq{polarisation} with covariant current
conservation turns out to be more restrictive. Employing $J_{{\mbox{\tiny
fluc}}}=\langle J_{{\mbox{\tiny fluc}}} \rangle$ implies that
\Eq{consistencyM} is only satisfied if in addition
\beq\label{con1}
0=\left[\bar D_\nu,\llangle \left[a_\mu,[a^\mu,a^\nu]\right]\rrangle\right]
\eeq
holds true. This is in accordance with neglecting cubic correlators
for the collision integrals.

Similarly, the consistent conservation of the fluctuation current
implies the consistency condition \Eq{consistencyF}, and holds if
\beq
\label{con2}
0= \left[ a_\mu, \langle J_{\mbox{\tiny fluc}}^\mu \rangle \right]\ .
\eeq
It is interesting to note that the consistent current conservation
relates the second moment approximation with the neglection of
correlators of gauge field fluctuations.  We conclude, that
\Eqs{polarisation} with \eq{con1} and \eq{con2} form a
gauge-consistent set of approximations.

This terminates the general discussion of a semi-classical transport
theory built upon a microscopical point particle picture. The
following sections discuss the weakly coupled plasmas close to thermal
equilibrium, and the techniques are put to work.

\newpage

\section{Plasmas close to equilibrium}
\label{Scales}

In the remaining part of the article, we employ the present formalism
to the classical and the quantum non-Abelian plasma close to
equilibrium.  Prior to this, we discuss briefly the relevant physical
scales for relativistic classical and quantum plasmas close to
equilibrium.  Here, we restore the fundamental constants $\hbar, c$
and $k_B$ in the formulas.

\subsection{Classical plasmas}
\label{ScalesClassical}

To discuss the relevant physical scales in the classical non-Abelian
plasma, it is convenient to discuss first the simpler Abelian case,
which has been considered in detail in the literature
\cite{Landau5,K}. At equilibrium the classical distribution function
is given by the relativistic Maxwell distribution,
\beq
\label{eq-class}
\bar f^{\rm eq} (p_0) =  \exp\left(\frac{\mu- p_0}{k_B T}\right) \ ,
\eeq
where $\mu$ is the chemical potential. The mean density of particles
${\bar N}$
is then deduced from the above distribution function. If we neglect the
masses of the particles ($m \ll T$), then
\beq
{\bar N} = 8 \pi \left(\frac{k_B T}{2 \pi \hbar c}  \right)^3
e^{\mu/k_BT} \ .
\eeq 
The value of the fugacity of the system $z=e^{\mu/k_BT}$ is then fixed
by knowing the mean density of the plasma \footnote{Note that the
  dependence on $\hbar$ of the mean density arises only because our
  momentum measure is $d^3 p/(2 \pi \hbar)^3$; it is just a
  normalisation constant.}.  The interparticle distance is then $\bar
r \sim {\bar N}^{-1/3}$. As we are considering a classical plasma, we
are assuming $\bar r \gg \lambda_{\rm dB}$, where $\lambda_{\rm dB}$
is the de Broglie wave length, $\lambda_{\rm dB} \sim \hbar/p$, with
$p$ some typical momenta associated to the particles, so that $p\sim
k_B T/c$.  The previous inequality is satisfied if $z \ll 1$, which is
the condition under which quantum statistical effects can be
neglected.

Another typical scale in a plasma close to equilibrium is the Debye
length $r_D$. The Debye length is the distance over which the
screening effects of the electric fields in the plasma are felt.  For
an electromagnetic plasma, the Debye length squared is given by
\cite{K}
\beq
r^2_D = \0{k_B T}{ 4 \pi \bar N  e^2} \ .
\label{ADeb-class}
\eeq
Notice that the electric charge contained in the above formula is a
dimensionful parameter: it is just the electric charge of the point
particles of the system.

In the classical case, and in the absence of the fundamental constant
$\hbar$, the only dimensionless quantities that can be constructed
from the basic scales of the problem are dimensionless ratios of the
basic scales of the problem. The most important one is the {\it plasma
  parameter} $\epsilon$. The plasma parameter is defined as the ratio
\cite{K}
\beq
\epsilon = \0{\bar r^3}{r^3_D} \ .      \label{pla-par}
\eeq
The quantity $1/\epsilon$ gives the number of particles contained in a
sphere of radius $r_D$. If $\epsilon \ll 1$ this implies that a large
number of particles are in that sphere, and thus a large number of
particles are interacting in this volume, and the collective character
of their interactions in the plasma cannot be neglected.  For the
kinetic description to make sense, $\epsilon$ has to be small
\cite{K}. This does not require, in general, that the interactions
have to be weak and treated perturbatively.

Let us now consider the non-Abelian plasma. The interparticle distance
is defined as in the previous case. The main difference with respect
to the Abelian case concerns the Debye length, defined as the distance
over which the screening effects of the non-Abelian electric fields in
the plasma are noticed. It reads
\beq
r^2_D = \0{ k_B T}{ 4 \pi \bar N  g^2 C_2} \ ,
\label{NADeb-class}
\eeq
where $C_2$, defined in \Eq{quadraticQ}, is a dimensionful quantity,
carrying the same dimensions as the electric charge squared in
\Eq{ADeb-class}. The coupling constant $g$ is a dimensionless
parameter. In the non-Abelian plasma one can also construct the plasma
parameter, defined as in \Eq{pla-par}.

It is interesting to note that there are two natural dimensionless
parameters in the non-Abelian plasma: $\epsilon$ and $g$.  The
condition for the plasma parameter being small translates into
\beq
\left(\frac{4 \pi C_2}{k_B T} \right)^{3/2} \bar N^{1/2} g^3 \ll 1 \ ,
\eeq
which is certainly satisfied for small gauge coupling constant $g \ll
1$.  But it can also be fulfilled for a rarefied plasma. Thus, one may
have a small plasma parameter {\it without} having a small gauge
coupling constant. This is an interesting observation, since the
inequalities $\epsilon \ll 1$ and $g \ll 1$ have different physical
meanings. A small gauge coupling constant allows us to treat the
non-Abelian interaction perturbatively, while $\epsilon \ll 1$ just
means having a collective field description of the physics occurring
in the plasma. In principle, these two situations are different. If we
knew how to treat the non-Abelian interactions {\it exactly}, we could
also have a kinetic description of the classical non-Abelian plasmas
without requiring $g \ll 1$.

\subsection{Quantum plasmas}
\label{ScalesQuantum}

Now we consider the quantum non-Abelian plasma, and consider the
quantum counterparts of all the above quantities, as derived from
quantum field theory. For a quantum plasma at equilibrium the one
particle distribution function is
\begin{mathletters}
\label{eq-quan0}
\step
\bea
\label{eq-quan0B}
\bar f^{\rm eq}_{\rm B} (p_0) 
&=&  \frac{1}{\exp{\left(\0{p_0-\mu}{k_B T}\right)}  - 1} \ ,
\zeile
\step
\label{eq-quan0F}
\bar f^{\rm eq}_{\rm F} (p_0) 
&=&  \frac{1}
{\exp{\left(\0{p_0 \mp \mu }{k_B T}\right)}  + 1} \ ,
\eea
\end{mathletters}%
where the subscripts `B' and `F' refer to the Bose-Einstein and
Fermi-Dirac statistics, respectively. In the fermionic distribution
function the $\mp$ sign refers to partices/antiparticles,
respectively.

Note that in the limit of low occupation numbers, one can recover the
classical distribution function from the quantum one. This happens for
large values of the fugacity. Also we should point out that a chemical
potential associated to a specific species of particles can only be
introduced if there is a conserved charge associated to them. Since it
is impossible to associate a global $U(1)$ symmetry to the gluons, one
cannot introduce a gluonic chemical potential.

In the remaining part of this article, we will mainly study the
physics of non-Abelian plasmas close to thermal equilibrium,
and put $\mu =0$. If we further neglect the masses of the particles,
then  the mean density is  
$\bar N \sim (k_B T/\hbar c)^3$. The interparticle distance
$\bar r \sim \bar N^{-1/3}$ becomes of the same order as the de 
Broglie wavelength, which is why quantum statistical effects cannot be
neglected in this case.

The value of the Debye mass is obtained from quantum field theory.  It
depends on the specific quantum statistics of the particles and their
representation of $SU(N)$. From the quantum Debye mass one can deduce
the value of the Debye length, which is of order
\beq
r^2 _D \sim \frac{1}{g^2} \left(\frac{\hbar c}{k_B T} \right)^2 \ .
\eeq
It is not difficult to check that the plasma parameter, defined as in
\Eq{pla-par}, becomes proportional to $g^3$.  Thus $\epsilon$ is small
{\it if and only if} $g \ll 1$.  This is so, because in a quantum
field theoretical formulation one does not have the freedom to fix the
mean density $\bar N$ in an arbitrary way, as in the classical case
({\it i.e.}, one cannot introduce a fugacity for gluons). This
explains why the kinetic description of a quantum non-Abelian plasma
is deeply linked to the small gauge coupling regime of the theory.

\newpage

\section{HTL effective theory}
\label{HTL}

In this section we consider the example of the HTL effective theory.
In physical terms, it describes the leading-order chromoelectrical
screening effects in a non-Abelian plasmas close to thermal
equilibrium \cite{LeBellac:1996}. The HTL-resummed gluon propagator no
longer has poles on the light cone, and the dispersion relation, which
is a complicated function of momenta, yields a screening mass -- the
Debye mass $m_D$ -- for the chromo-electric fields. Also, the HTL
polarisation tensor has an imaginary part which is responsible for the
absorption and emission of soft gluons by the particles, known as
Landau damping. The corresponding effective action for QCD is highly
non-local, but can be brought into a local form when written as a
transport equation.

Within the conceptual framework laid-out in the previous sections the
HTL effective theory can be obtained in a very simple manner
\cite{KLLM,KLLM2}. A prerequisite for a kinetic description to be
viable is a small plasma parameter $\epsilon\ll 1$. We shall ensure
this by assuming that the temperature is sufficiently high such that
the gauge coupling as a function of temperature obeys
\beq
g\ll 1\ .
\eeq 
We also assume that all particle masses $m$ are small compared to both
the temperature and the Debye mass which allows considering them as
massless.  It is then shown that the HTL effective theory emerges
within the simplest non-trivial approximation to the set of transport
equations derived in Section~\ref{Effective}, which is the {\it first
  moment approximation}. Hence, fluctuations will play no part in the
effective mean field dynamical equations. Solving the approximate
transport equations within the first moment approximation and to
leading order in the gauge coupling reproduces the HTL effective
theory.  In the present section we drop the bar on the mean fields for
notational simplicity.

\subsection{Non-Abelian Vlasov equations}
\label{HTL-Vlasov}

We begin with the set of mean field equations (\ref{NA-Macro}) and
neglect the effect of statistical fluctuations entirely, $\de f\equiv
0$. In that case, \Eqs{NA-Macro} become the non-Abelian Vlasov
equations \cite{Heinz:1983nx}
\begin{mathletters}\label{NAV-Macro}\step
\begin{eqnarray}
p^\mu\,D_\mu \,f&=& g\, p^\mu\, Q_a F^a_{\mu\nu}\, 
\partial _p^\nu\, f \ ,          \label{NAV-1}
\zeile\step
D_\mu  F^{\mu\nu} &=& J^\nu \ , \label{NAVJ-1} 
\end{eqnarray}
where the colour current is given by
\beq
\step
J^\mu_a (x) = 
g \sum_{\hbox{\tiny helicities}\atop\hbox{\tiny species}} 
                   \int dP dQ\, Q_a p^\mu f (x,p,Q) \ .
\eeq
\end{mathletters}%
We will omit the species and helicity indices on the distribution
functions, and in the sequel, we will also omit the above sum, in
order to keep the notation as simple as possible. Equation \eq{NAV-1}
is then solved perturbatively, as it admits a consistent expansion in
powers of $g$. Close to equilibrium, we expand the distribution
function as in \Eq{f-g} up to leading order in the coupling constant
\beq
\label{m-close-eq-f}
f(x,p,Q)=     f^{\rm eq}(p_0)
               + g f^{(1)}(x,p,Q) \ .
\eeq
In the strictly classical approach, the relativistic Maxwell
distribution \Eq{eq-class} at equilibrium is used, which is
semi-classically replaced by the corresponding quantum distributions
\Eqs{eq-quan0}.  Here, we consider massless particles with two
helicities as internal degrees of freedom.
 
It is convenient to rewrite the equations in terms of current
densities.  The momentum measure we use is
\beq
d P = \frac{d^4 p}{(2 \pi)^3}\, 2\Theta(p_0)\, \de(p^2)
\eeq
for massless particles.  Consider the current densities
\begin{mathletters}
\step
\begin{eqnarray}
J_{a_1\cdots a_n}^\rho(x,p)
&=& g\ p^\rho  \int dQ \,Q_{a_1}\cdots Q_{a_n} f(x,p,Q), \label{NA-Jp}
\zeile
\step
{\cal J}_{a_1\cdots a_n}^\rho(x,\vv)
&=&       \int d\tilde P\, J_{a_1\cdots a_n}^\rho(x,p) \ .\label{NA-Jv}
\end{eqnarray}
\end{mathletters}%
Here we introduced the vector $v^\mu=(1,{\bf v})$, where ${\bf v}$
describes the velocity of the particle with ${\bf v}^2=1$. The measure
$d\tilde P$ integrates over the radial components. It is related to
\Eq{Pmeasure} by $dP= d\tilde P d\Omega/4\pi$, and reads
\beq
  d\tilde P = 
\frac{1}{2 \pi^2}\, dp_0\,  d|\vp|\, |\vp|^2\, 2\Theta(p_0)\, \de(p^2) \ .
\eeq
The colour current is obtained by performing the remaining angle
integration
\beq
J^\mu_a(x)=\int \mbox{\small{$\frac{d\Omega}{4\pi}$}}{\cal J}^\mu_a(x,v).
\eeq 
For massless particles, a simple consequence of these definitions is
$v^\mu {\cal J}^0_a(x,v) = {\cal J}^\mu_a(x,v)$, a relation we will
use continously.

We now insert \Eq{m-close-eq-f} into \Eqs{NAV-Macro} and expand in
powers of $g$. The leading order term $p\cdot D f^{\rm eq.}(p_0)$
vanishes.  After multiplying \Eq{NAV-1} by $g Q_a p^\rho/p_0$, summing
over two helicities, and integrating over $d\tilde P dQ$, we obtain
for the mean current density at order $g$
\begin{mathletters}
\label{NAV-mean}
\step
\begin{eqnarray} 
\label{NAV-soft-mean}
v^\mu D_\mu {\cal J}^\rho(x,\vv) &=& m^2_D v^\rho v^\mu F_{0\mu}(x)
\label{NAV-curr}  
\ ,
\zeile
\step
D_\mu F^{\mu\nu}(x)  &=&  J^\nu(x) \ ,
\label{soft-YM0}
\end{eqnarray}
\end{mathletters}%
with the Debye mass
\beq\label{class-Debye}
m_D^2 = -  \frac{g^2  C_2}{\pi^2} 
        \int^\infty_0 dp p^2 \frac{d f^{\rm eq}(p)}{d p} \ . 
\eeq
In the classical case the Debye mass is given by the inverse of the
Debye length \Eq{NADeb-class}, as a function of the mean density.  In
the quark-gluon plasma, with gluons in the adjoint representation,
$C_2 =N$, and $N_F$ quarks and $N_F$ antiquarks in the fundamental
representation, $C_2 =1/2$, and all the particles carrying two
helicities, the Debye mass reads
\beq\label{quant-Debye}
m_D^2 = \frac{g^2 T^2}{3} \left( N + \frac{N_F}{2} \right) \ .
\eeq

From \Eq{NAV-soft-mean} we can estimate the typical momentum scale of
the mean fields. If the effects of statistical fluctuations are
neglected, the typical momentum scales associated to the mean current
and the mean field strength are of the order of the Debye mass $m_D$.
Here and in the sequel, we will refer to those scales as soft scales.
The momentum scales with momenta $\ll m_D$ will be referred to as
ultra-soft from now on.

The Boltzmann equation \eq{NAV-soft-mean} is consistent with current
conservation. This follows easily from the $\rho=0$ component of
\Eq{NAV-soft-mean}, which yields
\beq
D_\mu J^\mu
=\int \0{d\Omega}{4\pi}D_\mu {\cal J}^\mu(x,\vv)
=m^2_D F_{0\mu}\int \0{d\Omega}{4\pi} v^\mu
=0 \ .
\eeq
The last equation vanishes because both the angle average of $\vv$ and
the component $F^{00}$ vanish.

\subsection{Solution to the transport equation} 
\label{HTL-Amplitudes0}

In a first step we have found the Boltzmann equation to leading order
in the Vlasov approximation. In a second step, we are interested in
integrating-out the quasi-particle degrees of freedom. This amounts to
solving the Boltzmann equation and to express the induced current
explicitly as a functional of the soft gauge fields
\cite{Blaizot:1994be}.  The solution to \Eq{NAV-curr} is constructed
with the knowledge of the retarded Green's function
\beq
i v^\mu D_\mu \, G_{{\rm ret}}(x,y;\vv) = \delta^{(4)} (x-y) \ .
\eeq
It reads
\beq
\label{ret-GF}
G_{{\rm ret}}(x,y;\vv)_{ab} 
= -i \theta(x_0-y_0)  \delta^{(3)} \left({\bf x}-{\bf y} 
      - {\bf v}(x_0-y_0) \right) \, U_{ab} (x, y) \ .
\eeq
Here, we introduced the parallel transporter $U_{ab}[A]$ which is 
defined via a path-ordered exponential as
\begin{mathletters}
\step
\bea
\label{ParallelTransporterDef}
U(x,y)
&=&
P\,\exp\left(i g \int_y^xdz^\mu\,A_\mu(z)\right)\ ,
\zeile
\step
U_{ab}(x,x)&=& \delta_{ab}\ , 
\eea
\step
and obeys
\beq
\label{ParallelTransporter}
v^\mu D_\mu^x \, \left. U_{ab}(x,y)\right|_{y=x-v t} =0\ .
\eeq
\end{mathletters}%
Using \Eq{ret-GF}, one finds for the current density
\beq
\label{class-HTL-density}
{\cal J}^\mu_a (x,\vv) 
=  -m^2_D \,v^\mu v^\nu \, \int^{\infty}_0 d \tau\ 
U_{ab}(x,x-v \tau)  F_{\nu 0,b} (x-v\tau)  
\eeq
and for the HTL effective current
\beq\label{class-HTL-curr}
J^\mu_a (x) 
=  -m^2_D \int\!
\frac{d\Omega_{{\bf v}}}{4\pi}\,v^\mu v^\nu \,\int^{\infty}_0 d \tau\ 
U_{ab}(x,x-v \tau)   
F_{\nu 0,b} (x-v\tau)  \ .
\eeq
The above colour current agrees with the HTL colour current, if one
uses the value of the Debye mass for the quark-gluon plasma,
\Eq{quant-Debye}.

Inserting \Eq{class-HTL-curr} into \Eq{soft-YM0} yields an effective
theory for the soft gauge fields only. This final step can be seen as
integrating-out the particles from the dynamical equations for the
soft non-Abelian fields. The soft current $J[A]$ of
\Eq{class-HTL-curr} is related as $J(x)=-\de \Gamma_{\rm HTL}[A]/\de
A(x)$ to the generating functional $\Gamma_{\rm HTL}[A]$ for soft
amplitudes. Integrating the HTL current is known to give the HTL
effective action $\Gamma_{\rm HTL}[A]$ explicitly
\cite{Braaten:1990mz,Braaten:1990it,Braaten:1990az,Braaten:1992gm,Frenkel:1990br,Taylor:1990ia,Efraty:1992gk,Efraty:1993pd,KLLM2}.
A simple and elegant expression for the HTL effective action was given
in \cite{Braaten:1992gm}, and it reads
\beq\label{HTL-action}
\Gamma_{\rm HTL}[A] = \frac{m^2_D}{2} \int\!
\frac{d\Omega_{{\bf v}}}{4\pi}\, \int\! d^4x\, d^4y \,{\rm Tr}
\left(F_{\mu \nu}(x) \langle x| \frac{v^\nu v^\rho} {- (v \cdot D)^2 }
|y \rangle F_\rho ^\mu (y)  \right) \ .
\eeq
The leading-order effective action for soft gauge fields is then given
by adding the HTL effective action to the Yang-Mills one.

\subsection{Soft amplitudes}
\label{HTL-Amplitudes-1}

As a first application, we consider the polarisation tensor for soft
gauge fields. The HTL colour current can be expanded in powers of the
gauge fields as
\beq\label{SoftAmplitudes}
J^a_\mu[A]=
      \Pi^{ab}_{\mu\nu}\, A^\nu_b
+\012 \Pi^{abc}_{\mu\nu\rho}\, A^\nu_b\,A^\rho_c
+\ldots\ ,
\eeq 
where the expansion coefficients (or `soft amplitudes') correspond to
1PI-irreducible amplitudes in thermal equilibrium.

We solve the transport equation in momentum space
\cite{Zhang:1996nw,Manuel:1996td} in order to find the explicit
expressions of the HTLs.  Using the Fourier transform
\beq
{\cal  J}_a ^{\mu}(k,v) = \int d^4 x\, e ^{i k \cdot x}\,
 {\cal  J}_a ^{\mu}(x,v) 
\label{eq:3.1}
\eeq
we can write the  transport equation  in momentum space
\bea
& &v \cdot k \,  {\cal J}^{\mu} _a(k,v) +
i g f_{abc} \int \frac{d^4 q}{(2 \pi)^4} \,
v \cdot A^b (k-q)\,  {\cal J}^{\mu c} (q,v) \nonumber \\
& & =  - m_D^2 v^{\mu} \left[ v \cdot k \, A_0 ^a (k) 
- k_0\, v \cdot A^a (k) +
i  g f_{abc} \int \frac{d^4 q}{(2 \pi)^4} \,
v \cdot A^b (k-q) A_0 ^c (q) \right] \ .
\label{eq:3.2}
\eea
Now, after assuming that ${\cal J}_a^{\mu} (k,v)$ can be expressed as
an infinite power series in the gauge field $A^a _{\mu} (k)$,
\Eq{eq:3.2} can be solved iteratively for each order in the power
series.  We impose retarded boundary conditions by the prescription
$p_0 \rightarrow p_0 + i \epsilon$, with $\epsilon \rightarrow 0^+$.
The first order solution is
\beq
 {\cal J}^{\mu\, (1)}_a (k,v) =
  m_D^2 \, v^{\mu}
\left (k_0 \, \frac{ v \cdot A_a (k)} {v \cdot k}- A^0 _a (k) \right)
\ .
\label{cur1}
\eeq
Inserting \Eq{cur1} in \Eq{eq:3.2} allows solving for the second order
term in the series, which reads
\beq
 {\cal J}^{\mu \,(2)}_a (k,v) =
-i g  m_D^2 f_{abc}  \int \frac{d^4 q}{(2 \pi)^4} \, v^{\mu} q_0
 \frac{ v \cdot A^b (k-q) \, v \cdot A^c (q)}{(v \cdot k) (v \cdot q)}
\ .
\label{cur2}
\eeq
The $n$-th order term ($n>2$) can be expressed as a function of the
$(n-1)$-th one as
\beq
 {\cal J}^{\mu \,(n)}_a (k,v)=
 -i g f_{abc} \int \frac{d^4 q}{(2 \pi)^4}  \,
 \frac{ v \cdot A^b (k-q)} {v \cdot k} \, {\cal J}^{\mu \,(n-1)}_c (q,v)
\ .
\label{curn}
\eeq
The complete expression of the induced colour current 
 is thus given by
\bea
J^{\mu}_a (x) &=& \int \frac{d^4 k}{(2 \pi)^4}\, 
e ^{-i k \cdot x}\,
\sum_{n=1}^{\infty} J_a ^{\mu \,(n)}(k) \nonumber \\
&=& \int \frac{d \Omega_{\bf v}}{4 \pi} 
\int \frac{d^4 k}{(2 \pi)^4}\, e ^{-i k \cdot
x}\,
\sum_{n=1}^{\infty}  {\cal J}_a ^{\mu \, (n)}(x,v) \ .
\label{currto}
\end{eqnarray}
To obtain the corresponding $n$-point HTL amplitude, one only needs to
perform $n-1$ functional derivatives of the current with respect to
the vector gauge fields.  The leading order coefficient is given by
the HTL polarisation tensor.  This is given by
\beq\label{HTL-Polarisation}
\Pi^{\mu \nu}_{ab}(k) 
= \delta_{ab} m^2_D 
    \left( - g^{\mu 0} g^{\nu 0} 
           + k_0 \int\!\!\frac{d\Omega_{\bf v}}{4\pi}
                         \frac{v^\mu v^\nu}{k \cdot v} 
    \right) \ .
\eeq
It obeys $k_\mu \Pi^{\mu \nu}_{ab}(k) =0$ due to gauge invariance, and
agrees with the HTL polarisation tensor of QCD
\cite{LeBellac:1996,KLLM2}, if one uses the quantum Debye mass
\Eq{quant-Debye}.

The polarisation tensor has an imaginary part, due to
\beq
\frac{1}{k \cdot v+i0^+} = {\cal P}\frac{1}{k \cdot v} -i\pi\de(k \cdot v) \ .
\eeq
The imaginary part corresponds to Landau damping and describes the emission 
and absorption of soft gluons by the hard particles. It can be expressed as
\beq
{\rm Im}\, \Pi^{\mu \nu}_{ab} (k) 
= -\delta_{ab} m^2_D \pi  k_0 
   \int \!\!\frac{d\Omega_{\bf v}}{4\pi} 
    v^\mu v^\nu \,\delta (k \cdot v) \ .            \label{imag}
\eeq
Notice the appearance of the $\de$-function under the angle average.
Because of $\vv^2=1$ it implies that \Eq{imag} is only non-vanishing
for space-like momenta with $|\vk|\ge k_0$.  It is closely related to
fluctuations within the plasma due to the fluctuation-dissipation
theorem. We shall come back to this point in
Section~\ref{BeyondCorrelators}.

The polarisation tensor can be projected into their longitudinal ($L$)
and transverse ($T$) components as
\begin{mathletters}
\bea
 \Pi^{00}_{ab} (k_0, {\bf k}) & = &\delta_{ab} \, \Pi_{L}
(k_0, {\bf k}) \ ,\\[2mm]
 \Pi^{0i}_{ab} (k_0, {\bf k}) & = & \delta_{ab} \,
k_0 \, \frac{k^i}{|{\bf k}|^2}\, \Pi_{L} (k_0, {\bf k}) \ , \\
\Pi^{ij}_{ab} (k_0, {\bf k}) & = & \delta_{ab} \left[
 \left ( \delta^{ij}- \frac{k^i k^j}{|{\bf k}|^2} \right) \Pi_{T} (k_0,
{\bf k})+ \frac{k^i k^j} {|{\bf k}|^2} \, \frac{k_0^2 }{|{\bf k}|^2} \,
\Pi_{L} (k_0, {\bf k}) \right] \ ,
\eea
\label{resultreal}
\end{mathletters}
$\!\!$where 
\begin{mathletters}
\bea
\Pi_{L} (k_0, {\bf k}) & = & m^2 _{D} \left( \frac{k_0}{2|{\bf
k}|} \left(
 \,{\rm ln\,}\left|{\frac{k_0+|{\bf k}|}{k_0-|{\bf k}|}}\right| 
-i \pi \, \Theta(|{\bf k}|^2 -k_0^2) \right)
-1  \right) \ , \\
 \Pi_{T} (k_0, {\bf k}) & = &- m^2 _{D} \, \frac{k_0^2}{2 |{\bf
k}|^2} \left[ 1 + \frac12 \left( \frac{|{\bf k}|}{k_0} -
\frac{k_0}{|{\bf k}|} \right) \, \left( {\rm ln\,} \left|{\frac{k_0+
|{\bf k}|}{k_0-|{\bf k}|}}\right| -i \pi \, \Theta(|{\bf k}|^2 -k_0^2)
 \right) \, \right] \ .
\eea
\label{pipi}
\end{mathletters}%
Similarly, all HTLs, such as $\Pi_{\mu \nu \rho}^{abc}$, are obtained.

The poles of the longitudinal and transverse parts of the gluon
propagator give the dispersion laws for the collective excitations in
the non-Abelian plasma
\begin{mathletters}
\label{dispersion-laws}
\begin{eqnarray}
|{\bf k}|^2 
- {\rm Re}\, \Pi_L(k_0,{\bf k}) \Large|_{k_0=\omega_L({\bf k})} 
& = & 0 \ ,
\\
 k^2_0
- |{\bf k}|^2 
+ {\rm Re}\,\Pi_T(k_0,{\bf k})\Large|_{k_0=\omega_T({\bf k})}
& = & 0 \ .
\end{eqnarray}
\end{mathletters}%
The plasma frequency $\omega_{\rm pl}$ follows from
\Eq{dispersion-laws} as
\beq\label{PF}
\omega^2_{\rm pl}=
\01{3} m_D^2 \ .
\eeq
For generic external momenta the dispersion relations can only be
solved numerically. In turn, if the spatial momenta are much smaller
than the plasma frequency $|{\bf k}| \ll \omega_{\rm pl}$, solutions
to \Eq{dispersion-laws} can be expanded in powers of $|{\bf
  k}|^2/\omega^2_{\rm pl}$ as
\begin{mathletters}\label{DRunbroken-Expansion}
\begin{eqnarray}
\omega^2_L({\bf k}) 
& = &
\omega^2_{\rm pl}
\left[
1 
+ \035 \frac{|{\bf k}|^2}{\omega^2_{\rm pl}} 
+ {\cal O} (\frac{|{\bf k}|^4}{\omega^4_{\rm pl}}) 
\right]\ , \\
\omega^2_T({\bf k}) 
& = &
\omega^2_{\rm pl} 
\left[
1
+ \left(1 + \ \015 \right)
  \frac{|{\bf k}|^2}{\omega^2_{\rm pl}} 
+ {\cal O} (\frac{|{\bf k}|^4}{\omega^4_{\rm pl}})  \right]\,.
\end{eqnarray}
\end{mathletters}%

\subsection{Energy-momentum tensor}
\label{HTL-Energy}

For a second application, we come back to the conservation laws for
the energy momentum tensor in the Vlasov approximation. They have been
given in \Eq{em-conservation}, and simplify in the present
approximation to
\begin{mathletters}
\step
\label{em-conservation-Vlasov}
\begin{eqnarray}
\partial_\nu \Theta^{\mu \nu}(x) 
&=& - F^{\mu \nu}_a (x) J_{\nu a} (x) \ ,
\zeile
\step
\partial_\nu t^{\mu \nu} (x)
&=&\ \ \, F^{\mu \nu}_a (x) J_{\nu a} (x)
=\int\0{d\Omega}{4\pi}F^{\mu \nu}_a (x)  \, v_\nu\, {\cal J}_{0, a}(x,\vv)\ .
\label{p-em-Vlasov}
\end{eqnarray}
\end{mathletters}%
In the second line we have inserted the current density ${\cal
  J}(x,\vv)$ on the right-hand side. Following an observation due to
Blaizot and Iancu \cite{Blaizot:1994am}, it is possible to derive an
explicit expression for the $t^{\mu 0}$ components of the particle's
energy-momentum tensor. Indeed, making use of the Boltzmann equation
\eq{NAV-soft-mean}, we can substitute the term $F^{0\nu}\, v_\nu$ in
\Eq{p-em-Vlasov} to obtain
\beq
\partial_\mu t^{\mu 0} (x)
={m_D^2}\int \0{d\Omega}{4\pi} 
 \left[ v^\mu D_\mu\, W(x,\vv)\right]^a \, W_a(x,\vv)\ .
\eeq
Here, we found it convenient to introduce $W_a(x,\vv)\equiv m^{-2}_D
{\cal J}_{0, a}(x,\vv)$. Using the identity
$\partial_\mu(A^aB_a)=(D_\mu A)^aB_a+A^a (D_\mu B)_a$, we can
trivially integrate this equation to find, apart from an integration
constant
\beq
t^{\mu 0} (x)
=\01{2}{m_D^{2}}\int\0{d\Omega}{4\pi} \,
v^\mu \,W_a(x,\vv)\, W_a(x,\vv)\ .
\eeq
Combining it with the energy-momentum tensor of the mean fields
$T^{\mu\nu}=\Theta^{\mu\nu}+t^{\mu\nu}$, we obtain for the total
energy
\beq
\label{HTL-T00}
T^{00}(x)=
\012\left[{\bf E}_a(x)\cdot {\bf E}_a(x)
         +{\bf B}_a(x)\cdot {\bf B}_a(x)
         +m_D^{2}\int\0{d\Omega}{4\pi}\, W_a(x,\vv)\, W_a(x,\vv)\right]
\eeq
where we have introduced the colour electric field $E^i\equiv F^{i0}$
and the colour magnetic field $B^i\equiv \s012\epsilon^{ijk}F^{jk}$.
For the energy flux (or Poynting vector), the result reads
\beq
T^{i0}(x)=
 \left[{\bf E}_a(x)\, \times \,{\bf B}_a(x)\right]^i
+\01{2}{m_D^{2}}\int\0{d\Omega}{4\pi}\, v^i\, W_a(x,\vv)\, W_a(x,\vv)\ .
\eeq
The local expression \Eq{HTL-T00} for the energy has proven quite
useful for the integrating-out of modes at the Debye scale
\cite{Bodeker:1998hm}, and for certain lattice implementations
\cite{Bodeker:2000gx}.

\newpage

\section{Beyond the Vlasov approximation}
\label{Beyond}

In this section, the physics related to colour relaxation in a
non-Abelian plasma is studied. The HTL transport equation was found to
be collisionless, which implies that colour relaxation effects cannot
be described in this approximation.  The effects of collisions have to
be taken into account. The corresponding effective transport theory
was first derived in
\cite{Bodeker:1998hm,Bodeker:2000ud,Bodeker:1999ey} (see also the
discussion in Section~\ref{IntroWeak}).

Within the present approach the physics at length scales larger than
the inverse Debye mass can be probed once the gauge field modes with
momenta about $m_D$ have been integrated out, that is, incorporated
within the quasi-particle distribution function \cite{LM,LM2,LM4}. The
simplest approximation which includes the genuine effects due to
source fluctuations of the quasi-particle distribution function is the
{\it polarisation approximation} as discussed in
Section~\ref{EffectiveSystematic}. The polarisation approximation
requires that the two-particle correlation functions remain small
within a Debye volume. This is the case if the plasma parameter is
sufficiently small, which is assumed anyhow.
 
Here, a detailed derivation is given for both the mean field equations
at leading order beyond the HTL effective theory and an effective
theory for the ultra-soft gauge fields to leading logarithmic order.
We proceed in two steps. The first step consists in solving the
dynamical equations for the fluctuations as functions of initial
fluctuations of the particle distribution function. From this, a
collision term and a noise source for the effective mean field
Boltzmann equation are obtained. Here, it will be necessary to perform
a leading logarithmic approximation, assuming that the gauge coupling
is sufficiently small to give
\beq\label{LLO} 
g\ll 1\quad {\rm  and}\quad \0{1}{\ln 1/g}\ll 1\ .  
\eeq 
In a second step the quasi-particle degrees of freedom are integrated
out as well. This implies that the mean field transport equation has
to be solved. This expresses the induced current explicitly as a
functional of the ultra-soft gauge fields only. Ultimately, we shall
see that a simple Langevin-type dynamical equation emerges
\cite{Bodeker:1998hm}. Here, we extend the reasoning as presented in
\cite{LM,LM2}.
 
\subsection{Leading order dynamics}
\label{BeyondLeading}

We now allow for small statistical fluctuations $\delta f(x,p,Q)$
around \Eq{m-close-eq-f}, writing
\beq\label{f+df}
f(x,p,Q)=\bar f^{\rm eq.}(p_0)+ g \bar f^{(1)}(x,p,Q) +\delta f(x,p,Q)\ ,
\eeq 
and rewrite the approximations to (\ref{NA-Macro}) and (\ref{NA-fluc})
in terms of current densities and their fluctuations. Note that the
fluctuations $\delta f(x,p,Q)$ in the close-to-equilibrium case are
already of the order of $g$. This observation is important for the
consistent approximation in powers of the gauge coupling. As a
consequence, the term $g \bar f^{(1)}$ in \Eq{f+df} will now account
for the ultra-soft modes for momenta $\ll m_D$. Integrating-out the
fluctuations results in an effective theory for the latter.

As before, we obtain the dynamical equation for the mean current
density at leading order in $g$, after multiplying (\ref{NA-1}) by $g
Q_a p^\rho/p_0$, summing over two helicities, and integrating over
$d\tilde P dQ$. The result is
\begin{mathletters} 
\step                     \label{NA-fluc-mean}
\begin{eqnarray}                         \label{soft-mean}
  v^\mu \bar D_\mu \bar {\cal J}^\rho
+ m^2_D v^\rho v^\mu \bar F_{\mu0}
& = & \left\langle \eta^\rho\right\rangle 
     +\left\langle \xi^\rho\right\rangle
\ ,
\zeile\step 
    \bar D_\mu\bar F^{\mu\nu} 
   +\left\langle J_{\mbox{\tiny fluc}}^{\nu}\right\rangle 
&=&  \bar J^\nu \ .                       \label{soft-YM}
\end{eqnarray}
\end{mathletters}%
In a systematic expansion in $g$, we have to neglect cubic correlator 
terms as compared to quadratic ones, as they are suppressed explicitly 
by an additional power in $g$. Therefore, we find to leading order 
\begin{mathletters}
\label{NA-log-1}
\step 
\begin{eqnarray}
\eta^\rho_a  & = &
g \int d \tilde P\, \frac{p^\rho}{p_0} 
 (\bar D_\mu a_\nu-\bar D_\nu a_\mu)^b\,
  \partial^\nu_p \,\delta J^{\mu}_{ab}(x,p)  \ ,        
\label{NA-eta1}
\zeile \step 
\xi^\rho_a   & = &
  -gf_{abc}v^\mu\, a^b_{\mu}\ \delta {\cal J}^{c,\rho}\ ,
\label{NA-xi1}
\zeile\step
 J_{\mbox{\tiny fluc}}^{\rho,a}&=&  
 g f^{dbc}\! \left(\bar D_{\mu}^{ad}  a_{b}^{\mu} a_c^\rho  
+\delta^{ad}  a^b_\mu\! \left(\bar D^\mu a^\rho
-\bar D^\rho a^\mu\right)^c  \right) .                
\label{NA-Jfluc1}
\end{eqnarray}
\end{mathletters}%
The same philosophy is applied to the dynamical equations for the 
fluctuations. To leading order in $g$, the result reads
\begin{mathletters}\label{NA-log-2}\step 
\begin{eqnarray}
&&\left(v^\mu \bar D_\mu \,\delta{\cal J}^\rho \right)_a \, 
= - m^2_D v^\rho v^\mu \left(\bar D_\mu a_0-\bar D_0 a_\mu\right)_a 
  - g f_{abc}v^\mu a_\mu ^b \bar {\cal J}^{c,\rho}       \label{vD-dJa} 
\ ,
\zeile\step 
&&v^\mu \left(\partial_\mu\de_{ac}\de_{bd} + g\bar A_\mu^m
\left( f_{amc}\,\delta_{bd} + f_{bmd}\delta_{ac}\right) \right) 
\delta {\cal J}^{\rho}_{cd} 
=  g v^\mu a_\mu^m 
\left( f_{mac}\,\delta_{bd} +f_{mbd}\delta_{ac}\right)   
\bar{\cal J}^{\rho}_{cd} \ , \quad\quad\quad\quad  \label{vD-dJab}
\zeile\step 
&&\left(\bar D^2 a^\mu-\bar D^\mu(\bar D a)\right)_a+2 g f_{abc} 
\bar F_b^{\mu\nu}a_{c,\nu}
= \delta J_a^\mu  \ .                                       \label{dJa}
\end{eqnarray}
\end{mathletters}%
Notice that the dynamical equations \eq{vD-dJa} and \eq{dJa} can also
be obtained from the HTL effective equations \Eqs{NAV-mean} within the
present approximation. It suffices to expand \Eqs{NAV-mean} to linear
order in $\bar A_\mu\to \bar A_\mu+a_\mu$ and $\bar{\cal
  J}\to\bar{\cal J}+{\de \cal J}$, which gives \Eq{vD-dJa} for the
dynamics of ${\de \cal J}$ and \Eq{dJa} for the dynamics of $a_\mu$.

The typical momentum scale associated to the fluctuations can be
estimated from \Eq{NA-log-2}. We find that it is of the order of the
Debye mass $\sim m_D$, that is, of the same order as the mean fields
in \Eqs{NAV-mean}. This confirms explicitly the discussion made above.
The typical momentum scales associated to the mean fields in
\Eq{NA-fluc-mean} are therefore $\ll m_D$.

\subsection{Integrating-out the fluctuations}
\label{BeyondIntegrating}

We solve the equations for the fluctuations \Eqs{NA-log-2} with an
initial boundary condition for $\delta f$, and $a_\mu(t=0)=0$. Exact
solutions to \Eqs{vD-dJa} and (\ref{vD-dJab}) can be obtained.  It is
convenient to proceed as follows \cite{K}.  We separate the colour
current fluctuations into a source part and an induced part,
\beq
\label{sou+ind}
\delta {\cal J}^\mu =   \delta {\cal J}^\mu _{\mbox{\tiny source}} 
                      + \delta {\cal J}^\mu _{\mbox{\tiny induced}} \ . 
\eeq
The induced piece $\delta {\cal J}^\mu_{\mbox{\tiny induced}}$ is the
part of the current which contains the dependence on $a_\mu$, and thus
takes the polarisation effects of the plasma into account.  The source
piece $\delta {\cal J}^\mu _{\mbox{\tiny source}}$ is the part of the
current which depends only on the initial condition, given by the
solution of the homogeneous equation \Eq{homogeneous}. This splitting
will be useful later on since ultimately all the relevant correlators
can be expressed in terms of correlators of $\de {\cal
  J}^\mu_{{\mbox{\tiny source}}}$.

We start by solving the homogeneous differential equation
\beq\label{homogeneous}
v^\mu \bar D_\mu \,  \delta   {\cal J}^\rho (x,\vv) = 0 \ ,
\eeq 
with the initial condition $\de {\cal J}^\mu_a(t=0, {\bf x},\vv)$.  It
is not difficult to check, by direct inspection, that the solution to
the homogeneous problem is
\begin{mathletters}\label{dJ-expl}\step  
\beq
\delta {\cal J}^\rho_{a,\mbox{\tiny source}}  (x,\vv) 
 = \bar U_{ab} (x, x-v t) \, 
   \delta {\cal J}^\rho _b(t=0, {\bf x}-{\bf v} t, \vv) \ .
\eeq
The full solution of \Eq{vD-dJa} is now constructed using the retarded
Green's function \Eq{ret-GF}. For $x_0 \equiv t \geq 0$ the induced
piece can be expressed as
\begin{eqnarray}\step 
\delta {\cal J}^\rho_{a,\mbox{\tiny induced}} (x,\vv)
=  
- \int_0 ^{\infty} d \tau \, \bar U_{ab} (x, x_\tau)
  && \left[  \ m^2_D v^\rho v^\mu \left(\bar D_\mu a_0
        - \bar D_0 a_\mu\right)^b (x_\tau) \right.
\nonumber\\ &&       
\left. 
\ + g f_{bdc}  v^\mu a_\mu ^d (x_\tau) \bar {\cal J}^\rho_c(x_\tau, v)  
\right] 
\,. \label{dJ-expl-induced} 
\end{eqnarray}
\end{mathletters}%
We have  introduced 
\beq
x_\tau \equiv x - v \tau \ ,
\quad {\rm thus}\quad 
x_t=(0, {\bf x}-{\bf v} t)\ .
\eeq 
Since $a_\mu(t=0)=0$, one can check that the above current obeys the
correct initial condition.

Let us remark that the induced solution \Eq{dJ-expl-induced} can be
obtained directly from the explicit solution of the HTL current
density \Eq{class-HTL-density} by expanding it to linear order in
small deviations about the mean gauge fields. The right-hand side of
\Eq{class-HTL-density} depends only on the gauge fields $\bar
A(x_\tau)$. Linearising \Eq{class-HTL-density} to leading order about
the mean gauge field $\bar A(x_\tau)+\de A(x_\tau)$ yields
\bea
\de{\cal J}^\mu_{a,\mbox{\tiny induced}} (x,\vv) 
=  -m^2_D \,v^\mu v^\nu  \int^{\infty}_0 d \tau\ &&
\left[
\bar U_{ab}(x,x_\tau) \ \de \bar F_{\nu 0,b} (x_\tau) 
\right.
\nonumber\\
&&
\left.
+ \de \bar   U_{ab}(x,x_\tau) \ \bar F_{\nu 0,b} (x_\tau) \ .
\right]
\label{induced-HTL}
\eea
In order to evaluate \Eq{induced-HTL} explicitly, we need to know that
\beq
\de  F^b_{\mu \nu}(x_\tau)
=\left(\bar D_\mu\, \de A_\nu- \bar D_\nu\, \de A_\mu\right)^b (x_\tau)\ 
+O[(\de A)^2]\ .
\eeq
For the second term in \Eq{induced-HTL}, we make use of an identity
for the variation of the parallel transporter
\Eq{ParallelTransporterDef} with respect to the gauge fields $\de\bar
U(x,x_\tau)/\de A(x_{\sigma})$, namely
\beq\label{VarParallelTransporter}
\de\bar U_{ab}(x,x_\tau)=
\int _0^\infty d{\sigma}\bar U_{ac}(x,x_{\sigma})\,
\left[g\, f_{cde}\,v^\mu\,\de A_\mu(x_{\sigma})\right]\,
\bar U_{eb}(x_\sigma,x_\tau)\ .
\eeq
Using \Eq{VarParallelTransporter} and the explicit expression
\Eq{class-HTL-density}, we confirm that \Eq{induced-HTL} coincides
with the explicit result for the induced current \Eq{dJ-expl-induced},
if we replace $\de A$ by $a$.

The equation (\ref{vD-dJab}) can be solved in a similar way.  The
solution is
\begin{eqnarray}
&&\delta {\cal J}^\rho _{ab} (x,v)  =  
\bar U_{am} (x, x_t) \bar U_{bn} (x, x_t)
\, \delta {\cal J}^\rho _{mn} (x_t, v) 
\nonumber\zeile 
&&\quad\quad  - g\int_0 ^{\infty}d\tau \,
\bar U_{am} (x, x_\tau)
\bar U_{bn} (x, x_\tau)
\left(f_{mpc} \delta_{nd}  +  f_{npd} \delta_{mc}\right)
v^\mu a_\mu ^p (x_\tau)\bar {\cal J}^\rho_{cd}(x_\tau, v) \ . \quad\quad 
\end{eqnarray}
Now we seek solutions to \Eq{dJa} with the colour current of the
fluctuation as found above.  However, notice that this equation is
non-local in $a_\mu$, which makes it difficult to find exact
solutions. Nevertheless, one can solve the equation in an iterative
way, by making a double expansion in both $g\bar A$ and $g\bar {\cal
  J}$. This is possible since the parallel transporter $\bar U$ admits
an expansion in $g \bar A$, so that the current $\delta {\cal J}^\rho$
can be expressed as a power series in $g \bar A$
\beq
\delta {\cal J}^\rho =  \delta {\cal J}^{\rho (0)} 
                       +\delta {\cal J}^{\rho (1)} 
                       +\delta {\cal J}^{\rho (2)} + \cdots \ ,
\eeq
and thus \Eq{dJa} can be solved for every order in $g \bar A$. To
lowest order in $g \bar A$, using $\bar U_{ab} = \delta_{ab} + {\cal
  O}(g \bar A)$, \Eq{dJa} becomes
\beq
  \partial^\mu \left[ \partial_\mu a_{\nu, a}^{(0)} 
- \partial_\nu a_{\mu, a}^{(0)} \right] 
= \delta J_{\nu, a}^{(0)} \ .                       \label{Abelgeq} 
\eeq
Using the one-sided Fourier transform \footnote{The one-sided Fourier
  transform with respect to the time variable is defined as $F(\omega)
  = \int^{\infty}_0 dt \,e^{i \omega t} F(t)$} \cite{K}, and
\Eq{dJ-expl}, we find
\begin{eqnarray}
\delta  J _{a\,+}^{\mu \, (0)}(k)
&=& \Pi^{\mu \nu}_{ab}(k) a_{\nu,b}^{(0)} (k)
\nonumber \\ &&  
    - g f_{abc} \int\!\!\frac{d\Omega_{\bf v}}{4\pi}
        \frac{1}{-i\ k \cdot v} \int\frac{d^4q}{(2\pi)^4} 
        v^\rho a_{\rho}^{b(0)}(q) \, \bar{\cal J}^{\mu,c}(k-q,v) 
\nonumber \\ &&  
+\int\frac{d\Omega_{\bf v}}{4\pi}
 \frac{\delta{\cal J}^{\mu}_{a}(t=0,{\bf k},v)}{-i\, k\cdot v} \ ,
                                                             \label{dJ+}
\end{eqnarray}
where $\Pi^{\mu \nu}_{ab}(k)$ is the polarisation tensor
\Eq{HTL-Polarisation}. Retarded boundary conditions are assumed above,
with the prescription $k_0 \rightarrow k_0 + i 0^+$.

We solve \Eq{Abelgeq} iteratively  in momentum space for $a_\mu$
as an  infinite power series in $g\bar {\cal J}$,
\beq
a_\mu^{(0)}= a_\mu^{(0,0)}
            +a_\mu^{(0,1)}
            +a_\mu^{(0,2)}+\ldots
\eeq
where the second index counts the powers of the background 
current $g\bar {\cal J}$.

Notice that in this type of Abelianised approximation, the equation 
(\ref{Abelgeq}) has a (perturbative) Abelian gauge symmetry associated 
to the fluctuation $a_\mu$. This symmetry is only broken by the term 
proportional to $\bar {\cal J}$ in the current. It is an exact 
symmetry for the term $a_\mu ^{(0,0)}$ in the above expansion.  We 
will use this perturbative gauge symmetry in order to simplify the 
computations, and finally check that the results of the approximate 
collision integrals do not depend on the choice of the fluctuation gauge.
  
Using the one-sided Fourier transform, we find
the following results for the longitudinal fields, 
in the gauge ${\bf k} \cdot {\bf a}^{(0,0)} =0$, 
\begin{mathletters} \step                                        \label{aL+}
\begin{eqnarray}
a_{0,a\,+}^{(0,0)}(k)
&=& \frac{1}{{\bf k}^2-\Pi_L}
    \int\!\!\frac{d\Omega_{\bf v}}{4\pi}
     \frac{\delta{\cal J}_{0,a}(t=0,{\bf k},v)}{-i\ k \cdot v}\ , 
                                                            \label{aL0}
\zeile\step  
a_{0,a\,+}^{(0,1)}(k)
&=& \frac{-g f_{abc}}{{\bf k}^2-\Pi_L}
    \int\frac{d\Omega_{\bf v}}{4\pi}\frac{1}{-i\ k\cdot v}
    \int\frac{d^4q}{(2\pi)^4} v^\mu a_{\mu}^{b(0,0)}(q) \, 
    \bar{\cal J}_0^c(k-q,v) \ ,                             \label{aL1}
\end{eqnarray}
\end{mathletters}%
while we find
\begin{mathletters}
\step                                          \label{aT+}
\begin{eqnarray}
a_{i,a\,+}^{T(0,0)}(k)
&=&\frac{1}{-k^2+\Pi_T}
   \int\!\!\frac{d\Omega_{\bf v}}{4\pi}
   \frac{\delta{\cal J}^{T}_{i,a}(t=0,{\bf k},v)}{-i\ k \cdot v}\ , 
                                                            \label{aT0}
\zeile\step
a_{i,a\,+}^{T(0,1)}(k)
&=& \frac{-g f_{abc}}{-k^2+\Pi_T}P_{ij}^T({\bf k})
    \int\frac{d\Omega_{\bf v}}{4\pi}\frac{1}{-i\ k\cdot v}
    \int\frac{d^4q}{(2\pi)^4} v^\mu a_{\mu}^{b(0,0)}(q) \, 
     \bar{\cal J}_j^c(k-q,v) \ ,   \quad\quad \quad                         
\label{aT1}
\end{eqnarray}
\end{mathletters}%
for the transverse fields.  The functions $\Pi_{L/T}(k)$ are the
longitudinal/transverse polarisation tensors of the plasma,
$P_{ij}^T({\bf k})=\delta_{ij}-{k_ik_j}/{{\bf k}^2}$ the transverse
projector, and $a_i^T\equiv P^T_{ij}a_j$.

In the approximation $g \ll 1$, it will be enough to consider the
solution of leading (zeroth) order in $g \bar A$, and the zeroth and
first order in $g\bar {\cal J}$. The remaining terms are subleading in
the leading logarithmic approximation.  However, in principle all
tools are available to compute the complete perturbative series. If we
could solve \Eq{dJa} exactly, it would not be necessary to use this
perturbative expansion.

\subsection{Correlators and Landau damping}
\label{BeyondCorrelators}

With the explicit expressions obtained in \Eqs{dJ+}, \eq{aL+} and
\eq{aT+}, we can express all fluctuations in terms of initial
conditions $\de {\cal J}^\mu_a(t=0, {\bf x},v)$ and the mean fields.
From \Eq{average} one deduces the statistical average over colour
current densities $\de {\cal J}$. We expand the momentum
$\delta$-function in polar coordinates
\beq
\delta^{(3)}({\bf p}-{\bf p}') 
= \frac{1}{p^2} \,\delta(p - p') \, 
  \delta^{(2)}(\Omega_{\bf v }-\Omega_{\bf v'}) \ ,
\eeq
where $\Omega_{\bf v}$ represents the angular variables associated to 
the vector ${\bf v}= {\bf p} /|{\bf p}|$. After some simple integrations 
we arrive at 
\bea
\llangle\delta {\cal J}_\mu^a (t=0,{\bf x},v) \, 
        \delta {\cal J}_\nu^b (t=0,{\bf x}',v')\rrangle 
&=& 2 g^2 B_C \,C_2 \,\delta^{ab} \, v_\mu v'_\nu \, 
     \delta^{(3)}({\bf x}-{\bf x}') \,
     \delta^{(2)} (\Omega_{\bf v }-\Omega_{\bf v'})
\nonumber\\ && 
+\,{\tilde g}^{ab}_{2,\mu\nu}({\bf x},v;{\bf x}',v')\ , \label{cur-cor}
\eea
where $v^\mu= (1,\vv)$, and 
\beq                                                \label{class-f-cor}
B_C = \frac{2}{\pi} \int_0^\infty dp\, p^2\,\bar f^{\rm eq}(p) \ ,
\eeq
for classical statistics. For a quantum plasma, the value of the
constant $B_C$ is obtained from the quantum correlators 
\Eq{averageB} and \Eq{averageF}. For bosonic statistics it reads
\beq
B_C = \frac{2}{\pi} \int_0^\infty dp\, p^2\,\bar f^{\rm eq}_B(p)
\left(1 + {\bar f}^{\rm eq}_B(p) \right) \ ,
\eeq
while for fermionic statistics it is
\beq
B_C = \frac{2}{\pi} \int_0^\infty dp\, p^2\,\bar f^{\rm eq}_F(p)
\left(1 - {\bar f}^{\rm eq}_F(p) \right) \ .
\eeq
The function ${\tilde g}^{ab}_{2,\mu\nu}$ is obtained from the
two-particle correlation function $\tilde g_2$. Notice that we have
neglected the piece $g \bar f^{(1)}$ above, as this is subleading in
an expansion in $g$.  Since we know the dynamical evolution of all
fluctuations we can also deduce the dynamical evolution of the
correlators of fluctuations, with the initial condition \Eq{average}.
This corresponds to solving \Eq{quad-corr} in the present
approximation.

From the explicit solution \Eq{dJ-expl} and the average \Eq{cur-cor}
we then find, at leading order in $g$ and neglecting the non-local
term in \Eq{cur-cor},
\begin{eqnarray}
\label{source-corr}
 \llangle\delta {\cal J}_{\mbox{\tiny source}}^{a,\mu} (x,v) \ 
         \delta {\cal J}_{\mbox{\tiny source}}^{b,\nu} (x',v')\rrangle 
&=&
2 g^2 B_C \,C_2\,  
    \delta^{(3)}[{\bf x}-{\bf x}'-{\bf v}(t-t')] \,
    \delta^{(2)} (\Omega_{\bf v }-\Omega_{\bf v'}) 
\nonumber\\ &\times & 
     \,v^\mu v'^\nu\, 
\bar U^{ac}(x, x-vt)\,  \bar U^{bc}(x', x'-v't') \ .
\end{eqnarray}
Expanding the parallel transporter $\bar U$, and switching to momentum
space we find the spectral density to zeroth order in $ g \bar A$
\beq
\llangle \delta {\cal J}_{\mu}^a \ 
         \delta {\cal J}_{\nu}^b
\rrangle^{{\mbox{\tiny source}}\,(0)}_{k,v,v'} 
 = 2 g^2 B_C \,C_2 \, \delta^{ab} \,v_\mu v'_\nu \,
   \delta^{(2)} (\Omega_{\bf v }-\Omega_{\bf v'})
   \,(2 \pi)\,\delta (k \cdot v) \ .                    \label{spectral-dJ}
\eeq
This is the basic correlator reflecting the microscopic fluctuations
of the quasi-particle distribution function within the plasma, and it
is all that is required to derive the collision integral relevant for
colour conductivity.

As a simple and illustrative example let us reconsider the case of
Landau damping. As discussed in Section \ref{HTL-Amplitudes-1}, the
HTL polarisation tensor \Eq{HTL-Polarisation} has an imaginary part
which describes the absorption of soft gluonic fields by the hard
particles. This imaginary part is closely linked to fluctuations in
the gauge fields, which can be seen as follows. We compute the
correlator of two transverse fields $a$, and in particular consider
only the part which corresponds to the source fluctuations of the
particle distribution function. They yield the field-independent part
of the correlator. Hence, we compute the self-correlator of \Eq{aT0},
using \Eq{spectral-dJ}, and arrive at
\bea                                                
 \llangle a_{i,a}^{T(0,0)}(k)\,                     
          a_{j,b}^{T(0,0)}(q) \rrangle 
&=&  g^2 B_C \,C_2 \delta_{ab} 
  (2 \pi )^4 \delta^{(4)} (k+q)
\nonumber\\
&\times&  
\frac{P_{ik}^T({\bf k})P_{jl}^T({\bf k})}{|-k^2+\Pi_T|^2}  
  \int\!\!\frac{d\Omega_{\bf v}}{4\pi} v_k v_l \,
  \delta (k \cdot v)  \ .  \label{basic-corr}
\eea
Comparing \Eq{basic-corr} with \Eq{imag}, the above correlator  can 
be written as
\beq
 \llangle a_{i,a}^{T(0,0)}(k)\,  
          a_{j,b}^{T(0,0)}(q) \rrangle 
= \frac{4 \pi T}{k_0} 
  \frac{{\rm Im}\, \Pi_{ij,T}^{ab}(k)}{|-k^2+\Pi_T|^2}  
  (2 \pi)^3  \delta^{(4)} (k+q) \ .                 \label{first-FDT}
\eeq
Here, we have used the relation
\beq
{2 g^2 C_2 B_C}= 4 \pi T {m^2_D} \ .              \label{magical-rel}
\eeq
For a quark-gluon plasma, with gluons in the adjoint representation,
$C_2 = N$, and $N_F$ quarks and antiquarks in the fundamental
representation, $C_2= 1/2$, a similar relation can be written,
after summing over species of particles. Thus
\beq
\sum_{\hbox{\tiny species}} 
C_2 B_C= \frac{ 2 N}{\pi} \int^\infty_0 dp p^2 
\bar f^{\rm eq}_{\rm B} (1 + \bar f^{\rm eq}_{\rm B})
       + \frac{ 2 N_F}{\pi} \int^\infty_0 dp p^2 
\bar f^{\rm eq}_{\rm F} (1 - \bar f^{\rm eq}_{\rm F}) \ .
\eeq
And the relation  \Eq{magical-rel} now reads
\beq
2 g^2 \sum_{\hbox{\tiny species}} 
C_2 B_C = 4 \pi T m^2_D
\eeq
Equation (\ref{first-FDT}) is a form of the fluctuation-dissipation
theorem, which links the dissipative process, Landau damping, with
(induced) statistical fluctuations of the gauge fields.

\subsection{Domain of validity}
\label{Validity2on}

The present considerations are based on the polarisation
approximation, introduced in Section \ref{Effective}. It amounts to a
truncation of the hierarchy of dynamical equations for correlation
functions, neglecting two-particle correlations $g_2$. This has lead
to a closed set equations for the mean fields and their fluctuations.
It remains to be shown that two-particle correlations $g_2$ indeed
remain small (see \Eq{BasicGibbs} and \Eq{2-point-cor}).

Here, we compute two-particle correlators in the static limit, based
on the solution for the fluctuation dynamics as found in Section
\ref{BeyondIntegrating}. We shall find that these correlations are
only negligible when the distance between particles is sufficiently
large. This correlation length defines the domain of validity of our
equations.

For convenience, we consider the colour current density fluctuations
$\delta {\cal J}$. An analogous reasoning applies for the one-particle
distribution function. We split the colour current density
fluctuations as in \Eq{sou+ind} into a source (or free) part, and an
induced part. The statistical correlator of the source densities has
already been displayed in \Eq{source-corr} and \Eq{spectral-dJ}. We
just note that the equal time correlator \Eq{spectral-dJ} was deduced
from the local term of the initial time correlator of \Eq{average}
which is the free term. To obtain \Eq{spectral-dJ}, one has to
multiply the local term in \Eq{average} by colour charges, and
integrate over the charges and the modulus of the momentum. The time
evolution of $\delta J^{sou}$ is given by the free dynamical
equations.  The effect of interactions in the plasma are responsible
for two-particle correlations ${\tilde g}_2$ in \Eq{average}. In our
approach, this corresponds to the part of the correlators which can be
computed from $\delta J^{ind}$.  More specifically, we have
\beq
\llangle \delta {\cal J}_{0}^a \ 
         \delta {\cal J}_{0}^b
\rrangle  =  \llangle \delta {\cal J}_{0}^a \ 
         \delta {\cal J}_{0}^b
\rrangle^{{\mbox{\tiny sou}}} + 
 \llangle \delta {\cal J}_{0}^{a,{\mbox{\tiny ind}} } \ 
         \delta {\cal J}_{0}^{b,{\mbox{\tiny sou}}}  \rrangle 
 +  \llangle \delta {\cal J}_{0}^{a,{\mbox{\tiny sou}} } \ 
         \delta {\cal J}_{0}^{b,{\mbox{\tiny ind}}}  \rrangle
+ \llangle \delta {\cal J}_{0}^{a,{\mbox{\tiny ind}}} \ 
         \delta {\cal J}_{0}^{b,,{\mbox{\tiny ind}}}  \rrangle  \ .
\eeq
We identify the two-particle correlation function with the correlators
arising from $\delta J^{ind}$.  This part of the current takes into
account the interactions. Since, ultimately, $\delta J^{ind}$ is
expressed in terms of $\delta J^{sou}$ and the mean fields, we have a
general recipe to compute these correlation functions.  Here, we compute the
above correlators using the results given in Section
\ref{BeyondIntegrating}. From \Eq{spectral-dJ}, we have
\bea
\llangle \delta { J}_{0}^a \ 
         \delta { J}_{0}^b
\rrangle^{{\mbox{\tiny sou}},(0,0)}_k  
&= &  
\int \frac{d\Omega_\vv}{4\pi}\, \int \frac{d\Omega_\vv'}{4\pi}\,
\llangle \delta {\cal J}_{0}^a \ 
         \delta {\cal J}_{0}^b
\rrangle^{{\mbox{\tiny source}}\,(0,0)}_{k,v,v'} 
\nonumber\\
&= &    
g^2 B_C \,C_2 \, \delta^{ab}  \,
    \int \frac{d\Omega_\vv}{4\pi}  \,\delta (k \cdot v) \ .
\eea
To lowest order, we find
\begin{mathletters}
\bea
\llangle \delta { J}_{0}^{a,{\mbox{\tiny sou}} } \ 
         \delta { J}_{0}^{b,{\mbox{\tiny ind}}}  \rrangle_k^{(0,0)}
&=& 2 \pi T m_D^2 \, \delta^{ab} \frac{\Pi_L(k)}{{\bf k}^2-\Pi_L(k)}
\int \frac{d\Omega_\vv}{4\pi}  \,\delta (k \cdot v) \ ,
\\
\llangle \delta { J}_{0}^{a,{\mbox{\tiny ind}} } \ 
         \delta { J}_{0}^{b,{\mbox{\tiny ind}}}  \rrangle_k^{(0,0)}
&=& 2 \pi T m_D^2 \, \delta^{ab} \left(\frac{\Pi_L(k)}{{\bf k}^2-\Pi_L(k)}
\right)^2
\int \frac{d\Omega_\vv}{4\pi}  \,\delta (k \cdot v) \ .
\eea
\end{mathletters}%
Here, we have used \Eq{magical-rel}.  Collecting terms and taking the
static limit $k_0 =0$, we obtain
\beq
\llangle \delta { J}_{0}^a \ 
         \delta { J}_{0}^b
\rrangle^{(0,0)}_{\bf k} = 2 \pi T m_D^2 \, \delta^{ab}
\left(1 - 2 \frac{m^2_D}{{\bf k}^2 + m^2_D} + \frac{m^4_D}{
\left({\bf k}^2 + m^2_D\right)^2} \right) \ .
\eeq
The non-local pieces represent the effect of correlations in the
system. Taking the inverse Fourier transform, we find
\beq
\llangle \delta { J}_{0}^a \ 
         \delta { J}_{0}^b
\rrangle^{(0,0)}_{\bf r} = 2 \pi T m_D^2 \, \delta^{ab}
\left( \delta^{(3)}({\bf r}) - \frac{m_D^2}{2 \pi r} e^{-r m_D}
+ \frac{m^3_D}{8 \pi}e^{-r m_D} \right)  \ .
\eeq
In the static limit, the two-particle correlations are exponentially
suppressed at distances $r \gg 1/m_D$. This is the domain of validity
of the polarisation approximation.  Using the expansion introduced to
compute the correlators, one could equally compute the terms $\langle
\delta { J}_{0}^a \ \delta { J}_{0}^b \rangle^{(n,m)}$. However, those
terms are suppressed in the weak coupling expansion.  Notice also that
infrared (IR) problems in the computation of the correlation functions
do show up, due to the unscreened magnetic modes (see below). These IR
problems provide an additional limitation for the domain of validity
of the transport equations.  The IR problems are, supposedly, cured by
the non-perturbative appearance of a magnetic mass at order $g m_D$.

\subsection{Collision integrals}
\label{BeyondCollision}

We are now ready to compute at leading order in $g$ the collision 
integrals appearing on the right-hand side of \Eq{soft-mean}. 
We shall combine the expansions introduced earlier to expand the 
collision integrals in powers of $\bar {\cal J}$, while
retaining only the zeroth order in $g\bar A$,  
\beq
\langle \xi\rangle=\langle \xi^{(0)}\rangle
+\langle \xi^{(1)}\rangle+\langle \xi^{(2)}\rangle+\ldots\ ,
\eeq
and similarly for $\langle \eta\rangle$  and
$\langle J_{\mbox{\tiny fluc}}\rangle$. We  find that the induced 
current $\langle J_{\mbox{\tiny fluc}}^{(0)}\rangle$
vanishes, as do the fluctuation integrals $\langle \eta^{(0)}\rangle$ 
and $\langle \xi^{(0)}\rangle$.  The vanishing of 
$\langle J_{\mbox{\tiny fluc}}^{(0)}\rangle$
is deduced trivially from the fact that 
$\langle a_a^{(0,0)}  a_b^{(0,0)} \rangle \sim \delta_{ab}$,
while this correlator always appears contracted with the 
antisymmetric constants $f_{abc}$ in
$J_{\mbox{\tiny fluc}}$.  To check that $\langle \eta^{(0)}\rangle = 0$, 
one needs the statistical correlator
$\langle \delta J^\mu _a \delta J^\rho _{ab}\rangle$, which is 
proportional to $\sum_a d_{aab}=0$ for $SU(N)$. 
The vanishing of $\langle \eta^{(0)}\rangle$
is consistent with the fact 
that in the Abelian limit the counterpart of  $\langle \eta\rangle$
vanishes at equilibrium \cite{K}. 
Finally, $\langle \xi^{(0)}\rangle = 0$ due to a
contraction of $f_{abc}$ with a correlator symmetric in the colour indices.

In the same spirit we evaluate the terms in the collision integrals 
containing one 
$\bar{\cal J}$ field and no background gauge $\bar A$ fields. Consider
\begin{eqnarray}
\left\langle\xi^{(1)}_{\rho,a}(x,\vv)\right\rangle 
&\equiv &
\left.\left\langle\xi_{\rho,a}(x,\vv)\right\rangle
\right|_{\bar A=0,\,\,{\rm linear\, in\,}\bar J} 
\nonumber \\
&=& g f_{abc} v^\mu \bigg\{ 
-\left\langle a^{(0,1)}_{\mu,b} (x)\,
        \delta {\cal J}_{\rho,c}^{(0)}(x,\vv) \right\rangle 
\nonumber \\
&&\qquad\qquad
  + \, g f_{cde} v^\nu  \int^{\infty} _{0} \!\!\! d \tau 
   \bar{\cal J}_{\rho,e}( x_\tau, \vv) 
   \left\langle   a^{(0,0)}_{\mu,b} (x)\ a^{(0,0)}_{\nu,d} (x_\tau) 
\right\rangle \bigg\}. 
\quad \quad
\end{eqnarray}
Using the values for $a_\mu$ and $\delta {\cal J}^{(0)}$ as found
earlier, we obtain in momentum space\footnote{Note the typo in
  Eq.~(7.39) of \cite{LM2} where a factor $1/4 \pi$ is missing.} (see
Appendix~\ref{Correlators} for a detailed computation of the
correlators)
\begin{eqnarray}
\label{first-exp-coll}
&&
\left. 
\left\langle\xi^{(1)}_{\rho,a}(k,\vv)\right\rangle 
\right|_{\rm LLO}
=
-\frac{g^4 C_2 N B_C}{4 \pi} v_\rho
\!\int\!\frac{d\Omega_{{\bf v}'}}{4\pi}\,C_{\rm LLO} ({\bf v}, {\bf v}') 
\left[ \bar {\cal J}^{0}_a (k,\vv) -  \bar {\cal J}^{0}_a (k,\vv') \right] \ , 
\quad \quad \quad \end{eqnarray}
where
\beq\label{C}
C({\bf v}, {\bf v}')
= 
\int \frac{d^4 q}{(2 \pi)^4} 
\left|\frac{v_i P_{ij}^T(q) v'_j}{-q^2+\Pi_T}\right|^2 (2 \pi) 
\delta(q \cdot v) (2 \pi) \delta(q \cdot v')  
\eeq
still has to be evaluated within LLO.

To arrive at the above expression we have used the $SU(N)$ relation
$f_{abc} f_{abd} = N \delta_{cd}$. Within the momentum integral, we
have neglected in the momenta of the mean fields, $k$, in front of the
momenta of the fluctuations, $q$. As we discussed above, the momenta
associated to the background fields are much smaller than those
associated to the fluctuations which justifies this approximation to
leading order.  This is precisely what makes the collision integral,
which in principle contains a convolution over momenta, local in
$k$-space (resp. $x$-space). The only remaining non-locality stems
from the angle convolution of \Eq{C}.  Notice that we have only given
the part arising from the transverse fields $a$, as the one associated
to the longitudinal modes is subleading. This is easy to see once one
realises that the above integral is logarithmically divergent in the
infrared region, while the longitudinal contribution is finite. At
this point, we can also note that the collision integral computed this
way is independent of the perturbative Abelian gauge used to solve
equation (\ref{Abelgeq}).  This is so because the collision integral
computed this way can always be expressed in terms of the imaginary
parts of the polarisation tensors (\ref{imag}) in the plasma, which
are known to be gauge-independent.
 
In any case, the transverse polarisation tensor $\Pi_T$ vanishes at
$q_0 =0$, and the dynamical screening is not enough to make \Eq{C}
finite.  An IR cutoff must be introduced by hand in order to evaluate
the integral.  With a cutoff of order $\Lambda \approx g m_D$ we thus
find at logarithmic accuracy
\beq
C_{\rm LLO}({\bf v}, {\bf v}') 
= 
\frac{2}{\pi^2 m^2_D}  \ln\left(1/g \right) 
\frac{({\bf v}\cdot{\bf v}')^2}{\sqrt{1-({\bf v}\cdot{\bf v}')^2}}\ .
\eeq
The logarithmic dependence on the gauge coupling comes from the integral
\beq
\int^{m_D}_{\Lambda}\0{dq}{q}=\ln\0{m_D}{\Lambda}\ .
\eeq
The natural cut-off, as we shall see below, is rather given by the
hard gluon damping rate. At logarithmic accuracy, their difference has
no effect.

Using also the relation \Eq{magical-rel} we finally arrive at the 
collision integral to leading logarithmic accuracy,
\begin{mathletters}
\step
\begin{eqnarray}
\label{leading-log}
\left. 
\left\langle \xi^{(1)}_{\rho,a}(x,v)\right\rangle
\right|_{\rm LLO} 
&=&
 -  \frac{g^2}{4\pi} N T \ln\left(1/g \right) 
v_\rho\!\int\!
\frac{d\Omega_{{\bf v}'}}{4\pi}\, {I}(\vv,\vv')\bar{\cal J}^0_a(x,\vv'), 
\zeile\step
\label{IK}
{I}(\vv,\vv') &=&  \delta^{(2)}({\bf v}-{\bf v}')-{K}(\vv,\vv')
\zeile\step
{K}(\vv,\vv')&=&  \frac{4}{\pi}
\frac{({\bf v}\cdot{\bf v}')^2}{\sqrt{1-({\bf v}\cdot{\bf v}')^2}} \ ,
\end{eqnarray}
\end{mathletters}%
where we have introduced 
$ \delta^{(2)}({\bf v}-{\bf v}') 
\equiv 4 \pi  \delta^{(2)}(\Omega_{\bf v}- \Omega_{\bf v}') $, 
$\int \frac{d\Omega_\vv}{4 \pi} \delta^{(2)}({\bf v}-{\bf v}')=1$.
The collision integral has first been obtained in \cite{Bodeker:1998hm}, 
and subsequently in \cite{Arnold:1999cy,LM,LM2,Valle,Blaizot:1999xk}.

We can verify explicitly that the collision integral to leading 
logarithmic accuracy is consistent with gauge invariance. This should 
be so, as the approximations employed have been shown in 
Section~\ref{ConsistentApprox} on general grounds to be consistent with 
gauge invariance. Evaluating the correlator in \Eq{DJ} to
leading logarithmic accuracy yields 
\beq
\label{c-check}
\left. 
g f_{abc} \left\langle a^b_\mu (x) \de J^\mu_c (x) \right\rangle 
\right|_{\rm LLO} 
= - \frac{g^2}{4\pi} N T \ln\left(1/g \right) 
    \int \frac{d\Omega_{{\bf v}}}{4\pi}\,
         \frac{d\Omega_{{\bf v}'}}{4\pi}\, 
         {I}(\vv,\vv')\bar{\cal J}^0_a(x,\vv') \ ,
\eeq
which vanishes, because 
\beq\label{averageI}
\!\int\! \frac{d\Omega_{{\bf v}}}{4\pi}\, {I}(\vv,\vv') = 0 \ .
\eeq
We thus establish that $\bar D_\mu \bar J^\mu =0$, in accordance 
with \Eq{soft-YM} in the present approximation.

\subsection{Stochastic noise}
\label{BeyondStochastic}

The collision integral obtained above describes a dissipative process
in the plasma. In principle it could trigger the system to abandon
equilibrium \cite{K2}. Whenever dissipative processes are encountered,
it is important to also identify the stochastic source related to it.
This is the essence of the fluctuation-dissipation theorem
\cite{Landau5,K2}.  Phenomenologically, this is well known, and
sometimes used the other way around: imposing the
fluctuation-dissipation theorem allows one to identify a source for
stochastic noise with the strength of its self-correlator fixed by the
dissipative processes.

In the present formalism, it is possible to identify directly the
source for stochastic noise which prevents the system from abandoning
equilibrium as discussed in section~\ref{EffectiveCollisions}.  The
relevant noise term is given by the field-independent, that is, the
{\it source} fluctuations in $\xi^{(0)}$,
\beq
\xi^{\rho(0)}_a(x,\vv)\equiv 
\left.
-g f_{abc} \, v^\mu a^b_\mu(x)\de{\cal J}^{\rho,c}(x,\vv)
\right|_{\bar A=0,\,\bar J=0}\ .
\eeq
While its average vanishes, $\langle \xi^{(0)}(x,\vv) \rangle=0$, 
its correlator 
\beq
\llangle \xi^{\rho (0)}_a (x,\vv)\, \xi^{\sigma (0)}_b (y,\vv') \rrangle 
= g^2 f_{apc} f_{bde} v^\mu v'^\nu
  \llangle 
  a^p_\mu (x)\,\delta {\cal J}_{\mbox{\tiny source}}^{\rho, c}(x,\vv)\, 
  a^d_\nu (y)\,\delta {\cal J}_{\mbox{\tiny source}}^{\sigma, e}(y,\vv')
  \rrangle^{(0)}
\eeq
does not.  In order to evaluate this correlator we switch to
Fourier space. Within the second moment approximation we expand the
correlator $\langle \delta f \delta f \delta f \delta f \rangle$
into products of second order correlators 
$\langle \delta f \delta f\rangle \langle \delta f \delta f \rangle$
and find  
\begin{eqnarray}
&&\llangle \xi^{\rho (0)}_a (k,\vv)\,  \xi^{\sigma (0)}_b (p,\vv') \rrangle 
 =   g^2 f_{apc} f_{bde} v^\mu v'^\nu \int \frac{d^4 q}{(2 \pi)^4} 
      \int \frac{d^4 r}{(2 \pi)^4} 
\nonumber \zeile
&& \quad\quad \times\left \{\ \, 
\llangle a^{(0,0)}_{\mu p} (q)\,a^{(0,0)}_{\nu d} (r) \rrangle\,  
\llangle \delta {\cal J}_{\mbox{\tiny source}}^{(0) \rho, c} (k-q,\vv)\, 
         \delta {\cal J}_{\mbox{\tiny source}}^{(0) \sigma, e}(p-r,\vv') 
\rrangle \right. \nonumber 
\zeile && \quad\quad \quad + \left. 
\llangle a^{(0,0)}_{\mu p} (q)\,
         \delta {\cal J}_{\mbox{\tiny source}}^{(0)\sigma, e} (p-r,\vv') 
\rrangle\,  
\llangle \delta {\cal J}_{\mbox{\tiny source}}^{(0) \rho, c} (k-q,\vv)\, 
         a^{(0,0)}_{\nu d}(r) 
\rrangle 
\right \}
\end{eqnarray}
In the leading logarithmic approximation we retain only the contributions 
from the transverse modes. Evaluating the correlators leads to
\beq
\llangle \xi^{\mu,a}_{(0)} (x,\vv) \,
         \xi^{\nu,b}_{(0)} (y,\vv') \rrangle 
= \frac{g^6  N C_2^2  B^2_C} {(2 \pi)^3 m^2_D}\, 
  \ln{(1/g)}\, v^\mu v'^\nu\, {I}(\vv,\vv')\, 
  \delta^{ab}\, \delta^{(4)} (x-y) \ .               \label{noi-Bolt-0}
\eeq
After averaging over the angles of ${\bf v}$ and ${\bf v'}$, and 
using the relation \Eq{magical-rel}, the correlator becomes
\beq
\llangle \xi^{i,a}_{(0)} (x)\, \xi^{j,b}_{(0)}  (y) \rrangle 
= 2 T\, \frac{m^2_D}{3}\, \frac{g^2}{4 \pi}\, N T \ln{(1/g)}\, 
  \delta^{ab}\, \delta^{ij} \,\delta^{(4)} (x-y) \ .   \label{noi-Bolt}
\eeq
In particular, all correlators 
$\langle \xi^{0}_{(0)} (x) \xi^{\mu}_{(0)}(y) \rangle$ vanish. 
\Eq{noi-Bolt} identifies $\xi_{(0)}^i(x)$ as a source of 
white noise. The noise term has been derived in \cite{Bodeker:1998hm}, 
and subsequently in \cite{Arnold:1999cy,LM,LM2}.

The presence of this noise term does not interfere with the covariant 
current conservation confirmed at the end of the previous section. 
This can be seen as follows. The noise term enters \Eq{c-check} as the 
angle average over the $0$-component of $\xi^\mu_{(0)}(x,\vv)$. As we 
have established above, the logarithmically enhanced contribution from 
the noise source stems from its correlator \Eq{noi-Bolt-0}. Averaging 
the temporal component of \Eq{noi-Bolt-0} over the angles of $\bf v$, 
and using \Eq{averageI}, it follows that
\beq
\llangle
        \xi_{(0)}^{0,a}(x,\vv' )
        \int \frac{d\Omega_\vv}{4\pi}\,\xi^{0,b}_{(0)}(y,\vv)
\rrangle
=0  \ .                                               \label{no-noise}
\eeq 
We thus conclude, that the temporal component of the noise,
$\xi^0_{(0)}(x,\vv)$, has no preferred $\vv$-direction, which implies that 
$\int \frac{d\Omega}{4\pi}\xi^0_{(0)}(x,\vv)=0$  
in the leading logarithmic approximation. Thus, the mean current 
conservation is not affected by the noise term.

\subsection{Ultrasoft amplitudes}
\label{BeyondUltra}

After integrating-out the statistical fluctuations to leading
logarithmic order, we end up with the following set of mean field
equations \cite{Bodeker:1998hm} (from now on, we drop the bar to
denote the mean fields),
\begin{mathletters} 
\label{fin-set}
\step
\begin{eqnarray} 
v^\mu  D_\mu {\cal J}^\rho(x,\vv)
&=& - m^2_D v^\rho v^\mu  F_{\mu0} (x)
    - \gamma\  v^\rho\!\int\! \frac{d\Omega_{{\bf v}'}}{4\pi}\, 
      {I}(\vv,\vv') {\cal J}^0(x,\vv') 
    + \zeta^\rho(x,\vv)\, , \quad\quad \quad                                
\label{last-Bol}
\zeile\step 
D_\mu F^{\mu\nu}  
&=& J^\nu\equiv 
\int\!\frac{d\Omega_{{\bf v}}}{4\pi}\,{\cal J}^\nu(x,\vv)\, .
\end{eqnarray}
\end{mathletters}%
Here, we denote by $\zeta^\rho(x,\vv)$ the stochastic noise term
identified in the preceding section, its correlator given by
\Eq{noi-Bolt-0}.  We also introduced
\beq  \gamma = \frac{g^2}{4\pi} N T \ln\left(\01g \right)\ ,
\eeq 
which is identified as (twice) the damping rate for the ultra-soft
currents \cite{damp-rat-P}. We refer to \Eq{last-Bol} as a
Boltzmann-Langevin equation as it accounts for quasi-particle
interactions via a collision integral as well as for the stochastic
character of the underlying fluctuations in the distribution function.

In \Eq{fin-set}, both the quasi-particle degrees of freedom and the
ultra-soft gauge modes are present. In order to integrate-out the
quasi-particle degrees of freedom it is necessary to solve
\Eq{last-Bol} explicitly to obtain the current as a functional of the
ultra-soft gauge fields $J[A]$, from which a generating functional
$J(x)=-\de\Gamma[A]/\de A(x)$ could in principle be deduced in full
analogy to the HTL case. The ultra-soft amplitudes of the plasma can
be deduced from $J[A]$ itself \cite{Bodeker:2000ud,Blaizot:2000fq}, as
done in \Eq{SoftAmplitudes} for the soft amplitudes,
\beq\label{UltraSoftAmplitudes}
J^a_\mu[A]=
      \Pi^{ab}_{\mu\nu}\, A^\nu_b
+\012 \Pi^{abc}_{\mu\nu\rho}\, A^\nu_b\,A^\rho_c
+\ldots
\eeq 
With this in mind, we consider again the Boltzmann-Langevin equation
\eq{last-Bol} for the quasi-particle distribution function. It has
three distinct scale parameters: the temperature $T$, the Debye mass
$m_D$, and the damping term $\gamma$. In the leading logarithmic
approximation, these scales are well separated,
\beq
g^2T\ll\gamma\ll m_D\ll T\ ,
\eeq 
and at least logarithmically larger than the non-perturbative scale of
the magnetic mass. This is why \Eq{last-Bol} is dominated by different
terms, depending on the momentum range considered.  For hard momenta,
\Eq{last-Bol} is only dominated by the left-hand side, reducing it to
the (trivial) current of hard particles moving on world lines.  For
momenta about the Debye mass, the term proportional to $m_D^2$ becomes
equally important, while the noise term and the collision integral
remain suppressed by $\gamma/m_D$. The resulting current is then given
by \Eq{class-HTL-curr}, the HTL current. Momentum modes below the
Debye mass are affected by the damping term of the collision integral.
Close to the scale of the Debye mass, the higher order corrections to
\Eq{class-HTL-curr} are obtained as an expansion in $\gamma/v\cdot D$.
We write
\beq
{\cal J}^\mu(x,\vv) 
= \sum_{n=0}^{\infty}  {\cal J}^\mu_{(n)}(x,\vv) \ , \label{seriesJ}
\eeq
where the current densities ${\cal J}^\mu_{(n)}(x,\vv)$ obey the
differential equations
\begin{mathletters}\label{seriesA}\step
\begin{eqnarray} 
v^\mu  D_\mu {\cal J}^\nu_{(0)}(x,\vv)
  &=& - m^2_D  v^\nu v^\mu  F_{\mu0} (x)  + \zeta^\nu(x,\vv) \ ,
\zeile\step
v^\mu  D_\mu {\cal J}^\nu_{(n)}(x,\vv)
  &=&- \gamma  v^\nu \int\! \frac{d\Omega_{{\bf v}'}}{4\pi}\, 
                 {I}(\vv,\vv') {\cal J}^0_{(n-1)}(x,\vv') \ .
\eea
\end{mathletters}%
Apart from the noise term, the leading order term in this expansion,
${\cal J}^\nu_{(0)}(x,\vv)$, coincides with the HTL current
\Eq{class-HTL-curr}. All higher order terms ${\cal
  J}^\mu_{(n)}(x,\vv)$ are smaller by powers of $\sim (\gamma/v\cdot
D)^n$, and recursively given by
\begin{mathletters}\label{solutionA}\step
\bea
{\cal J}^\nu_{(0)}(x,\vv) 
&=& \int^{\infty}_0 d \tau U(x,x-v \tau)\, 
    \left\{ -m^2_D\, v^\nu v^\mu\,  F_{\mu0} (x-v\tau) 
            + \zeta^\nu(x-v\tau,\vv) \right\} \ ,\quad \quad \quad 
\zeile\step
{\cal J}^{\nu}_{(n)} (x,\vv) 
&=& -\gamma\,\int^{\infty}_0 d \tau
       U(x,x-v \tau)  v^\nu \int\! \frac{d\Omega_{{\bf v}'}}{4\pi}\, 
                 {\cal I}(\vv,\vv')\, {\cal J}^0_{(n-1)}(x,\vv') \ .
\eea
\end{mathletters}%
This recursive expansion is consistent with covariant current conservation. 
For every partial sum up to order $n$, we have
\beq
  D_\mu \left(  J^\mu_{(0)}
              + J^\mu_{(1)}
              + \ldots 
              + J^\mu_{(n)} \right) = 0\ .
\eeq
This expansion has been considered in \cite{LM2}.  It describes
correctly how the presence of the collision integral modifies the
ultra-soft current. From \Eq{solutionA}, all ultrasoft amplitudes of
the gauge fields at equilibrium can be deduced. The noise source does
not contribute when amplitudes like the coefficients in the expansion
\Eq{UltraSoftAmplitudes} are evaluated at equilibrium. Some explicit
results for this case have been given recently
\cite{Bodeker:2000ud,Blaizot:2000fq}.

The expansion \Eq{solutionA} has a limited domain of validity because
the effective expansion parameter grows large for both small
frequencies $k_0$ and small momenta $\vk$. This implies that the
overdamped regime where $v\cdot D\ll \gamma$ cannot be reached.
Alternatively, one can separate the local from the non-local part of
the collision integral to perform an expansion in the latter only.
The effective expansion parameter is then $\gamma/(v\cdot D +
\gamma)$, which has a better infrared behaviour.  It is expected that
the expansion is much better for the spatial than for the temporal
component of ${\cal J}^\rho(x,\vv)$. This is so, because the term
proportional to the non-local part ${K}(\vv,\vv')$ of the collision
integral in \Eq{last-Bol} gives no contribution to the dynamical
equations of the spatial component $J^i$ after angle averaging
\Eq{last-Bol} over the directions of $\vv$. However, for the dynamical
equation of the temporal component, this term precisely cancels the
local damping term, which is of course a direct consequence of current
conservation.

In this light, we decompose the current as in \Eq{seriesJ}, but
expanding effectively in $\gamma/(v\cdot D + \gamma)$. We find the
differential equations
\begin{mathletters}\label{seriesB}\step
\bea
(v^\mu  D_\mu +\gamma) {\cal J}^\rho_{(0)}(x,\vv)
&=& - m^2_D v^\rho v^\mu  F_{\mu0} (x) + \zeta^\rho(x,\vv)\ , 
\label{soft-current-B}
\zeile\step
(v^\mu  D_\mu+\gamma) {\cal J}^\rho _{(n)}(x,\vv)
&=& \gamma\  v^\rho\int\!\frac{d\Omega_{{\bf v}'}}{4\pi}\, 
          {K}(\vv,\vv') {\cal J}^0_{(n-1)}(x,\vv') .
\end{eqnarray}
\end{mathletters}%
The retarded Green's function $G_{{\rm ret}}$ obeys
\beq
i \left(v^\mu D_\mu + \gamma  \right) \, G_{{\rm ret}}(x,y;\vv) 
= \delta^{(4)} (x-y) \ ,
\eeq
and reads, for $t=x_0-y_0$,
\beq\label{green2}
G_{{\rm ret}}(x,y;\vv)_{ab} 
= -i \theta(t)  \delta^{(3)} \left({\bf x}-{\bf y}-{\bf v}t\right)
   \exp (-\gamma t)\,U_{ab}(x, y) \ .
\eeq
The iterative solution to the Boltzmann-Langevin equation 
is
\begin{mathletters}\label{solutionB}\step
\bea
\label{NA-Ohm-J}
{\cal J}^{\rho}_{(0)} (x,\vv) 
&=& \int^{\infty}_0 d \tau \exp (-\gamma \tau)\,
       U(x,x_\tau) \left\{  -m^2_D\,v^\rho v^j\, F_{j 0} (x_\tau)
       + \zeta^\rho(x_\tau,\vv)  \right\}\ ,
\zeile\step
{\cal J}^\rho _{(n)}(x,\vv) 
&=& \gamma \int^{\infty}_0 d \tau \exp (-\gamma \tau)\,
        U(x,x_\tau)\, v^\rho
        \int\ \frac{d\Omega_{{\bf v}'}}{4\pi}\, {K}(\vv,\vv') \,
                                     {\cal J}^0_{(n-1)}(x_\tau,\vv') \ .
\quad\quad\quad
\eea
\end{mathletters}%
This expansion in consistent with current conservation, if the angle 
average of ${\cal J}^0_{(n)}$ vanishes for some $n$. This follows
from taking the temporal  component of \Eqs{seriesB} and averaging 
the equation over ${\bf v}$, to find
\begin{mathletters}\step
\bea
D_0 J^0 _{(0)} + D_i J^i_{(0)} 
&=& - \gamma J^0 _{(0)}  \ , \zeile\step
D_0 J^0 _{(n)} + D_i J^i_{(n)} 
&=& \gamma J^0 _{(n-1)} - \gamma J^0 _{(n)}  \ 
\eea
for the individual contributions, and
\step\beq
    D_\mu \left(  J^\mu_{(0)} 
                + J^\mu_{(1)}
                + \ldots 
                + J^\mu_{(n)} \right)
  + \gamma J^0 _{(n)}   = 0   \ 
\eeq
\end{mathletters}%
for their sum, which is consistent if $\gamma J^0_{(n)}$ is vanishing 
for some $n$. 

To leading order, the ultra-soft colour current ${\cal J}^i_{(0)}(x,\vv)$ 
in \Eq{NA-Ohm-J} has the same functional dependence on the field strength 
and on the parallel transporter as the soft colour current 
\Eq{class-HTL-curr}. There is, however, 
an additional damping factor $\exp(-\gamma \tau)$ in the integrand.

\subsection{Langevin dynamics}
\label{BeyondLangevin}

Finally, we consider the overdamped regime (or quasi-local limit) of 
the above equations. This is the regime where $k_0\ll |\vk|\ll\gamma$.
Consider the mean field currents \Eq{solutionB}. 
The terms contributing to these currents are exponentially suppressed for 
times $\tau$ much larger than the characteristic time scale $1/\gamma$. 
On the other hand, the fields occurring in the integrand typically vary 
very slowly, that is on time scales $\ll 1/m_{\rm D}$. Thus, in the 
quasi-local limit we can perform the approximations
\begin{mathletters}\step
\bea
 U_{ab}(x,x-v \tau) &\approx& U_{ab}(x,x) = \delta_{ab} \ ,\zeile\step
F_{j 0}  (x-v\tau)  &\approx&   F_{j 0}  (x) \ .
\eea
\end{mathletters}%
In this case the remaining integration can be performed. The solution 
for the spatial current $J^i(x)$ stems entirely from the leading order 
term \Eq{NA-Ohm-J}. All higher order corrections vanish, because 
they are proportional to
\beq
\int\frac{d\Omega_\vv}{4\pi}\,\vv \,{K}(\vv,\vv')
\eeq
which vanishes.
For the complete set of gauge field equations in the quasi-local limit we 
also need to know $J^0(x)$. The iterative solution \Eq{solutionB} gives 
$J^0_{(n)}(x)=0$ to any finite order. Therefore, we use instead the
unapproximated dynamical equation for ${\cal J}^0(x,\vv)$, which yields, 
averaged 
over the directions of $\vv$, current conservation. In combination with 
the solution \Eq{Jsigma} for the spatial current, the Boltzmann-Langevin 
equation \eq{fin-set} becomes
\begin{mathletters}\label{BEQ}\step
\begin{eqnarray}
D_\mu  F^{\mu i} & = & \sigma  E^i  + \nu^i \ , \zeile\step
D_i E^i & = & J^0 \ , \zeile\step
D_0 J^{0}& =&  -\sigma J^0  - D_i \nu^i  \ .
\end{eqnarray}
\end{mathletters}%
In the limit, where the temporal derivative term $D_0 F^{0i} $ can be
neglected, one finally obtains for the spatial current from
\Eq{solutionB}
\begin{mathletters}\step
\bea
\label{Jsigma}
J^i_a         &=&  \sigma  E^i _a + \nu^i_a \ , \zeile\step 
\sigma        &=&  \frac{4 \pi m_D^2}{3Ng^2T \ln \left(1/g\right)}\ ,
\eea
where $\sigma$ denotes the colour conductivity of the plasma. The noise 
term reads
\bea
\label{l-noise}
\step \nu(x)&=&\01{\gamma}\int \0{d\Omega_\vv}{4\pi} \zeta(x,v)\ , 
\zeile\step 
\left\langle \nu^i_{a}(x)\ \nu^j_{b}(y)\right\rangle 
&= & 2\, T \,\sigma\, \delta^{ij}\, \delta_{ab}\, \delta^{(4)}(x-y) \ .
\eea
\end{mathletters}%
The noise term appearing in the Yang-Mills equation becomes white
noise within this last approximation. The fluctuation-dissipation
theorem is fulfilled because the strength of the noise-noise
correlator \Eq{l-noise} is precisely given by the dissipative term of
\Eq{Jsigma}. This is the simplest form of the FDT.  The colour
conductivity in the quasi-local limit has been obtained originally by
B\"odeker \cite{Bodeker:1998hm}.

It is worth pointing out that already in the leading logarithmic
approximation the noise term appearing in the Yang Mills equation is
not white, except in the local limit \Eq{l-noise}.  The noise in the
Boltzmann-Langevin equation, on the other hand, is white (see
\Eq{noi-Bolt}), when averaged over the directions of $\vv$.

\newpage

\section{Langevin approach}
\label{Langevin}

\subsection{Coarse-graining}
\label{LangevinCoarse}

In this section, we consider a more phenomenologically inspired
Langevin-type approach to fluctuations in non-Abelian plasmas
\cite{LM3}.  Let us step back to the original microscopic transport
equation \eq{NA-Micro}. The microscopic distribution function is a
strongly fluctuating quantity at scales associated to mean particle
distances, namely $\sim 1/T$ for a plasma close to thermal
equilibrium. In order to obtain an effective transport theory at
length scales much larger than the typical inter-particles distances,
one may wish to coarse-grain the distribution function and the
non-Abelian fields over scales characteristic for the problem under
investigation.  With coarse-graining, we have in mind a volume average
over a characteristic (physical) volume and/or over characteristic
time scales. Such a coarse-graining results in a coarse-grained
distribution function which is considerably smoother than the
microscopic one.

The appropriate physical volume depends on the particular physical
problem under investigation. As a guideline, one would like to have a
{\it scale separation} such that the coarse-graining scale is large as
compared to typical length scale at which two-particle correlations
grow large. This way, it is ensured that two- and higher-particle
correlators remain small within a coarse-graining volume. In addition,
the coarse-graining volume should be sufficiently large, such that the
remaining particle number fluctuations of the one-particle
distribution function within the coarse-graining volume remain
parametrically small. Finally, the coarse-graining scale should be
smaller than typical relaxation or damping scales of the problem under
investigation. It cannot be guaranteed on general grounds that these
requirements can be met. For a hot plasma close to equilibrium,
however, the appropriate coarse-graining scale is given by the Debye
radius.

Performing such a procedure with the microscopic transport equation
\eq{NA-f}, we expect to obtain a Boltzmann-Langevin-type of equation
for the coarse-grained one-particle distribution function
${f}(x,p,Q)$, namely
\begin{equation}    
\label{CoarseTrans}
p^\mu\left(\frac{\partial}{\partial x^\mu}
        -g f^{abc}A^{b}_\mu Q^c\frac{\partial}{\partial Q^a}
        -gQ_aF^{a}_{\mu\nu}\frac{\partial}{\partial p_\nu}\right) {f}(x,p,Q)
= C[{f}](x,p,Q) +\zeta(x,p,Q) \ . 
\end{equation}
Here, the transport equation contains an effective -- but not yet
specified -- collision term $C[f]$, and an associated source for
stochastic noise $\zeta$. In the collisionless limit $C=\zeta=0$, the
above set of transport equation reduces to those introduced by Heinz
\cite{Heinz:1983nx}. In the general case, however, the right-hand side
of \Eq{CoarseTrans} does not vanish due to effective interactions
(collisions) in the plasma, resulting in the term $C[f]$. In writing
\Eq{CoarseTrans}, we have already made the assumption that the
one-particle distribution function $f$ is a fluctuating quantity.
This is quite natural having in mind that $f$ describes a
coarse-grained microscopic distribution function for coloured point
particles, and justifies the presence of the stochastic source $\zeta$
in the transport equation. 

For non-charged particles, a similar (phenomenological) kinetic
equation has been considered by Bixon and Zwanzig \cite{Bixon1969}.
This philosophy and its extension for out-of-equilibrium situations
have also been applied to nuclear dynamics
\cite{Randrup:1990bb,Ayik:1990bb,Chomaz:1991yn}. For a recent review
of different approaches, we refer to \cite{Abe:1996yw}. Some recent
approaches which include stochastic noise sources within a
Schwinger-Dyson approach have been reported as well
\cite{Calzetta:2000xh}. For a stochastic interpretation of the
Kadanoff-Baym equations, see \cite{Greiner:1998vd}.

The coarse-grained transport equation is
accompanied by a corresponding Yang-Mills equation,
\begin{equation} 
\label{CoarseYM}
\left(D_\mu F^{\mu \nu} \right)_a = J^\nu_a (x)  
\equiv g \sum_{\hbox{\tiny helicities}\atop\hbox{\tiny species}} 
    \int dP \,dQ\, Q_a\, p^\nu\,f(x,p,Q) \ ,
\eeq
which contains, to leading order, the current due to the
quasi-particles, but no fluctuation-induced current.
 
Given the stochastic dynamical equation \Eq{CoarseTrans}, the question
arises as to what can be said on general grounds about the spectral
functions of $f(x,p,Q)$ and $\zeta(x,p,Q)$. For simplicity,  we shall
assume that the dissipative  processes are known close to equilibrium,
but no further information is given regarding the  underlying
fluctuations.  This way of proceeding is {\it complementary} to the
procedure of \cite{LM,LM2} as worked out in Sections \ref{Ensemble}
and \ref{Effective}, where  the right-hand side of \Eq{CoarseTrans} 
has been obtained from correlators of the microscopic statistical 
fluctuations.

Here, we show that the spectral function of the fluctuations close to
equilibrium can be obtained from the knowledge of the kinetic entropy
of the plasma. The spectral function of the noise source is shown to
be linked to the dissipative term in the effective transport
equation. This gives a well-defined prescription as to how the correct
source for noise can be identified {\it without} the detailed
knowledge of the underlying microscopic dynamics responsible for the
dissipation. The basic idea behind this approach relies on the essence
of the fluctuation-dissipation theorem. While this theorem is more
general, here we will only discuss the close to equilibrium situation.
According to the fluctuation-dissipation theorem, if a fluctuating
system remains close to equilibrium, then the dissipative process
occurring in it  are known. Vice versa, if one knows the dissipative
process in the system, one can describe the fluctuations without an
explicit knowledge of the microscopic structure or processes in the
system. The cornerstone of our  approach is the kinetic entropy of the
fluctuating system, which serves to identify the associated
thermodynamical forces. This leads to the spectral function for the
deviations from the non-interacting equilibrium.

\subsection{Classical dissipative systems}
\label{LangevinClassical}

Before entering into the discussion of plasmas, we will illustrate
this way of proceeding by  reviewing the simplest setting of
classical linear dissipative systems \cite{Landau5}.  A generalisation
to the more complex case of non-Abelian plasmas will then become a
natural step to perform.   We consider a classical homogeneous system
described by a set of variables $x_i$, where $i$ is a discrete index
running from $1$ to $n$.  These variables are normalised in such a way
that their mean values at equilibrium vanish. The entropy of the
system is a function of the quantities $x_i$, $S(x_i)$. If the system
is at equilibrium, the entropy reaches its maximum, and thus
$(\partial S/ \partial x_i)_{\rm eq} = 0$, $\forall i$. If the system
is taken slightly away from equilibrium, then one can expand the
difference $\Delta S =S- S_{\rm eq}$, where $S_{\rm eq}$ is the
entropy at equilibrium, in powers of $x_i$. If we expand up to
quadratic order, then
\begin{equation} 
\Delta S = \frac 12 \left(
\frac{ \partial^2 S}{ \partial x_i \partial x_j} \right)_{\rm eq} x^i x^j 
\equiv  - \frac 12 \beta_{ij} x^i x^j  \ .  
\end{equation} 
The matrix $\beta_{ij}$ is symmetric and
positive-definite, since the entropy reaches a maximum at equilibrium.
The thermodynamic forces $F_i$ are defined as the gradients of $\Delta S$
\begin{equation}
F_i = - \frac{ \partial \Delta S}{ \partial x_i}  \ .
\end{equation} 
For a system close to equilibrium  the thermodynamic forces are linear
functions of $x_i$, $F_i =  \beta_{ij} x^j$.
If the system is at equilibrium, the thermodynamic forces vanish. 
In more general situations the variables 
$x_i$ will evolve in time.
The time evolution of these variables is given as functions
of the thermodynamical forces. In a close to equilibrium case
one can expect that the evolution is linear in the forces
\begin{equation}
\frac{d x^i}{dt} = - \gamma^{ij} F_j \ ,
\end{equation} 
which, in turn, can be expressed as
\begin{equation}
\frac{d x^i}{dt} = - \lambda^{ij} x_j \ .
\end{equation} 
Within a phenomenological Langevin approach a white noise source is
added to account for the underlying fluctuations. Otherwise, the
system  would abandon equilibrium. Hence, we write instead
\begin{equation}
\frac{d x^i}{dt} = - \lambda^{ij} x_j + \zeta^i \ .
\end{equation} 
The first term on the right-hand side describes the  mean regression 
of the system towards equilibrium,  while the second term is the source
for stochastic noise.  The quantities $\gamma^{ij}$ are known as the
kinetic coefficients,  and it is not difficult to check that
$\gamma_{ij}= \lambda_{ik} \beta^{-1}_{kj}$.  From the value of the
coefficients $\beta_{ij}$ one can deduce the  equal time correlator
\begin{equation}
\left\langle x_i (t) x_j (t)\right\rangle =  \beta^{-1}_{ij} \ ,
\label{invbet}
\end{equation}
which is used to obtain Einstein's law
\begin{equation}
\label{einstein}
\left\langle x^i (t) F_j (t)\right\rangle = \delta^i _j  \ .
\end{equation}
After taking the time derivative of \Eq{einstein},  
assuming that the noise is white and Gaussian
\begin{equation}
\left\langle \zeta^i (t) \zeta^j (t') \right\rangle = \nu^{ij} \delta(t-t') \ ,
\end{equation}
we find that the strength of the noise self-correlator $\nu$ is determined
by the dissipative process
\begin{equation}
\nu^{ij} = \gamma^{ij} + \gamma^{ji} \ ,
\label{simpFDT}
\end{equation}
which is the fluctuation-dissipation relation we have been aiming at. 

\subsection{Non-Abelian plasmas as a classical dissipative system}
\label{LangevinFluctuations}

We now come back to the case of a non-Abelian plasma and generalise
the above discussion to the case of our concern.  We will consider the
non-Abelian plasma as a classical linear dissipative system, assuming
that we know the collision term in the transport equation. In order to
adopt the previous reasoning, we have to identify the dissipative term
in the transport equation, and to express it as a function of the
thermodynamical force obtained from the entropy. The deviation from
the equilibrium distribution is given here by
\begin{equation}
\Delta f(x,p,Q) = {f}(x,p,Q) - f_{\rm eq}(p_0)\ ,
\end{equation} 
and replaces the variables $x_i$ discussed above. 

\subsubsection*{1. Classical plasmas}

The entropy flux density for classical plasmas has been given in 
\Eq{Smicro} with \Eq{Sigma-cl} for the classical plasma. It reads 
explicitly
\beq
\label{Scl}
S_\mu (x) = - \int dP dQ\, p_\mu \, f(x,p,Q) 
\left( \ln{( f(x,p,Q) )} -1 \right) \ .
\end{equation}
The $\mu =0$ component of \Eq{Scl} gives the entropy density  of the
system.  The entropy itself is then obtained as $S = \int d^3x \, S_0
(x)$.

We shall now assume that the deviation of the mean particle number
from the equilibrium one is small within a coarse-graining volume.
This can always be arranged for at sufficiently small gauge coupling
$g\ll 1$, which ensures that fluctuations are parametrically
suppressed by the plasma parameter.  We then obtain $\Delta S$ just by
expanding the expression of the  entropy density in powers of $\Delta
f$ up to quadratic order. It is  important to take into account that
we will consider situations where  the small deviations from
equilibrium are such that both the particle  number and the energy
flux remain constant, thus
\begin{mathletters}
\begin{eqnarray}
\int  dP dQ \, p_0\ \ \,       \Delta f(x,p,Q) & = & 0 \ ,
\zeile
\int  dP dQ \, p_0p_\mu\,      \Delta f(x,p,Q) & = & 0 \, .
\end{eqnarray}
\end{mathletters}%
With these assumptions, one reaches 
\begin{eqnarray}
\Delta S_0 (x) & = &   
- \int dP dQ \, p_0 \,
\frac{(\Delta f(x,p,Q))^2}{2f_{\rm eq}(p_0)} \nonumber \\
               & = &  
- \int \0{d^3 p}{(2 \pi)^3} \, dQ   
\frac{(\Delta f(x,{\bf p},Q))^2}{2f_{\rm eq}({\omega_p} )} \ ,
\label{S0}
\end{eqnarray}
where we have taken into account the mass-shell  condition the second
line, with $p_0 = \omega_p =\sqrt{{\bf p}^2 + m^2}$.  Without loss of
generality, we will consider from now on the case of massless
particles, such that $\omega_p = p = |{\bf p}|$.

The {\it thermodynamic force} associated to $\Delta f$ is defined 
from the entropy as
\begin{equation}    
\label{F}
F(x,{\bf p},Q)  
= - \frac {\delta \Delta S}{\delta \Delta f(x,{\bf p},Q)} 
=  \0{1}{(2 \pi)^3} \frac{\Delta f(x,{\bf p},Q)}{f_{\rm eq}(p)} \ .
\end{equation}
We linearise the transport equation (\ref{CoarseTrans}) and express 
the collision integral close to equilibrium in terms of the 
thermodynamical force. Dividing \Eq{CoarseTrans} by $p_0$ and imposing 
the mass-shell constraint, we find
\begin{equation}  
\label{linearisedBeq}
   v^\mu D_\mu \Delta f 
-g v^\mu Q_aF^{a}_{\mu 0}\frac{d f_{\rm eq}}{d p} 
=   C[\Delta f](x,{\bf p},Q) 
  + \zeta(x,{\bf p},Q) \ , 
\end{equation}
where $v^\mu = p^\mu/p_0 =(1, {\bf v})$, with ${\bf v}^2 =1$.
We also introduced the shorthand $D_\mu \Delta f\equiv 
(\partial_\mu -g f^{abc}A_{\mu,b}Q_c\partial^Q_a)\Delta f$ 
as in \Eq{Df}.
It is understood that the collision integral has been linearised,
and we write it as
\begin{equation}\label{Cf} 
C_{\rm lin}[\Delta f] (t,{\bf x},{\bf p},Q) =  
\int  d^3 x' d^3 p' \, dQ'\,
{K}({\bf x},{\bf p},Q;{\bf x}',{\bf p'},Q') 
\Delta f(t,{\bf x}',{\bf p}',Q')\, ,  
\end{equation} 
with $t\equiv x_0$. For simplicity, we take the collision integral
local in time, but unrestricted otherwise.\footnote{Of course,  gauge
invariance imposes further conditions on both the collision  term and
the noise. However, these constraints are of no relevance  for the
present discussion.}   The thermodynamical force is linear in $\Delta
f$. The linearised collision integral can easily be expressed in terms
of $F$ as
\begin{equation}
\label{CF} 
C_{\rm lin}[F] (t,{\bf x},{\bf p},Q) =  
\int  d^3 x' d^3 p' \, dQ'\,
K({\bf x},{\bf p},Q;{\bf x}',{\bf p'},Q') 
\,(2 \pi)^3 \,f_{\rm eq}(p_0)F(t,{\bf x}',{\bf p}',Q')\, .  
\end{equation} 
According to the fluctuation-dissipation relation, the source of
stochastic noise has to obey
\begin{equation} 
\left\langle  \zeta(x,{\bf p},Q)  
          \zeta(x',{\bf p}',Q') \right\rangle  
=  
- \left(
 \frac{\delta C[F](x,{\bf p},Q)}{\delta F(x',{\bf p}',Q')}  
+\frac{\delta C[F](x',{\bf p}',Q')}{\delta F(x,{\bf p},Q)}\ 
\right)
\end{equation} 
in full analogy to \Eq{simpFDT}.  With the knowledge of the
thermodynamical force \Eq{F} and \Eq{Cf}, or simply using the explicit
expression \Eq{CF} for the linearised collision term, we arrive at
\begin{equation}\label{noisecorrelator} 
\left\langle \zeta(x,{\bf p},Q)  
         \zeta(x',{\bf p}',Q') \right\rangle  
=  
- (2 \pi)^3 \left(
 {f_{\rm eq}({p})}{K}({\bf x},{\bf p},Q;{\bf x}',{\bf p'},Q') 
              + {\rm sym.} \right) \delta ( t-  t')\ .  
\end{equation}
Here, symmetrisation in 
$({\bf x},{\bf p},Q)  \leftrightarrow ({\bf x}',{\bf p}',Q')$
is understood.

Notice that we can derive the equal-time correlator for the deviations
from the equilibrium distribution simply from the knowledge of the entropy 
and the thermodynamical force, exploiting Einstein's law in full analogy 
to the corresponding relation \Eq{invbet}. 
Using \Eq{F}  we find
\begin{equation} 
\label{outofeqcorr} 
\left\langle \Delta f(x,{\bf p},Q)  
         \Delta f(x',{\bf p}',Q') \right\rangle_{t=t'}  
= (2 \pi)^3
f_{\rm eq} (p) \delta^{(3)} ({\bf x} - {\bf x}')  
               \delta^{(3)} ({\bf p} - {\bf p}')  
               \delta( Q - Q')  \ .  
\end{equation} 
If the fluctuations
$\Delta f$ have vanishing mean value, then  \Eq{outofeqcorr}
reproduces the well-known result that the correlator  of fluctuations
at equilibrium is given by the equilibrium distribution
function. In order to make contact with the results of \cite{LM,LM2}
as discussed in Section \ref{Ensemble}, we
go a step further and consider the case where $\Delta f$
has a non-vanishing mean value to leading order in the gauge
coupling. Splitting  
\beq
\Delta f= g {\bar f}^{(1)} + \delta f
\eeq 
into a
deviation of the mean part  $\langle \Delta f\rangle= g {\bar
f}^{(1)}$ and a fluctuating part  $\langle \delta f\rangle=0$, and
using \Eq{outofeqcorr}, we obtain the equal time
correlator for the fluctuations $\delta f$ as  
\begin{eqnarray} 
\left\langle \delta f(x,{\bf p},Q)  
         \delta f(x',{\bf p}',Q') \right\rangle_{t=t'}  
&=& (2 \pi)^3
f_{\rm eq} (p) \delta^{(3)} ({\bf x} - {\bf x}')  
               \delta^{(3)} ({\bf p} - {\bf p}')  
               \delta( Q - Q')
\nonumber \\ && \label{fluccorr} 
- \left. g^2 {\bar f}^{(1)}(x,{\bf p},Q)
             {\bar f}^{(1)}(x',{\bf p}',Q')\right|_{t=t'}  \ .  
\end{eqnarray} 
This result agrees with the correlator obtained in \Eq{average} from the 
Gibbs ensemble average as defined in phase space in the limit where 
two-particle correlations are
small and given by products of one-particle correlators.

\subsubsection*{2. Quantum plasmas}

Up to now we have dealt with purely classical plasmas. On the same
footing, we can consider the soft and ultra-soft modes in a hot
quantum  plasma. These can be treated classically as their occupation
numbers  are large. The sole effect from their quantum nature reduces
to the different statistics, Bose-Einstein or Fermi-Dirac as opposed
to Maxwell-Boltzmann.  The corresponding quantum
fluctuation-dissipation theorem  reduces to an effective classical
one \cite{Landau5,K}.

A few changes are necessary to study hot quantum plasmas. As in
Section \ref{Ensemble},  we change the normalisation of $f$  by a
factor of $(2\pi\hbar)^3$ to obtain the standard normalisation for
the (dimensionless) quantum distribution function. Thus, the momentum
measure is also modified by the same factor, $dP= {d^4 p} 2\Theta
(p_0) \delta(p^2)/{(2\pi\hbar)^3}$  for massless particles, and $\hbar
=1$. To check the fluctuation-dissipation   relation in this case one
needs to start with the correct expression for the entropy for a
quantum plasma. The entropy flux density, as a function of $f(x,p,Q)$,
is  given by \Eq{Smicro} and \Eq{Sigma-qm}, to wit
\begin{equation} 
\label{Sq}
S_\mu (x) = - \int dP dQ\, p_\mu \, 
\Big( f \ln{ f} 
       \mp \left( 1 \pm f\right) \ln{(1 \pm f)}  
\Big) \ ,
\end{equation}
where the upper or lower sign applies for bosons or fermions.  From
the above expression of the entropy one can compute  $\Delta S$, and
proceed exactly as in the classical  case, expanding the entropy up to
quadratic order in the deviations from equilibrium. Thus, we obtain
the noise correlator
\bea
\left\langle \zeta(x, {\bf p}, Q ) 
             \zeta(x',{\bf p}',Q') 
\right\rangle 
&=& 
-(2\pi)^3 \delta (t-t')
\nonumber
\zeile
&&
\times
\left(
{f_{\rm eq}({p})}[1 \pm  f_{\rm eq}(p)]  
{K}({\bf x},{\bf p},Q;{\bf x}',{\bf p}',Q')+{\rm sym.}
\right) \ .
\label{noisecorrelatorQ}
\eea
Again, the spectral functions of the deviations from 
equilibrium are directly deduced from the entropy. As a result, we find
\bea
\left\langle \Delta f(x,{\bf p},Q) 
         \Delta f(x',{\bf p}',Q') \right\rangle_{t=t'} 
&= & 
(2\pi)^3 f_{\rm eq} (p)[1 \pm  f_{\rm eq}(p)]
\nonumber
\zeile
&& \times 
         \delta^{(3)} ({\bf x} - {\bf x}') 
         \delta^{(3)} ({\bf p} - {\bf p}') 
         \delta( Q - Q')  \ .
 \label{outofeqcorrQ}
\eea
Expanding  $\Delta f= g {\bar f}^{(1)} + \delta f$ as above, we obtain the
equal time correlator for $\delta f$, which agrees with the findings of 
\Eqs{averageB} and \eq{averageF} 
in the case where two-particle distribution functions can be 
expressed as products of one-particle distributions.

With the knowledge of the above spectral functions  for the fluctuations
in a classical or quantum plasma one can derive further spectral 
distributions for different physical quantities. In particular, we can 
find the correlations of the self-consistent  gauge field fluctuations
once the basic correlators as given above are known. This is how those
spectral functions were deduced in \cite{LM}.

\subsection{Application}
\label{LangevinApplication}

As a particular example of the above we consider B\"odeker's effective
kinetic equations which couple to the ultra-soft gauge field modes.
The linearised collision integral has been obtained to leading
logarithmic accuracy in Section \ref{BeyondCollision}.  We will first
consider the classical plasma  for particles carrying two
helicities. It is most efficient to write  the transport equation not
in terms of the full one-particle distribution function, but in terms
of the current density
\begin{equation} 
\label{current}
{\cal J}^\rho_a(x,{\bf v})
=  \frac{ g}{\pi^2} \, v^\rho \,\int dp\,dQ\, p^2\, Q_a\, 
\Delta f(x,{\bf p},Q)\, .
\end{equation}
(Notice that $f_{\rm eq}$ gives no contribution to the current.)
The current of \Eq{CoarseYM} follows after integrating over the angles
of ${\bf v}$,
$J_a^\mu (x) = \int \frac{ d\Omega}{4\pi} {\cal J}_a ^\mu(x,{\bf v})$.
Expressed in terms of \Eq{current}, the linearised Boltzmann-Langevin 
equation \Eq{linearisedBeq} becomes 
\begin{equation}    
\label{ultracur}
[v^\mu  D_\mu , {\cal J}^\rho](x,{\bf v})
=  -  m^2_D v^\rho v_\mu  F^{\mu0} (x)
   +  v^\rho C[{\cal J}^0](x,{\bf v})
   + \zeta^{\rho}(x,{\bf v})\, ,
\end{equation}
where  $m_D$ is the Debye mass
\begin{equation}\label{mD}
m_D ^2 = - \frac{g^2 C_2}{\pi^2} \int dp\, p^2 \, 
\frac{d f_{\rm eq}}{d p} \, ,
\end{equation}
and the quadratic Casimir $C_2$ has been defined in \Eq{quadraticQ}. 
The linearised collision integral is related to \Eq{Cf} by 
\bea 
C[{\cal J}_a^0](x,{\bf v})&= &
 \frac{g^2}{\pi^2}  \int d^3x'\, {d\Omega_{{\bf v}'}}
\, dp\, dp'\, dQ \,dQ'\, 
\nonumber
\zeile
&&\times
p^2  p'^2 \,Q_a\, 
{K}({\bf x},{\bf p}, Q;{\bf x}',{\bf p'},Q') \,
\Delta f(t,{\bf x}',{\bf p}',Q') 
\label{CK}
\eea
and corresponds precisely to the correlator \Eq{leading-log}, to wit
\begin{equation}    
\label{Col}
C[{\cal J}_a^0](x,{\bf v})=  
 - \gamma\  \int \frac{d\Omega_{{\bf v}'}}{4\pi}\, 
   {I}({\bf v},{\bf v}'){\cal J}_a^0(x,{\bf v}') \ ,
\end{equation}
where the kernel has been given in \Eq{IK},
\begin{equation}
{I}({\bf v},{\bf v}')=\delta^{(2)}({\bf v}-{\bf v}')-\frac{4}{\pi}
\frac{({\bf v}\cdot{\bf v}')^2}{\sqrt{1-({\bf v}\cdot {\bf v}')^2}}
\end{equation}
and $\gamma=g^2 N T \ln{(1/g)}/4 \pi$. Comparing \Eq{CK} with \Eq{Col} we 
learn that only the part of the kernel $K$ which is  symmetric under 
$({\bf x},{\bf p},Q)  \leftrightarrow ({\bf x}',{\bf p}',Q')$ 
contributes in the present case. This part can be expressed as
\begin{equation}
{K}({\bf x},{\bf p}, Q;{\bf x}',{\bf p'},Q') = 
-\gamma \, \frac{{I}({\bf v},{\bf v}')}{4\pi p^2}  
           \delta(p -p') \delta(Q-Q')\delta^{(3)}({\bf x}-{\bf x}')\, .
\end{equation} 
According to our findings above, the self-correlator of the
stochastic source for  the classical plasma obeys
\begin{eqnarray}
\left\langle \zeta^\mu_a (x,{\bf v})\, 
         \zeta^\nu_b (y,{\bf v}') \right\rangle 
&=& \frac{1}{(2\pi^2)^2} \, g^2\int dp\, dp'\, dQ\, dQ'\, 
p^2p'^2\,  Q_aQ'_b\,  v^\mu v'^\nu 
   \left\langle \zeta(x,{\bf p},Q)\, \zeta(y,{\bf p}',Q') \right\rangle 
\nonumber 
\zeile 
&=&
2\,\gamma\,T\,m^2_D 
   \,v^\mu v'^\nu\, {I}({\bf v},{\bf v}') \,
   \delta_{ab}\, \delta^{(4)} (x-y) \ .
\label{corrnoise}
\end{eqnarray} 
The  helicities of the particles have been taken into account as well.
In order to obtain \Eq{corrnoise}, we have made use of 
\Eqs{noisecorrelator}, (\ref{mD}) to (\ref{Col}), and of the relation 
$f_{\rm eq}=-T\, df_{\rm eq}/dp$ 
for the Maxwell-Boltzmann distribution.

The quantum plasma can be treated in exactly the same way. To 
confirm \Eq{corrnoise}, we only need to take into account the change 
of normalisation as commented above, and the relation
$f_{\rm eq}(1\pm f_{\rm eq})=-T\, df_{\rm eq}/dp$ for the 
Bose-Einstein and Fermi-Dirac distributions, respectively.

Using the explicit expression for the collision integral and the 
stochastic noise it is possible to confirm the covariant conservation 
of the current, $DJ=0$.

\subsection{Discussion}
\label{LangevinDiscussion}

We thus found that the correlator \Eq{corrnoise} is in full agreement
with the result of Section \ref{BeyondStochastic} for both the
classical or the quantum plasma. While this correlator has been
obtained in Section \ref{BeyondStochastic} from the corresponding
microscopic theory, here, it follows solely from the
fluctuation-dissipation theorem.  This way, it is established that the
effective Boltzmann-Langevin equation found in \cite{Bodeker:1998hm}
is indeed fully consistent with the  fluctuation-dissipation
theorem. More generally, the important observation is that the
spectral functions as derived here from the entropy and the
fluctuation-dissipation relation do agree with those obtained in
Section \ref{Ensemble} from a microscopic phase space  average. This
guarantees, on the other hand, that the formalism developed in
Sections \ref{Ensemble} -- \ref{Consistent} is consistent with the
fluctuation-dissipation theorem.

In the above discussion we have considered the stochastic noise as
Gaussian and Markovian. This is due to the fact that the small-scale
fluctuations (those within a coarse-graining volume) are to leading
order well separated from the typical relaxation scales  in the
plasma. Within the microscpoic approach,  these characteristics can be
understood ultimately as a consequence of an expansion in a small
plasma parameter (or a small gauge coupling).  More precisely, the
noise follows to be Gaussian  due to the polarisation approximation,
where higher order correlators beyond quadratic ones can be neglected.
The Markovian character of the noise follows because the ultra-soft
modes are well separated from the soft ones, and suppressed in the
collision integral to leading  order. This way, the collision term and
the correlator of stochastic noise are both local in $x$-space. Going
beyond  the logarithmic approximation, we expect from the explicit
computation  in Section \ref{BeyondCollision} that the coupling of the
soft and the ultra-soft modes makes the collision term {\it non-local}
in coordinate space. This non-trivial memory kernel should also result
in a non-Markovian, but still Gaussian, source for stochastic noise.

The present line of reasoning can in principle be extended to other
approaches.  Using the phenomenological derivation of \Eq{Col} from
\cite{Arnold:1999cy,Arnold}, the same arguments as above justify  the
presence of a noise source with  \Eq{corrnoise} in the corresponding
Boltzmann equation
\cite{Arnold:1999cy,Arnold,Blaizot:1999xk,Valle}. It might also be
fruitful to follow a similar line based on the entropy within a
quantum field theoretical language. An  interesting proposal to
self-consistently include the noise within a Schwinger-Dyson approach
has been made recently in \cite{Calzetta:2000xh}. Along these lines,
it might be feasible  to derive the source for stochastic noise
directly from the quantum field theory \cite{Blaizot:1999xk}.

While we have concentrated the discussion on plasmas close to thermal
equilibrium,  it is known that a fluctuation-dissipation theorem  can
be formulated as well for stationary and stable systems out of
equilibrium \cite{K}.  More generally, we have exploited the fact that
the entropy production vanishes. All the information which links the
dissipative characteristics with correlators of fluctuations can be
deduced from \Eq{Mean-Smu} in the case where the entropy production
vanishes.

This approach can also be  extended to take non-linear effects into
account \cite{Sitenko1982}.  Both the out-of-equilibrium situations
and the non-linear effects  can be treated, in principle, with the
general formalism as discussed in Sections \ref{Ensemble} --
\ref{Effective}.

\newpage

\section{Dense quark matter}
\label{DenseMatter}

Until now, the applications of the formalism have been focussed on
different aspects of hot non-Abelian plasmas close to thermal
equilibrium.  In this section we consider the regime of high baryonic
density and low temperature. In this part of the QCD phase diagram,
quarks form Cooper pairs due to the existence of attractive
interactions among them.  A colour superconducting phase then arises,
typically characterised by the Anderson-Higgs mechanism and the
existence of an energy gap associated to the fermionic quasiparticles.

For gluon momenta much larger than the fermionic gap, the effects of
diquark condensation are negligible. For scales larger than the gap,
but still smaller than the chemical potential, one finds to leading
order an effective theory for the gauge fields totally analogous to
the HTL theory, the so called hard dense loop (HDL) theory. This
effective theory can be derived from classical transport theory
\cite{Manuel:1996td}.

For gluon momenta of the order of, or smaller than, the fermionic gap,
the effects of Cooper pairing cannot be neglected. The kinetic
equations have to take into account the modification of the
quasiparticle dispersion relations, which ultimately reflect the fact
that the ground state of QCD in the superconducting phase is not the
same as in a normal phase.  We present a kinetic equation for the
gapped quasiparticles in a two flavour colour superconductor
\cite{Litim:2001je}.

\subsection{Normal phase}
\label{HDL-subsection}

We consider first the regime where the effects of Cooper pairing of
quarks can be neglected, and discuss the kinetic equations associated
to the normal phase of dense quark matter. At very high baryonic
density, the non-Abelian plasma is ultradegenerate.  The fermionic
equilibrium distribution function, neglecting the effects of Cooper
pairing, is given by
\beq
\label{ultrad-equi}
f^{\rm eq}(p_0) = \Theta (\mu -p_0) \ ,
\eeq
where $\Theta$ is the step function, and $\mu$ is the quark chemical
potential.  This distribution function describes a system where all
the fermionic energy levels are occupied, according to Pauli's
principle, up to the value of the Fermi energy $p_0 = \mu$. At zero
temperature there are no real gluons in the system.

The Vlasov approximation presented in Section~\ref{HTL} can be applied
to this non-thermal situation. The main input is the above equilibrium
distribution function, which affects both the value of the Debye mass
and the relevant scales of the system. The Debye mass reads
\beq
m^2_D = N_F \frac{ g^2 \mu^2}{2 \pi^2} \ ,
\eeq  
for $N_F$ different quark flavours. This is to be compared with $m^2_D
\sim g^2 T^2$ as found for a quantum plasma at high temperature.

Formally, the Vlasov approximation to a non-Abelian plasma at either
high temperature, or high baryonic density but vanishing temperature
in the normal phase, look almost identical.  The colour currents
obtained in the two cases are the same, only the explicit value of the
Debye mass differs. Roughly speaking, one could say that the role
played by $T$ in the HTL effective theory is now played by $\mu$. Now
$\mu$ is the hard scale, while $g \mu$ is the soft one.  As in the
thermal case, in the ultradegenerate limit $\delta^{(n)} J/\delta A_1
\dots \delta A_n |_{A=0}$ generates a $n+1$-point amplitude, which
looks formally the same as the $n+1$-point HTL.  Due to this
similarity, these amplitudes were called hard dense loops (HDL) in
\cite{Manuel:1996td} (see also \cite{Vija:1995is}).

The above considerations neglect the fact that quarks
form Cooper pairs, which modify both the shape of the quasiparticle
distribution function and also the underlying kinetic equations.
However, the Vlasov approximation remains a valid description for
specific momentum scales of the plasma.  The HDL effective theory can
be derived as well from quantum field theory.  An explicit computation
of the gluon self-energy in the superconducting phase of QCD, for
$N_F=2$ and $N_F=3$ \cite{Rischke:2000qz,Rischke:2000ra}, shows that
to leading order it reduces to the HDL value in the limit when the
gluon momentum $p$ obeys $\mu \gg p \gg \Delta$, where $\Delta$ is the
value of the gap.

\subsection{Superconducting phases}
\label{Superconducting}

Let us now consider the case when diquark condensates are formed,
modifying the ground state of QCD. The possible phases of QCD depend
strongly on the number of quark flavours participating in the
condensation. We will briefly review the two mostly studied phases.
These are the idealizations of considering pairing of either two or
three massless quark flavours.  More realistic situations should
consider effects due to non-vanishing quark masses , which may lead to
an even richer phase diagram. We also restrict the discussion to quark
condensation in the lowest angular momentum channel, the spin zero
condensates, as this channel is energetically favored.

For two light quark flavours the diquark condensate is such that the
$SU_c(3)$ group is broken down to $SU_c(2)$
\cite{Alford:1998zt,Rapp:1998zu}. Thus, five gluons get a mass, while
there are three gluons which remain massless, and exhibit confinement.
Furthermore, not all the quarks participate in the condensation. More
speficially, if we consider up and down quarks of colours red, green
and blue, one of the colours, say the blue one, does not participate in
the condesation process. Then the blue up and down quarks are gapless.
The condensate is such that if one could neglect the effects of the
quantum anomaly, which can be done at asymptotically large densities,
it would also break a global axial $U_A(1)$. Thus a (pseudo)
Nambu-Goldstone mode, similar to the $\eta'$ meson, is also present.
This meson becomes heavy as soon as one reduces the density of the
system, and its mass can be computed using instanton techniques.

For three light quark flavours the pattern of symmetry breaking induced
by the condensates is much more involved, as the condensates lock the
colour and flavour symmetry transformations (colour-flavour locking or CFL
phase) \cite{Alford:1999mk}.  They break spontaneously both colour,
chiral and baryon number symmetry $SU_c(3) \times SU_L(3) \times SU_R
(3) \times U_B(1) \rightarrow SU_{c+L+R}(3) \times {\cal Z}_2$. As a
result, all the gluons become massive, while there are nine
Nambu-Goldstone bosons, eight associated to the breaking of chiral
symmetry, and one associated to the breaking of baryon number
symmetry. At asymptotically large densities, when the effects of the
quantum anomaly can be neglected, there is an extra (pseudo)
Nambu-Goldstone boson associated to the breaking of $U(1)_A$. In the
CFL phase all the quarks of all flavours and colours participate in the
condensation.  The light modes are then the Nambu-Goldstone bosons,
which dominate the long distance physics of the superconductor.

If electromagnetic interactions are taken into account, then both the
2SC and CFL diquark condensates break spontaneously the standard
electromagnetic symmetry. However, a linear combination of the
original photon and a gluon remains massless in both cases. This new
field plays the role of the ``in-medium'' photon in the supercondutor.

Using standard techniques in BCS theory, it is possible to compute, in
the weak coupling limit $g \ll 1$, the microscopic properties of the
2SC and CFL superconductors. This concerns in particular the fermionic
gap $\Delta$ associated to the quarks, as well as the gluon Meissner
masses $m_M$. In weak coupling there is a hierarchy of scales $\Delta
\ll m_M \ll \mu$. Furthermore, it is also possible to compute the
relevant properties of the (pseudo) Nambu-Goldstone modes, which
acquire masses due to explicit chiral symmetry breaking effects of
QCD.  The propagation properties of the ``in-medium'' photon has been
obtained as well \cite{Litim:2001mv}. While all these computations
rely on a weak gauge coupling expansion, which might be unrealistic
for the astrophysical settings of interest, they provide both a
qualitative and semi-quantitative insight of the main microscopic
porperties of quark matter. These studies may be complemented with
others based on QCD-inspired models, which might be pushed to the
regime of more moderate densities, and thus large couplings. It would
be desirable that the microscopic properties of quark matter could
also be computed numerically. At present, no reliable numerical
algorithms are available for such a study.

\subsection{Quasiparticles in the 2SC phase}
\label{Quasiparticles}

To be specific, we restrict the remaining considerations to the case
of two massless quark flavours, the 2SC phase. It is shown that
coloured quasiparticle excitations of the 2SC condensate can be
formulated in terms of a simple transport equation.

The low energy physics of a two-flavour colour superconductor is
dominated by its light degrees of freedom. At vanishing temperature,
these are the massless gauge bosons, the gapless quarks and a
(pseudo-) Goldstone boson, similar to the $\eta'$ meson. However, the
gapless quarks and the $\eta'$ meson are neutral with respect to the
unbroken $SU(2)$ subgroup. In turn, the condensate, although neutral
with respect to the unbroken $SU(2)$, polarises the medium since their
constituents carry $SU(2)$ charges. Hence, the dynamics of the light
$SU(2)$ gauge fields differs from the vacuum theory. This picture has
recently been introduced by Rischke, Son and Stephanov
\cite{Rischke:2000cn}. Their infrared effective theory for momenta $k
\ll \Delta$ is
\begin{equation}\label{RSS}
S_{\rm eff}^{T=0} = \int d^4 x \left ( \frac{\epsilon}{2} \,
{\bf E}_a \cdot {\bf E}_a - \frac{1}{2\lambda} \,{\bf B}_a \cdot {\bf B}_a
\right) \ ,
\label{Seff-T0}
\end{equation}
where $E_i^a \equiv F_{0i}^a$ and $B_i^a \equiv \frac12 \epsilon_{ijk}
F_{jk}^a$ are the $SU(2)$ electric and magnetic fields. The constants
$\epsilon$ and $\lambda$ are the dielectric susceptibility and
magnetic permeability of the medium. To leading order, $\lambda = 1$
and $\epsilon = 1+ g^2 \mu^2/(18 \pi^2 \Delta^2)$
\cite{Rischke:2000cn}. As a consequence, the velocity of the $SU(2)$
gluons is smaller than in vacuum. This theory is confining, but the
scale of confinement is highly reduced with respect to the one in
vacuum with $\Lambda'_{\rm QCD} \sim \Delta \exp{(-\frac{2 \sqrt{2}
    \pi}{11} \frac{\mu}{g \Delta})}$ \cite{Rischke:2000cn}. Due to
asymptotic freedom, it is expected that perturbative computations are
reliable for energy scales larger than $\Lambda'_{\rm QCD}$.

At non-vanishing temperature, thermal excitations modify the low
energy physics.  The condensate melts at the critical temperature $T_c
\approx 0.567 \Delta_0$ \cite{Pisarski:2000bf} ($\Delta_0$ is the gap
at vanishing temperature). We restrict the discussion to temperatures
within $\Lambda'_{\rm QCD} \ll T <T_c$, which provides the basis for
the perturbative computations below. In this regime, the main
contribution to the long distance properties of the $SU(2)$ fields
stems from the thermal excitations of the constituents of the diquark
condensate. The thermal excitations of the massless gauge fields
contribute only at the order $g^2T^2$ and are subleading for
sufficiently large $\mu$.  Those of the gapless quarks and of the
$\eta'$ meson do not couple to the $SU(2)$ gauge fields.

The thermal excitations due to the constituents of the diquark
condensate display a quasiparticle structure. This implies that they
can be cast into a transport equation. To that end, and working in
natural units $k_B = \hbar = c=1$, we introduce the on-shell
one-particle phase space density $f(x,{\bf p},Q)$, $x^\mu=(t,{\bf
  x})$, describing the quasiparticles. The distribution function
depends on time, the phase space variables position ${\bf x}$,
momentum ${\bf p}$, and on $SU(2)$ colour charges $Q_a$, with the
colour index $a =1,2$ and $3$. The quasiparticles carry $SU(2)$ colour
charges simply because the constituents of the condensate do. The
on-shell condition for massless quarks $m_q=0$ relates the energy of
the quasiparticle excitation to the chemical potential and the gap as
\beq
\label{energy} 
p_0
\equiv \epsilon_p = \sqrt{(p-\mu)^2 + \Delta^2(T)} \ .
\eeq 
The gap is both
temperature and momentum-dependent. From now on, we can neglect its
momentum dependence which is a subleading effect.  The velocity of the
quasiparticles is given by
\beq\label{velocity}
{\bf v}_p 
\equiv \frac{ \partial \epsilon_p}{\partial {\bf p}} =
\frac{|p-\mu|}{\sqrt{(p-\mu)^2 + \Delta^2(T)}}\, {\hat {\bf p} \ ,} 
\eeq
and depends on both the chemical potential and the gap. For
$\Delta=0$, the quasiparticles would travel at the speed of light.
However, in the presence of the gap $\Delta\neq 0$, their propagation
is suppressed, $v_p \equiv |{\bf v}_p|\leq 1$.

The one-particle distribution function $f(x,{\bf p},Q)$ obeys a very
simple transport equation, given by
\begin{equation}\label{transport}
\left[D_t + {\bf v}_p \cdot {\bf D} -g Q_a \left({\bf E}^a +
{\bf v}_p \times {\bf B}^a \right) \frac{\partial}{\partial {\bf p}}
    \right] f = C[f] \,.
\end{equation}
Here, we have introduced the short-hand notation of \Eq{Df} for the
covariant derivative acting on $f$.  The first two terms on the
left-hand side of \Eq{transport} combine to a covariant drift term
$v_p^\mu D_\mu$, where $v_p^\mu = (1, {\bf v}_p)$ and $D_\mu =(D_t,
{\bf D})$.  The terms proportional to the colour electric and magnetic
fields provide a force term. The right-hand side of \Eq{transport}
contains a (yet unspecified) collision term $C[f]$.

Notice that \Eq{transport} has the same structure as the (on-shell)
transport equation valid for the unbroken phase of a non-Abelian
plasma. All what changes here are the energy and velocity of the
quasiparticles.  Once the temperature of the system is increased, and
the diquark condensates melt and the gap vanishes, $\Delta(T)
\rightarrow 0$, we recover the transport equation for the unbroken
phase.

The thermal quasiparticles carry an $SU(2)$ charge, and hence provide
an $SU(2)$ colour current. It is given by
\begin{equation}\label{current-2SC}
J^\mu_a(x)=g \sum_{\hbox{\tiny helicities}\atop\hbox{\tiny species}}
\int \frac{d^3 p}{(2\pi)^3}
dQ \ v^\mu_p\, Q_a\, f(x,{\bf p},Q) \ . 
\end{equation}
Below, we simply omit a species or helicity index on $f$, as well as
the explicit sum over them. We use the same definition for the colour
measure as in \Eq{col-mes}.  The colour current \Eq{current-2SC} is
covariantly conserved for $C[f]=0$. For $C[f] \neq 0$ a covariantly
conserved current implies certain restrictions in the form of the
collision term.

\subsection{Vlasov approximation}
\label{HSL}

We will now study  the Vlasov approximation, or collisionless
dynamics $C[f] =0$ of the colour superconductor close to thermal
equilibrium and to leading order in the gauge coupling. Consider the
distribution function 
\beq
f(x,{\bf p},Q)= f^{\rm eq}(p_0)+ g
f^{(1)}(x,{\bf p},Q) \ .
\eeq
 Here
\beq
f^{\rm eq.}(p_0) = \frac{1}
{\exp(\epsilon_p/T) +1}
\eeq
 is the fermionic equilibrium distribution
function and $g f^{(1)}(x,{\bf p},Q)$ describes a slight deviation
from equilibrium. For convenience, we also introduce the colour density
\begin{equation}\label{currentdensity}
J_{a}(x,{\bf p})= g  \!\int\! dQ Q_{a} f(x,{\bf p},Q) \ ,
\end{equation}
from which the induced colour current of the medium \Eq{current-2SC}
follows as $J^\rho _a(x) = \int \s0{d^3p}{(2\pi)^3}v_p^\rho
J_{a}(x,{\bf p})$. Expanding the transport equation \eq{transport} to
leading order in $g$, and taking the two helicities per quasiparticle
into account, we find the transport equation for the colour density as
\begin{equation}\label{transport-current}
\left[D_t + {\bf v}_p \cdot {\bf D}\right]   J(x,{\bf p})  =
2 g^2   \,  {\bf v}_p \cdot  {\bf E}(x)\, \frac{d f^{\rm eq}}{d
\epsilon_p} \ .
\end{equation}
The solution of the transport equation reads
\begin{eqnarray}
\label{current-solution}
J^\mu_a(x) =
2 g^2 
\!\int\! 
\frac{d^3 p\, d^4y}{(2\pi)^3}\,
v_p^\mu \, \langle x|\frac{1}{(v_p \cdot D)}|y\rangle_{ab} \,
{\bf v}_p \cdot {\bf E}_b (y)
\frac{d f^{\rm eq}}{d \epsilon_p} 
\ .
\end{eqnarray}
After having solved the transport equation, the relevant information
concerning the low energy effective theory is contained in the
functional $J[A]$. Notice that the above derivation is analogous to
the derivation of the HTL and HDL effective theories from kinetic
theory. Owing to this resemblance, we call the diagrams which are
derived from \Eq{current-solution} as {hard superconducting loops}
(HSL). The HSL effective action follows from \Eq{current-solution} by
solving $J[A] = -{\delta \Gamma_{\rm HSL}[A]}/{\delta A}$ for
$\Gamma_{\rm HSL}[A]$, and all HSL diagrams can be derived by
performing functional derivatives to the effective action (or the
induced current). We thus reach to the conclusion that the low energy
effective theory for a two-flavour colour superconductor at finite
temperature reads $S_{\rm eff}^T = S_{\rm eff}^{T=0} + \Gamma_{\rm
  HSL}$ to leading order in $g$. This theory is effective for modes
with $k\ll \Delta$.

Let us have a closer look into the induced current, which we formally
expand as
$J^a_\mu[A]
=
 \Pi^{ab}_{\mu\nu} A^\nu_b
+\s012\Gamma^{abc}_{\mu\nu\rho} A^\nu_b  A^\rho_c
+\ldots$ 
in powers of the gauge fields. The most relevant information on the
thermal effects is contained in the thermal polarisation tensor
$\Pi^{ab}_{\mu\nu}$. Using \Eq{current-solution}, we find
\beq
\Pi^{\mu \nu}_{ab}(k) 
= 
2 g^2  \delta_{ab}  
\int \frac{d^3 p}{(2 \pi)^3} 
\frac{d f^{\rm eq}}{d \epsilon_p}
\left( g^{\mu 0} g^{\nu 0} - k_0  \frac{v_p^\mu v_p^\nu}{k \cdot
v_p}\right) 
\,.
\eeq
It obeys the Ward identity $k_\mu \Pi^{\mu \nu}_{ab}(k) =0$. With
retarded boundary conditions $k_0 \rightarrow k_0 + i 0^+$, the
polarisation tensor has an imaginary part,
\beq
{\rm Im}\, \Pi^{\mu \nu}_{ab} (k) 
= \delta_{ab}\, 2 \pi g^2   k_0 
   \int \!\!\ \frac{d^3 p}{(2 \pi)^3} 
\frac{d f^{\rm eq}}{d \epsilon_p}
    v_p^\mu v_p^\nu \,\delta (k \cdot v_p) \,,            
\eeq
which corresponds to Landau damping. Performing the angular
integration, we obtain for the longitudinal and transverse projections
of the polarisation tensor
\begin{mathletters}\label{pipi-2SC}
\begin{eqnarray}
\Pi_{L} (k_0,{\bf k}) & = & 
\frac{g^2 }{ \pi^2} 
\int^\infty _0 \!\!\! dp\, p^2 
\frac{d f^{\rm eq}}{d \epsilon_p} 
\!\! \left[ 1 - \frac{1}{2}\frac{k_0}{|{\bf k}| v_p}
\label{Pi-L}
\left(\,{\rm ln\,}\left| \frac{k_0+ 
|{\bf k}| v_p}{k_0- |{\bf k}| v_p} \right| 
-i \pi \, \Theta( |{\bf k}|^2 v_p^2 -k_0^2) \right)  \right] \ , 
\\
\Pi_{T} (k_0,{\bf k}) & = & 
\frac{g^2 }{2 \pi^2}\frac{k_0^2}{|{\bf k}|^2}  
\int^\infty _0 \!\!\! dp\, p^2 
\frac{d f^{\rm eq.}}{d \epsilon_p}
   \!\!  \left[ 1 + \frac12 \left( \frac{|{\bf k}| v_p}{k_0} -
\frac{k_0}{|{\bf k}| v_p} \right) 
 \times\right. \nonumber \\
\label{Pi-T}
 & & \left. 
\quad\quad\quad\quad\quad\quad\quad\quad\quad\quad\quad\quad
\left( {\rm ln\,} \left|{\frac{k_0+
|{\bf k}| v_p}{k_0- |{\bf k}| v_p}}\right|  
-i \pi \, \Theta( |{\bf k}|^2 v_p^2 -k_0^2)
 \right) \, \right] \ ,
\end{eqnarray}
\end{mathletters}%
where $\Theta$ is the step function.  We first consider the real part
of the polarisation tensor. From \Eq{pipi-2SC}, and in the limit $k_0\to
0$, we infer that the longitudinal gauge bosons acquire a thermal
mass, the Debye mass, while the transverse ones remain massless. The
(square of the) Debye mass is given by
\begin{equation}\label{Debye}
m^2_D = -\frac{g^2 }{ \pi^2} \int^\infty _0 \!\!\! dp\,p^2 
\frac{d f^{\rm eq.}}{d \epsilon_p} \equiv M^2
\, I_0\left(\0{\Delta}{T},\0{T}{\mu}\right)\,. 
\end{equation}
For convenience, we have factored-out the Debye mass $M$ of the
ultradegenerate plasma in the normal phase, $M^2 \equiv {g^2 
  \mu^2}/{ \pi^2}$. The dimensionless functions
\begin{equation}
I_n \left(\0{\Delta}{T},\0{T}{\mu}\right)=
-\frac{1}{\mu^2}\int^\infty _0 \!\!\! dp\,
{p^2}\frac{d f^{\rm eq.}}{d \epsilon_p} 
v_p^n 
\end{equation}
obey $I_n\ge I_{n+1} >0$ for all $n$ due to $v_p \leq 1$. Equality
holds for vanishing gap. For the physically relevant range of
parameters $T < \Delta \ll \mu$, the functions $I_n$ are $ \ll 1$. In
particular, it is easy to see that $I_n(\infty,0)=0$: there is no
Debye screening for the $SU(2)$ gluons at $T=0$ in the superconducting
phase. In the limit where $\Delta/T \gg 1$, and to leading order in
$T/\mu\ll 1$, the Debye mass reduces to
\begin{equation}\label{Debye2}
m^2_D = M^2 \, 
\sqrt{2 \pi \frac{\Delta}{T}}
\exp(-{\Delta}/{T}) \,. 
\end{equation} 
The dispersion relations for the longitudinal and transverse gluons follow
from the poles of the corresponding propagators,
\begin{mathletters}\label{DR}
\begin{eqnarray}
\epsilon |{\bf k}|^2 
- {\rm Re}\, \Pi_L(k_0,{\bf k}) \Large|_{k_0=\omega_L({\bf k})} 
& = & 0 \ ,
\\
\epsilon k^2_0
- \01{\lambda}|{\bf k}|^2 
+ {\rm Re}\,\Pi_T(k_0,{\bf k})\Large|_{k_0=\omega_T({\bf k})}
& = & 0 \ .
\end{eqnarray}
\end{mathletters}%
Here, the terms containing $\Pi_{L,T}$ are due to the fermionic
quasi-particles, while the terms containing $\epsilon$ and $\lambda$
are the leading order contributions from the effective theory at
$T=0$, introduced in \Eq{RSS}.  At vanishing temperature, $\Pi_{T,L}
=0$, and only the transverse gluon propagates, but with velocity $v =
1/\sqrt{\epsilon \lambda} \ll 1$.  At non-vanishing temperature, a
plasmon or longitudinal mode also propagates.  Neglecting higher order
corrections in $k_0$ to the polarisation tensor at $T=0$, the plasma
frequency $\omega_{\rm pl}$ follows from \Eq{DR} as
\beq\label{PlasmaFrequency}
\omega^2_{\rm pl}=
\01{3\epsilon} M^2 \, I_2\left(\0{\Delta}{T},\0{T}{\mu}\right)\,.
\eeq
For generic external momenta the dispersion relations can only be
solved numerically. In turn, if the spatial momenta are much smaller
than the plasma frequency $|{\bf k}| \ll \omega_{\rm pl}$, solutions
to \Eq{DR} can be expanded in powers of $|{\bf k}|^2/\omega^2_{\rm
  pl}$ as
\begin{mathletters}\label{DR-Expansion}
\begin{eqnarray}
\omega^2_L({\bf k}) 
& = &
\omega^2_{\rm pl}
\left[
1 
+ \035\frac{I_4}{I_2} \frac{|{\bf k}|^2}{\omega^2_{\rm pl}} 
+ {\cal O} (\frac{|{\bf k}|^4}{\omega^4_{\rm pl}}) 
\right]\ , \\
\omega^2_T({\bf k}) 
& = &
\omega^2_{\rm pl} 
\left[
1
+ \left(\01{\epsilon\lambda}+ \015\frac{I_4}{I_2} \right)
  \frac{|{\bf k}|^2}{\omega^2_{\rm pl}} 
+ {\cal O} (\frac{|{\bf k}|^4}{\omega^4_{\rm pl}})  \right]\,.
\end{eqnarray}
\end{mathletters}%
Apart from the fact that the $T=0$ transverse mode does not propagate
at the speed of light in vacuum, the ratios of the functions $I_n$
measure the departure of the dispersion relations of the gluons in the
2SC phase with respect to the unbroken phase (see
\Eq{DRunbroken-Expansion}). The quantity $v^2_* =I_4/I_2$ has the
intuitive interpretation of a mean velocity squared of the
quasiparticles of the system. An approximate form of the HSL
polarisation tensor \Eq{pipi-2SC} could be given in terms of this mean
velocity (see \cite{Braaten:1993jw} for the use of a similar
approximation).

Let us now consider the imaginary part of \Eq{pipi-2SC}, which
describes Landau damping. Since $v_p \leq 1$, we conclude that Landau
damping only occurs for $k^2_0 \leq |{\bf k}|^2$. Hence, plasmon and
transverse gluon excitations are stable as long as $\omega_{L,T}({\bf
  k})> |{\bf k}|$.  Furthermore, we notice that the imaginary part of
\Eq{pipi-2SC} is logarithmically divergent: the quasiparticle velocity
vanishes for momenta close to the Fermi surface, which is an immediate
consequence of the presence of a gap, cf.~\Eq{velocity}. This
divergence does not appear in the real part, because the logarithm
acts as a regulator for the $1/v_p$ factor. To leading order in
$T/\mu$, and in the region of small frequencies $k_0^2 \ll |{\bf
  k}|^2$, we find at logarithmic accuracy, and for all values of
$\Delta/T$,
\begin{mathletters}\label{ImPi}
\begin{eqnarray}
{\rm Im}\, \Pi_L (k_0,{\bf k}) &=& 
- 2\pi M^2 \frac{k_0}{|{\bf k}|}\0{\Delta}{T}
\frac{  \ln{(|{\bf k}|/k_0)}}{(e^{\Delta/T} +1)(e^{-\Delta/T} +1)} \ ,
\\ 
{\rm Im}\, \Pi_T(k_0,{\bf k}) &=&  
\pi M^2 
\frac{k_0}{|{\bf k}|}
\left[
\frac{1}{e^{\Delta/T} +1} 
-2\frac{k^2_0}{|{\bf k}|^2}  \0{\Delta}{T}
 \frac{ \ln{|{\bf k}|/k_0}}{(e^{\Delta/T} +1)(e^{-\Delta/T} +1)} 
\right] \ .
\end{eqnarray}
\end{mathletters}%
For small frequencies, Landau damping is dominated by the logarithmic
terms, which are proportional to the gap.  Once the gap vanishes,
subleading terms in $\s0{\Delta}{T}$, not displayed in \Eq{ImPi}, take
over and reduce ${\rm Im} \Pi$ to known expressions for the normal
phase.

Finally, we explain how the polarisation tensor, as obtained within
the present transport theory, matches the computation of
$\Pi^{\mu\nu}$ for external momenta $k_0, |{\bf k}| \ll \Delta$ to
one-loop order from quantum field theory. The one-loop gluon
self-energy for a two-flavour colour superconductor has been computed
by Rischke, and the polarisation tensor for the unbroken $SU(2)$
subgroup is given in Eq.~(99) of \cite{Rischke:2000qz}. It contains
contributions from particle-particle, particle-antiparticle and
antiparticle-antiparticle excitations.  The particle-antiparticle
contribution to $\Pi^{00}$ and $\Pi^{0i}$ at low external momenta, and
the antiparticle-antiparticle excitations are subleading. The
particle-particle contributions divide into two types. The first ones
have poles for gluonic frequencies $k_0 = \pm \left(\epsilon_p +
  \epsilon_{p-k} \right)$ and an imaginary part once $k_0$ exceeds the
Cooper pair binding energy $2 \Delta$.  These terms are related to the
formation or breaking of a Cooper pair, and suppressed for low
external gluon momenta.  The second type of terms, only non-vanishing
for $T \neq 0$, have poles at $k_0 = \pm \left(\epsilon_p -
  \epsilon_{p-k} \right)$. For $|{\bf k}| \ll \Delta$ we approximate
it by $k_0 \approx \pm \frac{\partial \epsilon_p}{\partial {\bf p}}
\cdot {\bf k}$. The prefactor, a difference of thermal distribution
functions, is approximated by $f^{\rm eq}(\epsilon_p) - f^{\rm
  eq}(\epsilon_{p-k})\approx \frac{\partial \epsilon_p}{\partial {\bf
    p}} \cdot {\bf k} \,\frac{d f^{\rm eq}}{d \epsilon_p}$. After
simple algebraic manipulations we finally end up with the result given
above. We conclude that this part of the one-loop polarisation tensor
describes the collisionless dynamics of thermal quasiparticles for a
two-flavour colour superconductor.  The same type of approximations
can be carried out for $\Pi^{ij}$ to one-loop order.  There, apart
from the HSL contributions, additional terms arise due to
particle-particle and particle-antiparticle excitations, cf.~Eq.~(112)
of \cite{Rischke:2000qz}. We have not evaluated these terms
explicitly. However, we expect them to be subleading or vanishing, as
otherwise the Ward identity $k_\mu \Pi^{\mu \nu}_{ab}(k) =0$ is
violated. For $T=0$, this has been confirmed in \cite{Rischke:2000qz}.

\subsection{Discussion}
\label{Super-Discussion}

We have introduced a transport equation for the gapped quarks of
two-flavour colour superconductors. Its simple structure is based on
the quasiparticle behaviour of the thermal excitations of the
condensate, in consistency with the underlying quantum field theory.
We have constructed a low temperature infrared effective theory of the
superconductor.  To leading order, we found Landau damping, and Debye
screening of the chromo-electric fields. Beyond leading order,
chromo-magnetic fields are damped because they scatter with the
quasiparticles. The damping rate is related to the colour
conductivity.  It should be possible to compute the rate from the
transport equation~(\ref{transport}), amended by the relevant
collision term.  The latter can be derived, for example, using the
methods discussed in the preceeding sections.

We have neither discussed the transport equations for gapless quarks
nor for the $\eta'$ meson, because they do not carry $SU(2)$ charges.
However, their excitations are light compared to the gapped
quasiparticles, and dominant for other transport properties such as
thermal and electrical conductivities or shear viscosity. The
corresponding set of transport equations will be discussed elsewhere.

It would be very interesting to study the transport equations in a
three-flavour colour superconductor \cite{Alford:1999mk}.  For $N_f=3$
the quark-quark condensate breaks the $SU(3)$ gauge group completely,
as well as some global flavour symmetries.  Transport phenomena should
then be dominated by the Goldstone modes associated to the breaking of
the global symmetries.  The corresponding transport equations will be
substantially different for the two and three flavour case.

\newpage

\section{Summary and outlook}
\label{Discussion}

We have reviewed a new approach to the transport theory of non-Abelian
plasmas. The formalism relies on a semi-classical approximation and
considers, on the microscopic level, a system of classical {\it
  coloured} point particles interacting through classical non-Abelian
fields. It is assumed that the typical length scales of the
particle-like degrees of freedom are much smaller than those
associated to the classical non-Abelian fields. This scale separation
is at the root of the present formalism. The inclusion of stochastic
fluctuations due to the particles is also of crucial importance, as
well as the ensemble average in phase space, which takes the colour
charges as dynamical variables into account. On the macroscopic level,
the formalism results in a set of {\it effective} transport equations
for the quasi-particle distribution function, the mean gauge fields,
and their fluctuations.  The formalism is consistent with the
non-Abelian gauge symmetry.

Approximations have to be employed in order to obtain, or to solve,
the effective transport equations. For the integrating-out of
fluctuations, systematic expansions schemes, consistent with the
non-Abelian gauge symmetry, have been worked out.  Ultimately, the
procedure corresponds to the derivation of collision integrals, noise
sources and fluctuation-induced currents for effective transport
equations. The compatibility of the approach with the
fluctuation-dissipation theorem was established as well.  Of course,
reliable physical predictions based on the formalism are only as good
as the approximations inherent to the approach. This concerns most
notably the quasiparticle picture and the separation of scales.
However, for a weakly coupled plasma close to equilibrium, these
assumptions are satisfied. Although the Wong particle picture relies
on a high dimensional representation for the colour charges,
observables involving the quadratic or cubic Casimir are reproduced
correctly to leading order.

Interesting applications of the formalism concern hot and weakly
coupled plasmas close to thermal equilibrium.  We have reviewed how
the seminal hard thermal loop effective theory is deduced, based on
the simplest approximation compatible with gauge invariance and
neglecting fluctuations. This step corresponds to the integrating-out
of hard modes with $p\sim T$ to leading order in the gauge coupling.
Further, the simplest approximation which includes the genuine effects
due to fluctuations was shown to reproduce B\"odeker's effective
theory at leading logarithmic order.  This corresponds to
integrating-out the soft modes with $p\sim gT$ to leading logarithmic
order. These applications exemplify the efficiency of the formalism.
As an aside, we note that the effective theories for both classical
and quantum plasmas are identical, except for the value of the Debye
mass.  It is intriguing that a simple semi-classical transport theory
is able to correctly reproduce not only the dynamics of soft
non-Abelian fields with momenta about the Debye mass, but as well the
dynamics of the ultra-soft gluons at leading logarithmic order. These
findings imply a link beyond one-loop between the present formalism
and a full quantum field theoretical treatment.

A number of possible applications of the formalism to weakly coupled
thermal plasmas are worth being mentioned. We have reviewed the
computation of the colour conductivity to leading logarithmic order.
In principle, it should be possible to extend the analysis to higher
order by solving the dynamical equations for the fluctuations
iteratively.  A further important application concerns colourless
excitations of the plasma. These are responsible for most of the bulk
or hydrodynamical properties of the medium, described by transport
coefficients such as viscosities, electrical or thermal
conductivities. The main contributions to these transport coefficients
arise from hard and soft degrees of freedom.  A computation of
transport coefficients within the present formalism is a feasable
task, bearing in mind the efficiency of the formalism.  Despite the
fact that transport coefficients have already been obtained in the
literature to leading logarithmic order, it is worthwhile to derive
them from the present formalism, and to even extent the existing
results to higher order.

More generally, the formalism leads to an equally good description of
other physical systems, where the relevant thermal excitations can be
described by quasiparticles, and typical length scales associated to
the gauge fields are much larger than those of the quasiparticles. As
an example, we have reviewed an application to the physics of dense
quark matter in a colour superconducting phase with two massless quark
flavours. Based on a semi-classical transport equation for fermionic
quasiparticle excitations, we obtained the hard-superconducting loop
effective action for the $SU(2)$ gauge fields. It describes the
physics of Debye screening and Landau damping for the unbroken
non-Abelian gauge fields in the presence of a condensate. It will be
interesting to use this formalism for the study of transport
coefficients in colour superconducting matter.

All applications of the formalism have been done for weakly coupled
systems close to, or slightly out of, thermal equilibrium. It would be
interesting to understand if a semi-classical description is viable
for strongly coupled plasmas, or for plasmas fully out of equilibrium.
A kinetic description of a plasma requires a small plasma parameter.
For quantum plasmas, the gauge coupling and the plasma parameter are
deeply linked, since a small gauge coupling implies a small plasma
parameter, and vice versa. For classical plasmas, the plasma parameter
remains an independent parameter and can be made small even for large
gauge couplings. This observation may lead to a kinetic description of
strongly coupled classical plasmas. In principle, the formalism also
applies to plasmas out of equilibrium, simply because the ensemble
average does not rely on whether the system is in equlibrium or not.
Hence, the formalism provides an interesting starting point for
applications to out-of-equilibrium plasmas or to the physics of heavy
ion collisions.

\newpage

\acknowledgements

This review grew out of the habilitation thesis (University of
Heidelberg, 2000) by one of us (D.F.L.).  We are deeply indebted to
U.~Heinz and R.~Pisarski for stimulating discussions and for the
continuing support of our work.  We also thank D.~B\"odeker for useful
discussions.  Furthermore, D.F.L.~thanks G.~Aarts, J.~M.~Pawlowski,
M.~G.~Schmidt, P.~Watts and C.~Wetterich for discussions and helpful
comments on the manuscript.  C.M.~is specially grateful to R.~Jackiw
for introducing her to the subject of transport theory, and to her
former collaborators Q.~Liu and C.~Lucchesi.

This work has been supported by the European Community through the
Marie-Curie fellowships HPMF-CT-1999-00404 and HPMF-CT-1999-00391.

\newpage


\setcounter{section}{0}
\renewcommand{\thesection}{\Alph{section}}
\renewcommand{\theequation}{\Alph{section}\arabic{equation}}

\section{Sample computation for correlators}
\label{Evaluation of the collision integral}

In this Appendix we explain in more detail how to compute the
collision integral $\langle \xi \rangle$ which appears in the
transport equation \Eq{soft-mean}. We write the collision
integral in momentum space
\beq
\langle \xi^\rho_a (k,\vv)   \rangle = -g\, f_{abc}
\int \0{d^4p}{(2\pi)^4} \llangle
v^\mu a_{\mu,b}(p)\, \de {\cal J}_c^{\rho}(k-p,\vv) 
\rrangle \ .
\eeq
One first has to solve the dynamical equations for $a$ and $\de {\cal
  J}$, in order to express them in terms of the initial conditions.
This program has been carried out in Section~\ref{BeyondIntegrating},
where $a$ and $\de {\cal J}$ have been solved in a series in $g {\bar
  A}$ and $g {\bar J}$. The correlators are computed analogously.

We begin with the zero order contribution.  It is easy to show that
the first term in the series vanishes.  This is because $\langle
a^{(0,0)}_a a^{(0,0)}_b \rangle \propto \delta_{ab}$, and this
correlator is contracted with the antisymmetric tensor $f_{abc}$.

For the first correction in $g {\bar J}$, one needs to evaluate the
correlators $\langle a^{(0,0)}_b \de {\cal J}_c^{\rho,(0,1)} \rangle$,
and $\langle a^{(0,1)}_b \de {\cal J}_c^{\rho,(0,0)} \rangle$. We will
illustrate how to compute the contribution of the first term. The
second one is computed in a similar way. We need to evaluate
\beq
\label{app2}
 -g\, f_{abc}
\int \0{d^4p}{(2\pi)^4} \llangle
v^\mu a^{(0,0)}_{\mu,b}(p)\, \de {\cal J}_c^{\rho,(0,1)}(k-p,\vv) 
\rrangle \ ,
\eeq
where
\beq
\de {\cal J}_c^{\rho,(0,1)}(k-p,\vv) = -g f_{cde}
\frac{1}{-i (k-p) \cdot v} \int \0{d^4q}{(2\pi)^4}\,
v^\nu a_{\nu,d}^{(0,0)} (q) \bar {\cal J}_e^{\rho} (k-p-q,v) \ .
\eeq
Therefore, we have to evaluate the correlator $\llangle
a^{(0,0)}_{\mu,b}(q)\, a^{(0,0)}_{\nu,d}(p) \rrangle$, which has been
computed in \Eq{basic-corr} for the transverse components of the gauge
fields. These are the ones which give the leading order contribution
to \Eq{app2}. Using the values of \Eq{basic-corr}, and the $SU(N)$
relation $f_{abc} f_{cbe} = - N \delta_{ac}$, we find
\beq
- g^4 N B_c C_2
\int \0{d^4p}{(2\pi)^4}
\int \0{d\Omega_{\bf v'}}{4\pi} 
\left|\0{\vv_i P^T_{ik}(p)\vv_k'}{p^2+\Pi_T}\right|^2
\de(p\cdot v') \frac{\bar {\cal J}_a^{\rho}(k,\vv)}{-i (k-p) \cdot v} \ .
\eeq
Using retarded boundary conditions, we split 
\beq
\frac{1}{-i (k-p) \cdot v}= i {\cal P} \left(\frac{1}{-i (k-p) \cdot v}
\right) + \pi \delta \left( (k-p) \cdot v) \right).
\eeq
The term which goes with the principal value will be neglected,
because it gives a contribution which is damped at asymptotically
large times \cite{Landau10}. In the argument of the $\delta$-function,
we neglect ultrasoft momenta in front of the soft ones ($k \ll p$). We
thus end up with
\beq
- \frac{g^4 N B_c C_2}{4 \pi} v^\rho 
\int \0{d^4p}{(2\pi)^4}
\int \0{d\Omega_{\vv'}}{4\pi} 
\left|\0{\vv_i P^T_{ik}(p)\vv_k'}{p^2+\Pi_T}\right|^2
\, (2 \pi) \,\de(p\cdot v') (2 \pi) \, \de(p \cdot v)
\bar {\cal J}_a^{0}(k,\vv) \,.
\eeq
This is the first term in \Eq{first-exp-coll}.  The second term is
computed in a similar way, after evaluating the $\langle a^{(0,1)}_b
\de {\cal J}_c^{\rho,(0,0)} \rangle$ correlator. Notice that in order
to find a local collision integral, the separation of scales soft and
ultrasoft is a key ingredient.

\section{Correlators of Wigner functions in the classical limit}
\label{Correlators}

We have restricted our study to the use of classical and
semi-classical methods applied to non-Abelian plasmas. In this
Appendix we present a formal justification of the use of the quantum
correlators given in \Eqs{averageB} -- \eq{averageF}.  We follow here
the arguments and reasoning of \cite{Tsytovich:1989}.

To simplify the analysis we will only consider the Abelian case.  The
central quantity for a classical transport theory is the one-particle
distribution function $f$.  This function is split into its mean value
and the fluctuations around it as
\beq
f(x,p) = {\bar f}(x,p) + \delta f(x,p)  \ . 
\eeq 
We consider the  case where the system is homogeneous, 
thus ${\bar f}$ does not depend on $x^\mu$.
We also neglect the effect of
interactions. Then the fluctuations obey the equation
\beq
\label{clas-fluc-eq}
v^\mu \partial_\mu \, \delta f(x,p) = 0 \ ,
\eeq 
where $v^\mu = (1, {\bf v})$ is the particle four velocity.  The
correlation function of fluctuations was deduced in Section
\ref{Ensemble}, when in the Abelian case, the phase space variables
are just $z=({\bf x},\vp)$.

Now we turn to the quantum generalisation of the previous formalism.
The second quantisation representation of $f(p)$ is the particle
occupation number averaged over an statistical ensemble. In a quantum
formulation, the occupation number is given by the operator ${\hat
  a}^\dagger _{\vp} {\hat a}_\vp$ averaged over the vacuum state,
where ${\hat a}^\dagger _\vp$ is the creation operator of a particle
with momentum ${\bf p}$, and ${\hat a}_\vp$ is the annihilation
operator of a particle with momentum ${\bf p}$.  Therefore, we can
identify
\beq
\phi_\vp = \int d^3 p' \, \langle {\hat a}^\dagger _\vp 
{\hat a}_{\vp'} 
\rangle \ , \qquad  \langle {\hat a}^\dagger _\vp {\hat a}_{\vp'} 
\rangle =  \phi_\vp \,\delta^{(3)} ( {\bf p} - {\bf p}') \ .
\eeq
It is useful to introduce the operator
\beq
\label{q-op-1}
{\hat \psi}^{(0)}_\vp (t) = {\hat a}_{\vp} \, e^{-i E_p t} \ ,
\eeq
where $E_p$ is the particle energy. We thus have
\beq
\label{basic-q0-ope}
\phi_\vp = \int d^3 p' \langle {\hat \psi}^{(0) \dagger}_\vp
{\hat \psi}^{(0)}_{\vp'} \rangle \ .
\eeq
The quantum analogue of the classical distribution function is the
Wigner operator, defined as 
\beq
\langle {\hat f} (x,p) \rangle =  \int {d^4 v \, e^{-i p \cdot v} 
\langle \psi^\dagger (x + \frac12 v) \psi(x-\frac 12 v) \rangle}
\eeq
We will work with the Fourier transform of the Wigner operator,
\beq
{\hat f}^{(0)}_{\vp,\vk} (t) = {\hat \psi}^{(0) \dagger}_{\vp- \vk/2} (t)
{\hat \psi}^{(0)}_{\vp+\vk/2} (t) \ .
\label{FT-Wigner}
\eeq
One can define  the fluctuation operator
\beq
\delta {\hat f}^{(0)}_{\vp,\vk} (t) = {\hat f}^{(0)}_{\vp,\vk} (t)
- \langle {\hat f}^{(0)}_{\vp,\vk} (t) \rangle =
{\hat \psi}^{(0) \dagger}_{\vp- \vk/2} (t)
{\hat \psi}^{(0)}_{\vp+\vk/2} (t) - \langle 
{\hat \psi}^{(0) \dagger}_{\vp- \vk/2} (t)
{\hat \psi}^{(0)}_{\vp+\vk/2} (t) \rangle \ .
\eeq
According to the definition (\ref{q-op-1}), this operator obeys
\beq
\label{quan-fluc-eq}
\frac{ \partial}{\partial t} \delta {\hat f}^{(0)}_{\vp,\vk} (t)
+ i \left(E_{p+k/2} -E_{p-k/2} \right) 
\delta {\hat f}^{(0)}_{\vp,\vk} (t) = 0 \ .
\eeq
For $\vk \ll \vp$ we have
\beq
E_{p+k/2} -E_{p-k/2} \simeq {\bf k} \cdot \frac{\partial E_p}{\partial
{\bf p}} = {\bf k} \cdot {\bf v} \ ,
\eeq
and then Eq. (\ref{quan-fluc-eq}) agrees with Eq. (\ref{clas-fluc-eq}).
The solution of Eq. (\ref{quan-fluc-eq}) is
\bea
\delta {\hat f}^{(0)}_{\vp,\vk} (t) &  = & 
\delta {\hat f}^{(0)}_{\vp,\vk} (0)
\exp{\left\{-i \left( E_{p+k/2} -E_{p-k/2} \right) t \right\}} \ , \\
\delta {\hat f}^{(0)}_{\vp,\vk} (0) & = &  {\hat a}^\dagger _{\vp-\vk/2}
{\hat a}_{\vp'+\vk/2} -
\langle {\hat a}^\dagger _{\vp-\vk/2} {\hat a}_{\vp'+\vk/2} 
\rangle \ , \\
\delta {\hat f}^{(0)}_{\vp,\vk, \omega} & = & 
\delta {\hat f}^{(0)}_{\vp,\vk} (0)
\delta \left( \omega - E_{p+k/2} + E_{p-k/2} \right) \ .
\eea
One can now evaluate the correlator of fluctuation operators. 
One finds
\bea
\langle \delta {\hat f}^{(0)}_{\vp,\vk, \omega} 
\delta {\hat f}^{(0)}_{\vp',\vk', \omega'} \rangle & = & 
\delta \left( \omega - E_{p+k/2} + E_{p-k/2} \right)
 \delta \left( \omega' - E_{p'+k'/2} + E_{p'-k'/2} \right) \\
& \times & \left( \langle 
{\hat a}^\dagger_{\vp - \vk/2} {\hat a} _{\vp+\vk/2}
{\hat a}^\dagger_{\vp' -\vk'/2} {\hat a} _{\vp'+\vk'/2} \rangle
- \langle 
{\hat a}^\dagger_{\vp -\vk/2} {\hat a} _{\vp+\vk/2} \rangle
\langle {\hat a}^\dagger_{\vp' -\vk'/2} {\hat a}_{\vp'+\vk'/2} 
\rangle \right)
\ .
\nonumber
\eea
If one  decomposes the average of four
operators into products of the possible averaged values of pairs of
operators, and furthermore one uses the commutation/anticommutation
relations of the creation and annihilation operators, one then arrives to
\bea
\label{quant-corr-B}
\langle \delta {\hat f}^{(0)}_{\vp,\vk, \omega} 
\delta {\hat f}^{(0)}_{\vp',\vk', \omega'} \rangle & = &  
\phi_{\vp- \vk/2} \left(1 - \phi_{\vp + \vk/2} \right) 
\delta^{(3)} ( {\bf p} - {\bf p'}) \delta^{(4)} ( k + k' )
\delta \left(\omega -  E_{p+k/2} + E_{p-k/2} \right) \ ,
\\
\label{quant-corr-F}
\langle \delta {\hat f}^{(0)}_{\vp, \vk, \omega} 
\delta {\hat f}^{(0)}_{\vp', \vk', \omega'} \rangle & = &  
\phi_{\vp- \vk/2} \left(1 + \phi_{\vp +\vk/2} \right) 
\delta^{(3)} ( {\bf p} - {\bf p'}) \delta^{(4)} (  k +  k ')
\delta \left(\omega -  E_{p+k/2} + E_{p-k/2} \right) \ ,
\eea
where $k =( \omega, \vk)$, and Eq. (\ref{quant-corr-B}) refers to the
correlators for particles obeying bosonic statistics, while
Eq.~(\ref{quant-corr-F}) refers to particles obeying fermionic
statistics.  Note that in the limit $\vk \ll \vp$, the above
correlators reduce to
\beq
\label{clas-limi-quan}
\langle \delta {\hat f}^{(0)}_{\vp, \vk, \omega} 
\delta {\hat f}^{(0)}_{\vp', \vk', \omega'} \rangle_{B/F}  =   
\phi_{\vp}^{B/F} \left(1 \mp \phi_{\vp}^{B/F} \right) 
\delta^{(3)} ( {\bf p} - {\bf p'}) \delta^{(4)} (  k +  k ')
\delta \left(\omega -  {\bf k} \cdot {\bf v}\right) \ . 
\eeq
This expression corresponds to the Fourier transform of $\langle
\delta {\hat f}^{(0)}(x,p) \delta {\hat f}^{(0)}(x',p)
\rangle_{t=t'\neq 0}$.  It can be deduced from the initial time
correlators given in \Eqs{averageF} and \eq{averageB}, in the Abelian
limit, if the dynamical evolution of the fluctuations is given by
\Eq{clas-fluc-eq}.  Furthermore, for low occupation numbers, $\phi_\vp
\ll 1$, one recovers the corresponding classical limit.  Note that the
factors of $(2 \pi)$ of difference between \Eq{clas-limi-quan} and
\Eqs{averageB} and \eq{averageF} can be fixed by choosing the proper
normalisation of the momentum measure.

The considerations given above are only valid for free particles.
Modifications are necessary in order to include the effects of
interactions, in which case the time dependence of the operator ${\hat
  \psi}_{\vp} (t)$ will be a more complicated than in
\Eq{basic-q0-ope}.  Furthermore, it should be kept in mind that to
describe the system of relativistic particles in a covariant way, one
should introduce positive and negative energy states. Internal degrees
of freedom, such as spin, are to be introduced as internal indices as
well. The above discussion has also been restricted to the Abelian
case.  It provides a formal justification of why semi-classical
methods can be used to study the physics of specific momentum scales
in the plasma.  With the same tools, we could study the non-Abelian
case as well, only by enlarging the phase-space of the particles.
Additional technical difficulties are then encountered.

\newpage



\begin{thebibliography}{999}

\bibitem{Abbott} L.~F.~Abbott, {\it The Background Field Method Beyond
    One Loop}, Nucl.\ Phys.\ {\bf B185} (1981) 189.
  
\bibitem{Abe:1996yw}
Y.~Abe, S.~Ayik, P.~G.~Reinhard and E.~Suraud,
{\it On stochastic approaches of nuclear dynamics},
Phys.\ Rept.\  {\bf 275} (1996) 49.

\bibitem{AFS} A.~Alekseev, L.~Faddeev and S.~Shatashvili, J.\ Geom.\ 
  Phys.\ {\bf 3} (1989) 1.


\bibitem{Alford:2001dt} M.~G.~Alford, {\it Color superconducting quark
    matter}, hep-ph/0102047.


\bibitem{Alford:1998zt} M.~Alford, K.~Rajagopal and F.~Wilczek, {\it
    QCD at finite baryon density: Nucleon droplets and color
    superconductivity}, Phys.\ Lett.\ {\bf B422} (1998) 247
  [hep-ph/9711395].


\bibitem{Alford:1999mk} M.~Alford, K.~Rajagopal and F.~Wilczek, {\it
    Color-flavor locking and chiral symmetry breaking in high density
    QCD}, Nucl.\ Phys.\ {\bf B537} (1999) 443 [hep-ph/9804403].


\bibitem{Arnold:1999uz} P.~Arnold, {\it An effective theory for
    $\omega \ll k \ll gT$ color dynamics in hot non-Abelian plasmas},
  Phys.\ Rev.\ {\bf D62} (2000) 036003 [hep-ph/9912307].

\bibitem{Arnold:2000dr} P.~Arnold, G.~D.~Moore and L.~G.~Yaffe, {\it
    Transport coefficients in high temperature gauge theories. I:
    Leading-log results}, JHEP {\bf 0011} (2000) 001 [hep-ph/0010177].

\bibitem{Arnold:1999cy} P.~Arnold, D.~T.~Son and L.~G.~Yaffe, {\it
    Effective dynamics of hot, soft non-Abelian gauge fields: Color
    conductivity and $\log(1/\alpha)$ effects}, Phys.\ Rev.\ {\bf D59}
  (1999) 105020 [hep-ph/9810216].

\bibitem{Arnold} P.~Arnold, D.~T.~Son and L.~G.~Yaffe, {\it
    Longitudinal subtleties in diffusive Langevin equations for
    non-Abelian plasmas}, Phys.\ Rev.\ {\bf D60} (1999) 025007
  [hep-ph/9901304].

\bibitem{Arnold:1998gh} P.~Arnold and L.~G.~Yaffe, {\it Effective
    theories for real-time correlations in hot plasmas}, Phys.\ Rev.\ 
  {\bf D57} (1998) 1178 [hep-ph/9709449].


\bibitem{Arnold:1999ux} P.~Arnold and L.~G.~Yaffe, {\it
    Non-perturbative dynamics of hot non-Abelian gauge fields: Beyond
    leading log approximation}, Phys.\ Rev.\ {\bf D62} (2000) 125013
  [hep-ph/9912305].

\bibitem{Arnold:1999uy} P.~Arnold and L.~G.~Yaffe, {\it High
    temperature color conductivity at next-to-leading log order},
  Phys.\ Rev.\ {\bf D62} (2000) 125014 [hep-ph/9912306].



\bibitem{Ayik:1990bb}
S.~Ayik and C.~Gregoire,
{\it Transport Theory Of Fluctuation Phenomena In Nuclear Collisions},
Nucl.\ Phys.\ {\bf A513} (1990) 187.


\bibitem{Bailin:1984bm} D.~Bailin and A.~Love, {\it Superfluidity and
    superconductivity in relativistic fermion systems}, Phys.\ Rept.\ 
  {\bf 107} (1984) 325.



\bibitem{Bak:1994dj} D.~Bak, R.~Jackiw and S.~Pi, {\it NonAbelian
    Chern-Simons particles and their quantization,} Phys.\ Rev.\ {\bf
    D49} (1994) 6778 [hep-th/9402057].


\bibitem{Balachandran:1977ya} A.~P.~Balachandran, P.~Salomonson,
  B.~Skagerstam and J.~Winnberg, {\it Classical Description Of
    Particle Interacting With Nonabelian Gauge Field}, Phys.\ Rev.\ 
  {\bf D15} (1977) 2308.



\bibitem{Balachandran:1978ub} A.~P.~Balachandran, S.~Borchardt and
  A.~Stern, {\it Lagrangian And Hamiltonian Descriptions Of Yang-Mills
    Particles,} Phys.\ Rev.\ {\bf D17} (1978) 3247.



\bibitem{Balescu1960} R.~Balescu, {\it Irreversible processes in
    ionized gases}, Phys.~Fluids {\bf 3} (1960) 52.

\bibitem{Balescu1975} R.~Balescu, {\it Equilibrium and Non-Equilibrium
    Statistical Mechanics}, (Wiley, New York, 1975).

\bibitem{Barducci:1977xq} A.~Barducci, R.~Casalbuoni and L.~Lusanna,
  {\it Classical Scalar And Spinning Particles Interacting With
    External Yang-Mills Fields}, Nucl.\ Phys.\ {\bf B124} (1977) 93.

\bibitem{Barrois:1977xd} B.~C.~Barrois, {\it Superconducting quark
    matter}, Nucl.\ Phys.\ {\bf B129} (1977) 390.

\bibitem{Barrois:1979pv} B.~C.~Barrois, {\it Nonperturbative effects
    in dense quark matter} UMI 79-04847.

\bibitem{Baym:1997gq} G.~Baym and H.~Heiselberg, {\it The electrical
    conductivity in the early universe,} Phys.\ Rev.\ {\bf D56}
  (1997) 5254 [astro-ph/9704214].

\bibitem{Baym:1990uj} G.~Baym, H.~Monien, C.~J.~Pethick and
  D.~G.~Ravenhall, {\it Transverse Interactions And Transport In
    Relativistic Quark - Gluon And Electromagnetic Plasmas,} Phys.\ 
  Rev.\ Lett.\ {\bf 64} (1990) 1867.

\bibitem{Bixon1969} M.~Bixon and R.~Zwanzig, {\it Boltzmann-Langevin
    equation and hydrodynamic fluctuations}, Phys.~Rev.~{\bf 187}
  (1969) 267.

\bibitem{Blaizot:1993zk} J.~P.~Blaizot and E.~Iancu, {\it Kinetic
    equations for long wavelength excitations of the quark - gluon
    plasma}, Phys.\ Rev.\ Lett.\ {\bf 70} (1993) 3376
  [hep-ph/9301236].

\bibitem{Blaizot:1994be} J.~P.~Blaizot and E.~Iancu, {\it Soft
    collective excitations in hot gauge theories}, Nucl.\ Phys.\ {\bf
    B417} (1994) 608 [hep-ph/9306294].

\bibitem{Blaizot:1994da} J.~Blaizot and E.~Iancu, {\it NonAbelian
    excitations of the quark - gluon plasma}, Phys.\ Rev.\ Lett.\ {\bf
    72} (1994) 3317 [hep-ph/9401210].

\bibitem{Blaizot:1994am} J.~Blaizot and E.~Iancu, {\it Energy momentum
    tensors for the quark - gluon plasma}, Nucl.\ Phys.\ {\bf B421}
  (1994) 565 [hep-ph/9401211].

\bibitem{Blaizot:1999xk} J.~Blaizot and E.~Iancu, {\it A Boltzmann
    equation for the {QCD} plasma}, Nucl.\ Phys.\ {\bf B557} (1999)
  183 [hep-ph/9903389].

\bibitem{Blaizot:2000fq} J.~Blaizot and E.~Iancu, {\it Ultrasoft
    amplitudes in hot QCD}, Nucl.\ Phys.\ {\bf B570} (2000) 326
  [hep-ph/9906485].

\bibitem{Blaizot:2001nr} J.~Blaizot and E.~Iancu, {\it The quark-gluon
    plasma: Collective dynamics and hard thermal loops},
  hep-ph/0101103.


\bibitem{Bodeker:1998hm} D.~B\"odeker, {\it On the effective dynamics
    of soft non-abelian gauge fields at finite temperature}, Phys.\ 
  Lett.\ {\bf B426} (1998) 351 [hep-ph/9801430].

\bibitem{Bodeker:2000ud} D.~B\"odeker, {\it Diagrammatic approach to
    soft non-Abelian dynamics at high temperature}, Nucl.\ Phys.\ {\bf
    B566} (2000) 402 [hep-ph/9903478].

\bibitem{Bodeker:1999ey} D.~B\"odeker, {\it From hard thermal loops to
    Langevin dynamics}, Nucl.\ Phys.\ {\bf B559} (1999) 502
  [hep-ph/9905239].

\bibitem{Bodeker:1999} D.~B\"odeker, {\it Effective theories for hot
    non-Abelian dynamics}, in: Fifth Workshop on Quantum
  Chromodynamics, edited by H.M.~Fried, B.~M\"uller and Y.~Gabellini
  (World Scientific, Singapore 2000).


\bibitem{Bodeker:2001pa} D.~B\"odeker, {\it Non-equilibrium field
    theory}, Nucl.\ Phys.\ Proc.\ Suppl.\ {\bf 94} (2001) 61
  [hep-lat/0011077].

\bibitem{Bodeker:2000da} D.~B\"odeker, {\it A local Langevin equation
    for slow long-distance modes of hot non-Abelian gauge fields},
  Phys.\ Lett.\ {\bf B516} (2001) 174 [hep-ph/0012304].

\bibitem{Bodeker:1995pp} D.~B\"odeker, L.~McLerran and A.~Smilga, {\it
    Really computing nonperturbative real time correlation functions},
  Phys.\ Rev.\ {\bf D52} (1995) 4675 [hep-th/9504123].

\bibitem{Bodeker:2000gx} D.~B\"odeker, G.~D.~Moore and K.~Rummukainen,
  {\it Chern-Simons number diffusion and hard thermal loops on the
    lattice}, Phys.\ Rev.\ {\bf D61} (2000) 056003 [hep-ph/9907545].

\bibitem{Braaten:1990kk} E.~Braaten and R.~D.~Pisarski, {\it
    Resummation And Gauge Invariance Of The Gluon Damping Rate In Hot
    QCD}, Phys.\ Rev.\ Lett.\ {\bf 64} (1990) 1338.

\bibitem{Braaten:1990mz} E.~Braaten and R.~D.~Pisarski, {\it Soft
    Amplitudes In Hot Gauge Theories: A General Analysis}, Nucl.\ 
  Phys.\ {\bf B337} (1990) 569.

\bibitem{Braaten:1990it} E.~Braaten and R.~D.~Pisarski, {\it
    Calculation Of The Gluon Damping Rate In Hot QCD}, Phys.\ Rev.\ 
  {\bf D42} (1990) 2156.

\bibitem{Braaten:1990az} E.~Braaten and R.~D.~Pisarski, {\it Deducing
    Hard Thermal Loops From Ward Identities}, Nucl.\ Phys.\ {\bf B339}
  (1990) 310.

\bibitem{Braaten:1992gm} E.~Braaten and R.~D.~Pisarski, {\it Simple
    effective Lagrangian for hard thermal loops}, Phys.\ Rev.\ {\bf
    D45} (1992) 1827.

\bibitem{Braaten:1993jw} E.~Braaten and D.~Segel, {\it Neutrino energy
    loss from the plasma process at all temperatures and densities},
  Phys.\ Rev.\ {\bf D48} (1993) 1478 [hep-ph/9302213].

\bibitem{Brandt:1995mv} F.~T.~Brandt, J.~Frenkel and J.~C.~Taylor,
  {\it High temperature QCD and the classical Boltzmann equation in
    curved space-time}, Nucl.\ Phys.\ {\bf B437} (1995) 433
  [hep-th/9411130].

\bibitem{Brown:1979bv} L.~S.~Brown and W.~I.~Weisberger, {\it Vacuum
    Polarization In Uniform Nonabelian Gauge Fields}, Nucl.\ Phys.\ 
  {\bf B157} (1979) 285.

\bibitem{Brown:1999xq} L.~S.~Brown and L.~G.~Yaffe, {\it Effective
    field theory for quasi-classical plasmas,} Phys.\ Rept.\ {\bf 340}
  (2001) 1 [physics/9911055].

\bibitem{Calzetta:2000xh} E.~Calzetta and B.~L.~Hu, {\it Stochastic
    dynamics of correlations in quantum field theory: From
    Schwinger-Dyson to Boltzmann-Langevin equation}, Phys.\ Rev.\ {\bf
    D61} (2000) 025012 [hep-ph/9903291].

\bibitem{Carrington:2001ms} M.~E.~Carrington, H.~Defu and R.~Kobes,
  {\it Nonlinear response from transport theory and quantum field
    theory at finite temperature}, Phys.\ Rev.\ D {\bf 64} (2001)
  025001 [hep-ph/0102256].

\bibitem{Chan:1995zw} H.~M.~Chan, J.~Faridani and S.~T.~Tsou, {\it
    Equations of motion of Dirac - like topological charges in
    Yang-Mills fields,} Phys.\ Rev.\ {\bf D51} (1995) 7040.

\bibitem{Chomaz:1991yn}
P.~Chomaz, G.~F.~Burgio and J.~Randrup,
{\it Inclusion of fluctuations in nuclear dynamics},
Phys.\ Lett.\ {\bf B254} (1991) 340.

\bibitem{D'Hoker:1996ax} E.~D'Hoker and D.~G.~Gagn\'e, {\it Worldline
    path integrals for fermions with scalar, pseudoscalar and vector
    couplings}, Nucl.\ Phys.\ {\bf B467} (1996) 272 [hep-th/9508131].

\bibitem{D'Hoker:1996bj} E.~D'Hoker and D.~G.~Gagn\'e, {\it Worldline
    Path Integrals for Fermions with General Couplings,} Nucl.\ Phys.\ 
  {\bf B467} (1996) 297 [hep-th/9512080].

\bibitem{Dietrich:2000ex} D.~D.~Dietrich, G.~C.~Nayak and W.~Greiner,
  {\it Phase space description of the quark and gluon production from
    a space-time dependent chromofield,} hep-th/0007139.

\bibitem{Efraty:1992gk} R.~Efraty and V.~P.~Nair, {\it The Secret
    Chern-Simons action for the hot gluon plasma}, Phys.\ Rev.\ Lett.\ 
  {\bf 68} (1992) 2891 [hep-th/9201058].

\bibitem{Efraty:1993pd} R.~Efraty and V.~P.~Nair, {\it Chern-Simons
    theory and the quark - gluon plasma}, Phys.\ Rev.\ {\bf D47}
  (1993) 5601 [hep-th/9212068].

\bibitem{Elze:1990gm} H.~Elze, {\it Transport Equations And QCD
    Collective Modes In A Selfconsistent Covariant Background Gauge},
  Z.\ Phys.\ {\bf C47} (1990) 647.


\bibitem{Elze:1989un} H.~Elze and U.~Heinz, {\it Quark - Gluon
    Transport Theory}, Phys.\ Rept.\ {\bf 183} (1989) 81.

\bibitem{Frenkel:1990br} J.~Frenkel and J.~C.~Taylor, {\it High
    Temperature Limit Of Thermal QCD}, Nucl.\ Phys.\ {\bf B334} (1990)
  199.

\bibitem{Gibbons:1982} J.~Gibbons, D.~D.~Holm and B.~Kupershmidt, {\it
    Gauge Invariant Poisson Bracktes for Chromohydrodynamics}, Phys.\ 
  Lett.\ {\bf A90} (1982) 281.

\bibitem{Greiner:1998vd} C.~Greiner and S.~Leupold, {\it Stochastic
    interpretation of Kadanoff-Baym equations and their relation to
    Langevin processes}, Annals Phys.\ {\bf 270} (1998) 328
  [hep-ph/9802312].

\bibitem{Greiner:1997dx} C.~Greiner and B.~M\"uller, {\it Classical
    Fields Near Thermal Equilibrium}, Phys.\ Rev.\ {\bf D55} (1997)
  1026 [hep-th/9605048].

\bibitem{Grigoriev:1988bd} D.~Y.~Grigoriev and V.~A.~Rubakov, {\it
    Soliton Pair Creation At Finite Temperatures. Numerical Study In
    (1+1)-Dimensions}, Nucl.\ Phys.\ {\bf B299} (1988) 67.

\bibitem{GLW:1980} S.R. de Groot, W.A. van Leeuwen, and Ch.G. van
  Weert, {\it Relativistic Kinetic Theory}, (North Holland, Amsterdam,
  1980).

\bibitem{Gross} D.~J.~Gross, R.~D.~Pisarski and L.~G.~Yaffe, {\it QCD
    And Instantons At Finite Temperature}, Rev.\ Mod.\ Phys.\ {\bf 53}
  (1981) 43.

\bibitem{Heinz:1983nx} U.~Heinz, {\it Kinetic Theory For Nonabelian
    Plasmas}, Phys.\ Rev.\ Lett.\ {\bf 51} (1983) 351.

\bibitem{Heinz:1984my} U.~Heinz, {\it A Relativistic Colored Spinning
    Particle In An External Color Field}, Phys.\ Lett.\ {\bf B144}
  (1984) 228.

\bibitem{Heinz:1985yq} U.~Heinz, {\it Quark - Gluon Transport Theory.
    Part 1. The Classical Theory}, Annals Phys.\ {\bf 161} (1985) 48.

\bibitem{Heinz:1986qe} U.~Heinz, {\it Quark - Gluon Transport Theory.
    Part 2. Color Response And Color Correlations In A Quark - Gluon
    Plasma}, Annals Phys.\ {\bf 168} (1986) 148.

\bibitem{Heinz:1988fg} U.~Heinz, {\it The Approach To Equilibrium In A
    Quark - Gluon Plasma}, Z.\ Phys.\ {\bf C38} (1988) 203.

\bibitem{Heinz:1989cq} U.~Heinz, {\it Nonequilibrium Dynamics In
    Finite Temperature QCD}, Physica {\bf A158} (1989) 111.

\bibitem{Heiselberg:1994vy} H.~Heiselberg, {\it Viscosities of quark -
    gluon plasmas,} Phys.\ Rev.\ {\bf D49} (1994) 4739
  [hep-ph/9401309].

\bibitem{Heiselberg:1994px} H.~Heiselberg, {\it Color, spin and flavor
    diffusion in quark - gluon plasmas,} Phys.\ Rev.\ Lett.\ {\bf 72}
  (1994) 3013 [hep-ph/9401317].

\bibitem{Holm:1984hg} D.~D.~Holm and B.~A.~Kupershmidt, {\it
    Relativistic Chromodynamics And Yang-Mills Vlasov Plasma,} Phys.\ 
  Lett.\ {\bf A105} (1984) 225.

\bibitem{Hosoya:1985xm} A.~Hosoya and K.~Kajantie, {\it Transport
    Coefficients Of QCD Matter}, Nucl.\ Phys.~{\bf B250} (1985) 666.

\bibitem{Hu:1997sf} C.~R.~Hu and B.~M\"uller, {\it Classical lattice
    gauge field with hard thermal loops}, Phys.\ Lett.\ {\bf B409}
  (1997) 377 [hep-ph/9611292].

\bibitem{Huet:1997sh} P.~Huet and D.~T.~Son, {\it Long range physics
    in a hot non-Abelian plasma}, Phys.\ Lett.\ {\bf B393} (1997) 94
  [hep-ph/9610259].

\bibitem{Iancu:1998sg}
E.~Iancu,
{\it Effective theory for real-time dynamics in hot gauge theories},
Phys.\ Lett.\  {\bf B435} (1998) 152.

\bibitem{Jackiw:1993zr} R.~Jackiw and V.~P.~Nair, {\it High
    temperature response functions and the nonAbelian Kubo formula},
  Phys.\ Rev.\ {\bf D48} (1993) 4991 [hep-ph/9305241].

\bibitem{Jalilian-Marian:2001ad} J.~Jalilian-Marian, S.~Jeon and
  R.~Venugopalan, {\it Wong's equations and the small x effective
    action in QCD,} Phys.\ Rev.\ {\bf D63} (2001) 036004
  [hep-ph/0003070].

\bibitem{Jalilian-Marian:1999xt} J.~Jalilian-Marian, S.~Jeon,
  R.~Venugopalan and J.~Wirstam, {\it Minding one's P's and Q's: From
    the one loop effective action in quantum field theory to classical
    transport theory}, Phys.\ Rev.\ {\bf D62} (2000) 045020
  [hep-ph/9910299].

\bibitem{Jeon:1995if} S.~Jeon, {\it Hydrodynamic transport
    coefficients in relativistic scalar field theory}, Phys.\ Rev.\
  {\bf D52} (1995) 3591 [hep-ph/9409250].

\bibitem{Jeon:1996zm} S.~Jeon and L.~G.~Yaffe, {\it From Quantum Field
    Theory to Hydrodynamics: Transport Coefficients and Effective
    Kinetic Theory}, Phys.\ Rev.\ {\bf D53} (1996) 5799
  [hep-ph/9512263].

\bibitem{Johnson:1989qm} K.~Johnson, {\it Functional Integrals For
    Spin}, Annals Phys.\ {\bf 192} (1989) 104.
  
\bibitem{Joyce:1996zt}
M.~Joyce, T.~Prokopec and N.~Turok,
{\it Nonlocal electroweak baryogenesis. Part 2: The Classical regime},
Phys.\ Rev.\ D {\bf 53} (1996) 2958
[hep-ph/9410282].

\bibitem{Kadanoff1962} L.~P.~Kadanoff and G.~Baym, {\it Quantum
    Statistical Mechanics}, (Benjamin, New York, 1962).

\bibitem{Kapusta:1989tk} J.~I.~Kapusta, {\it Finite Temperature Field
    Theory}, (University Press, Cambridge, 1989).

\bibitem{KLLM} P.~F.~Kelly, Q.~Liu, C.~Lucchesi and C.~Manuel, {\it
    Deriving the hard thermal loops of QCD from classical transport
    theory}, Phys.\ Rev.\ Lett.\ {\bf 72} (1994) 3461
  [hep-ph/9403403].

\bibitem{KLLM2} P.~F.~Kelly, Q.~Liu, C.~Lucchesi and C.~Manuel, {\it
    Classical transport theory and hard thermal loops in the quark -
    gluon plasma}, Phys.\ Rev.\ {\bf D50} (1994) 4209
  [hep-ph/9406285].
  
\bibitem{K0} Yu. L. Klimontovich, {\it Kinetic Theory of Nonideal
    Gases and Nonideal Plasmas}, (Pergamon, Oxford, 1982).
  
\bibitem{K} Yu. L. Klimontovich, {\it Statistical Physics}, (Harwood
  Academic, Chur, Switzerland, 1986).
  
\bibitem{K2} Yu. L. Klimontovich, {\it Statistical Theory of Open
    Systems}, (Kluwer Academic Publishers, Dordrecht, 1995).

\bibitem{Kosyakov:1998qi} B.~P.~Kosyakov, {\it Exact solutions in the
    Yang-Mills-Wong theory}, Phys.\ Rev.\ {\bf D57} (1998) 5032
  [hep-th/9902039].

\bibitem{Landau5} L.D.~Landau and E.M.~Lifshitz, {\it Statistiscal
    Physics}, Part 1, 3rd edition, (Pergamon Press, Oxford, 1980).

\bibitem{Landau9} L.D.~Landau and E.M.~Lifshitz, {\it Statistiscal
    Physics}, Part 2, (Pergamon Press, Oxford, 1980).

\bibitem{Landau10} E.M.~Lifshitz and L. P.~Pitaevskii, {\it Physical
    Kinetics}, (Pergamon Press, Oxford, 1981).

\bibitem{LeBellac:1996} M.~Le Bellac, {\it Thermal Field Theory},
  (University Press, Cambridge, 1991).
  
\bibitem{Lenard1960} A.~Lenard, {\it On Bolgoliubov's kinetic equation
    for a spatially homogeneous palsma}, Ann.~Phys.~{\bf 3} (1960)
  390.

\bibitem{Linde:1980tu} A.~D.~Linde, {\it Confinement Of Monopoles At
    High Temperatures: A Solution Of The Primordial Monopole Problem},
  Phys.\ Lett.\ {\bf B96} (1980) 293.

\bibitem{Litim:Habil} D.~F.~Litim, {\it Semi-classical transport
    theory for non-Abelian plasmas}, habilitation thesis (University
  of Heidelberg, May 2000).

\bibitem{Litim:2000uj} D.~F.~Litim, {\it Aspects of semi-classical
    transport theory for QCD}, hep-ph/0010259.

\bibitem{LM} D.~F.~Litim and C.~Manuel, {\it Mean field dynamics in
    non-Abelian plasmas from classical transport theory}, Phys.\ Rev.\ 
  Lett.\ {\bf 82} (1999) 4981 [hep-ph/9902430].

\bibitem{LM2} D.~F.~Litim and C.~Manuel, {\it Effective transport
    equations for non-Abelian plasmas}, Nucl.\ Phys.\ {\bf B562}
  (1999) 237 [hep-ph/9906210].

\bibitem{LM3} D.~F.~Litim and C.~Manuel, {\it Fluctuations from
    dissipation in a hot non-Abelian plasma}, Phys.\ Rev.\ {\bf D61}
  (2000) 125004 [hep-ph/9910348].

\bibitem{LM4} D.~F.~Litim and C.~Manuel, {\it Deriving effective
    transport equations for non-Abelian plasmas}, hep-ph/0003302.


\bibitem{Litim:2001je} D.~F.~Litim and C.~Manuel, {\it Transport
    theory for a two-flavor color superconductor}, Phys.\ Rev.\ Lett.\ 
  {\bf 87} (2001) 052002 [hep-ph/0103092].


\bibitem{Litim:2001mv}
D.~F.~Litim and C.~Manuel,
{\it Photon self-energy in a color superconductor},
Phys.\ Rev.\ {\bf D64} (2001) 094013
[hep-ph/0105165].

\bibitem{Manuel:1996td}
C.~Manuel,
{\it Hard Dense Loops in a Cold Non-Abelian Plasma},
Phys.\ Rev.\  {\bf D53} (1996) 5866
[hep-ph/9512365].

\bibitem{Manuel:1998is} C.~Manuel, {\it Magnetic screening at finite
    temperature}, Annals Phys.\ {\bf 263} (1998) 238 [hep-ph/9612494].

\bibitem{Markov} Y.~A.~Markov and M.~A.~Markova, {\it The
    Balescu-Lenard collision term for a quark plasma: Classical
    model}, Theor.\ Math.\ Phys.\ {\bf 103} (1995) 444.

\bibitem{Moore:1995si} 
G.~D.~Moore and T.~Prokopec, 
{\it How fast can the wall move? A Study of the
electroweak phase transition dynamics},
Phys.\ Rev.\ D {\bf 52} (1995) 7182
[hep-ph/9506475]. 

\bibitem{Moore:1998sn} G.~D.~Moore, C.~Hu and B.~M\"uller, {\it
    Chern-Simons number diffusion with hard thermal loops}, Phys.\ 
  Rev.\ {\bf D58} (1998) 045001 [hep-ph/9710436].

\bibitem{Moore:2000zk} G.~D.~Moore, {\it The sphaleron rate:
    B\"odeker's leading log}, Nucl.\ Phys.\ {\bf B568} (2000) 367
  [hep-ph/9810313].

\bibitem{Mrowczynski:1989np}
S.~Mr\'owczy\'nski,
{\it Kinetic Theory Approach To Quark - Gluon Plasma Oscillations},
Phys.\ Rev.\  {\bf D39} (1989) 1940.

\bibitem{Nair:1993rx} V.~P.~Nair, {\it Hard thermal loops, gauged WZNW
    action and the energy of hot quark - gluon plasma}, Phys.\ Rev.\ 
  {\bf D48} (1993) 3432 [hep-ph/9307326].

\bibitem{Nair:1994xs} V.~P.~Nair, {\it Hamiltonian analysis of the
    effective action for hard thermal loops in QCD}, Phys.\ Rev.\ {\bf
    D50} (1994) 4201 [hep-th/9403146].

\bibitem{Nayak:1997ex} G.~C.~Nayak and V.~Ravishankar, {\it
    Pre-equilibrium evolution of non-abelian plasma}, Phys.\ Rev.\ 
  {\bf D55} (1997) 6877 [hep-th/9610215].

\bibitem{Nayak:1998kp} G.~C.~Nayak and V.~Ravishankar, {\it
    Pre-equilibrium evolution of quark-gluon plasma}, Phys.\ Rev.\ 
  {\bf C58} (1998) 356 [hep-ph/9710406].

\bibitem{Pisarski:1989vd} R.~D.~Pisarski, {\it Scattering Amplitudes
    In Hot Gauge Theories}, Phys.\ Rev.\ Lett.\ {\bf 63} (1989) 1129.

\bibitem{damp-rat-P} R.~D.~Pisarski, {\it Damping rates for moving
    particles in hot QCD}, Phys.\ Rev.\ {\bf D47} (1993) 5589.

\bibitem{Pisarski:1997cp} R.~D.~Pisarski, {\it Nonabelian Debye
    screening, tsunami waves, and worldline fermions}, hep-ph/9710370.

\bibitem{Pisarski:2000bf} R.~D.~Pisarski and D.~H.~Rischke, {\it Gaps
    and critical temperature for color superconductivity}, Phys.\ 
  Rev.\ {\bf D61} (2000) 051501 [nucl-th/9907041].

\bibitem{Rajagopal:2000wf} K.~Rajagopal and F.~Wilczek, {\it The
    Condensed Matter Physics of QCD,} hep-ph/0011333.

\bibitem{Randrup:1990bb}
J.~Randrup and B.~Remaud,
{\it Fluctuations In One Body Dynamics},
Nucl.\ Phys.\ {\bf A514} (1990) 339.

\bibitem{Rapp:1998zu} R.~Rapp, T.~Schafer, E.~V.~Shuryak and
  M.~Velkovsky, {\it Diquark Bose condensates in high density matter
    and instantons}, Phys.\ Rev.\ Lett.\ {\bf 81} (1998) 53
  [hep-ph/9711396].

\bibitem{Rischke:2000qz} D.~H.~Rischke, {\it Debye screening and
    Meissner effect in a two-flavor color superconductor}, Phys.\ 
  Rev.\ {\bf D62} (2000) 034007 [nucl-th/0001040].

\bibitem{Rischke:2000ra} D.~H.~Rischke, {\it Debye screening and
    Meissner effect in a three-flavor color superconductor}, Phys.\ 
  Rev.\ {\bf D62} (2000) 054017 [nucl-th/0003063].

\bibitem{Rischke:2000cn} D.~H.~Rischke, D.~T.~Son and M.~A.~Stephanov,
  {\it Asymptotic deconfinement in high-density QCD}, Phys.\ Rev.\ 
  Lett.\ {\bf 87} (2001) 062001 [hep-ph/0011379].

\bibitem{Schrieffer} J.~R.~Schrieffer, {\it Theory of
    Superconductivity}, (W.A. Benjamin, New York, 1964).

\bibitem{Selikhov} A.~V.~Selikhov, {\it Collision terms for QGP of
    Lenard-Balescu type}, Phys.\ Lett.\ {\bf B268} (1991) 263, Erratum
  Phys.~Lett.~{\bf B285} (1992) 398.

\bibitem{Selikhov2} A.~Selikhov and M.~Gyulassy, {\it Color diffusion
    and conductivity in a quark - gluon plasma}, Phys.\ Lett.\ {\bf
    B316} (1993) 373 [nucl-th/9307007].

\bibitem{SG} A.~V.~Selikhov and M.~Gyulassy, {\it QCD Fokker-Planck
    equations with color diffusion}, Phys.\ Rev.\ {\bf C49} (1994)
  1726.

\bibitem{Silin:1960} V.~P.~Silin, {\it On the electromagnetic
    properties of a relativistic plasma} Sov.~Phys.~JETP {\bf 11}
  (1960) 1136.

\bibitem{Sitenko1982} A.~G.~Sitenko, {\it Fluctuations and Non-Linear
    Wave Interactions in Plasmas}, (Pergamon, Oxford, 1982).

\bibitem{Smilga:1997cm} A.~V.~Smilga, {\it Physics of thermal QCD},
  Phys.\ Rept.\ {\bf 291} (1997) 1 [hep-ph/9612347].

\bibitem{Son:1999uk} D.~T.~Son, {\it Superconductivity by long-range
    color magnetic interaction in high-density quark matter}, Phys.\ 
  Rev.\ {\bf D59} (1999) 094019 [hep-ph/9812287].

\bibitem{Strassler:1992zr} M.~J.~Strassler, {\it Field theory without
    Feynman diagrams: One loop effective actions,} Nucl.\ Phys.\ {\bf
    B385} (1992) 145 [hep-ph/9205205].

\bibitem{Taylor:1990ia} J.~C.~Taylor and S.~M.~Wong, {\it The
    Effective Action Of Hard Thermal Loops In QCD}, Nucl.\ Phys.\ {\bf
    B346} (1990) 115.

\bibitem{Tsytovich:1989} V.~N.~Tsytovich, {\it Radiative-resonant
    collective wave-particle interactions}, Phys.~Rept.~{\bf 178}
  (1989) 261.

\bibitem{Valle} M.~A.~Valle Basagoiti, {\it Collision terms from
    fluctuations in the HTL theory for the quark-gluon plasma},
  hep-ph/9903462.

\bibitem{Vija:1995is} H.~Vija and M.~H.~Thoma, {\it Braaten-Pisarski
    method at finite chemical potential}, Phys.\ Lett.\ {\bf B342}
  (1995) 212 [hep-ph/9409246].

\bibitem{Winter} J.~Winter, {\it Covariant extension of the Wigner
    transformation to non-Abelian Yang-Mills symmetries for a Vlasov
    equation approach to the quark-gluon plasma}, J.~Phys.~(Paris)
  {\bf 45} (1984) C6-53.

\bibitem{Wong} S.~K.~Wong, {\it Field And Particle Equations For The
    Classical Yang-Mills Field And Particles With Isotopic Spin},
  Nuovo Cim.\ {\bf 65A} (1970) 689.

\bibitem{Zhang:1996nw} X.~Zhang and J.~Li, {\it Non-Abelian Kubo
    formula and the multiple time-scale method}, Annals Phys.\ {\bf
    250} (1996) 433 [hep-th/9605120].

\bibitem{Zinn-Justin:1989mi} J.~Zinn-Justin, {\it Quantum Field Theory
    And Critical Phenomena}, (Clarendon, Oxford, 1989).

\end{thebibliography}
\end{document}